\documentclass[english]{article}
\usepackage[T1]{fontenc}
\usepackage[latin9]{inputenc}
\usepackage{color}
\usepackage{mathrsfs}
\usepackage{amsmath}
\usepackage{amssymb}
\usepackage{esint}

\makeatletter
\newcommand{\lyxaddress}[1]{
	\par {\raggedright #1
	\vspace{1.4em}
	\noindent\par}
}

\usepackage{fullpage}

\makeatother

\usepackage{babel}
\begin{document}
\title{General relativity as a biconformal gauge theory}
\author{James T. Wheeler$^{\dagger}$}
\maketitle
\begin{abstract}
We consider the conformal group of a space of dim n = p + q, with
SO(p,q) metric. The quotient of this group by its homogeneous Weyl
subgroup gives a principal fiber bundle with 2n-dim base manifold
and Weyl fibers. The Cartan generalization to a curved 2n-dim geometry
admits an action functional linear in the curvatures. Because symmetry
is maintained between the translations and the special conformal transformations
in the construction, these spaces are called biconformal; this same
symmetry gives biconformal spaces overlapping structures with double
field theories, including manifest T-duality. We establish that biconformal
geometry is a form of double field theory, showing how general relativity
with integrable local scale invariance arises from its field equations.
While we discuss the relationship between biconformal geometries and
the double field theories of T-dual string theories, our principal
interest is the study of the gravity theory. We show that vanishing
torsion and vanishing co-torsion solutions to the field equations
overconstrain the system, implying a trivial biconformal space. Wih
co-torsion unconstrained, we show that (1) the torsion-free solutions
are foliated by copies of an n-dim Lie group, (2) torsion-free solutions
generically describe locally scale-covariant general relativity with
symmetric, divergence-free sources on either the co-tangent bundle
of n-dim (p,q)-spacetime or the torus of double field theory, and
(3) torsion-free solutions admit a subclass of spacetimes with n-dim
non-abelian Lie symmetry. These latter cases include the possibility
of a gravity-electroweak unification. It is notable that the field
equations reduce all curvature components to dependence only on the
solder form of an n-dim Lagrangian submanifold, despite the increased
number of curvature components and doubled number of initial independent
variables.
\end{abstract}

\lyxaddress{Keywords: general relativity, conformal, biconformal, scale invariant
general relativity, double field theory, T-duality, string theory}

$^{\dagger}$James T Wheeler, Utah State University Department of
Physics, 4415 Old Main Hill, Logan, UT 84322-4415, jim.wheeler@usu.edu

\tableofcontents{}

\newpage{}

\section{Introduction}

\subsection{Biconformal spaces and the biconformal action}

It was shown in the 1950s and 1960s that general relativity may be
cast as a Lorentz \cite{Utiyama} or Poincaré \cite{Kibble1} gauge
theory. Subsequent approaches \cite{Dirac,MacDowellMansouri,Neeman,ReggeN,IvanovI,Ivanov,NCG}
refined the methods and extended the initial symmetry to Weyl, deSitter
and conformal. A systematic approach to the resulting gauge theories
of gravity shows that it is possible to formulate general relativity
in several ways \cite{WheelerGaugeTheories}.

\subsubsection{Conformally based theories of gravity}

Generally, the use of conformal symmetries for gravity theories (and
in the MacDowell-Mansouri case, de Sitter) leads to actions functionals
which are quadratic in the curvature and apply only to 4-dimensional
spacetimes. This is because conformal scaling by $e^{\phi}$ changes
the volume form by $e^{n\phi}$ in $n$-dimensions. In four dimensions
this factor may be offset by two factors of the curvature, but in
even dimension $n=2k$, we require $k$ factors of the curvature to
make the action dimensionless. Various techniques allow these quadratic
theories to nonetheless reduce to general relativity \cite{MacDowellMansouri,Ivanov,NCG,WW}.
The most studied case is that of Weyl (conformal) gravity, in which
the difficulty of higher order field equations has been alternatively
exploited and overcome. When only the metric is varied, the field
equations are fourth order, and include solutions not found in general
relativity \cite{Schmidt,Smigla,SultanaKazanasEtAl}. Mannheim \cite{Mannheim}
attempts to use the additional scaling properties to explain galactic
rotation curves (but also see \cite{Flanagan}). Alternatively, it
has been shown in \cite{Weyl grav as GR} that by reformulating Weyl
gravity as a gauge theory and varying all of the gauge fields, the
additional field equations give the integrability conditions needed
to reduce the order and exactly reproduce locally scale invariant
general relativity.

There are two exceptions to these higher-order curvature requirements,
which have actions linear in the curvature and can be formulated in
any dimension. In the first of the two approaches, Dirac \cite{Dirac}
builds on previous work with scalar-tensor theories \cite{Jordan,Fierz,Brans Dicke,Brans},
achieving a curvature-linear action by including a scalar field to
help offset the scaling of the volume form. Dirac considers a Weyl
geometry in which the curvature is coupled to the Weyl vector and
a scalar field. Generalizing his action functional to $n$-dimensions,
it takes the form
\[
S=\int\left(\kappa^{2}g^{\alpha\beta}R_{\alpha\beta}-\frac{\beta}{2}\kappa^{2\left(\frac{n-4}{n-2}\right)}g^{\alpha\mu}g^{\beta\nu}\Omega_{\alpha\beta}\Omega_{\mu\nu}-\frac{4\left(n-1\right)}{n-2}g^{\alpha\beta}D_{\alpha}\kappa D_{\beta}\kappa-\lambda\kappa^{\frac{2n}{n-2}}\right)\sqrt{-g}d^{n}x
\]
where $\kappa$ is a scalar field of conformal weight $w_{\kappa}=-\frac{n-2}{2}$
and $\Omega_{\alpha\beta}=W_{\alpha,\beta}-W_{\beta,\alpha}$ is the
dilatational curvature. The only occurrence of the Weyl vector is
in the dilatational curvature, $\Omega_{\alpha\beta}$, which must
vanish or be unmeasurably small to avoid unphysical size changes.
Dirac interpreted $\Omega_{\alpha\beta}$ as the electromagnetic field
and the time dependence of the scalar field $\kappa$ as a time-dependent
gravitational constant. The electromagnetic interpretation is untenable,
but solutions with $\Omega_{\alpha\beta}=0$ or with other interpretations
remain to be explored.

The second curvature-linear action arises when the volume element
is like that of a phase space, since the ``momentum'' directions
have opposite conformal weight from the ``space'' dimensions. Such
a space arises naturally from the quotient of the conformal group
by its Weyl subgroup ($SO\left(p,q\right)$ and dilatations), with
the most general curvature-linear action built from the $SO\left(p,q\right)$
and dilatational curvatures \cite{WW} being
\begin{equation}
S=\int e_{ac\cdots d}^{\quad\quad be\cdots f}\left(\alpha\boldsymbol{\Omega}_{\;\;b}^{a}+\beta\delta_{\;b}^{a}\boldsymbol{\Omega}+\gamma\mathbf{e}^{a}\wedge\mathbf{f}_{b}\right)\land\mathbf{e}^{c}\land\cdots\land\mathbf{e}^{d}\land\mathbf{f}_{e}\land\cdots\land\mathbf{f}_{f}\label{Action}
\end{equation}
Here, $\boldsymbol{\Omega}_{\;\;b}^{a}$ is the curvature of the $SO\left(p,q\right)$
gauge field, $\boldsymbol{\Omega}$ is the dilatational curvature,
and $\alpha,\beta$ and $\gamma$ are dimensionless constants. The
differential forms $\mathbf{e}^{a}$ and $\mathbf{f}_{a}$ are the
gauge fields of translations and special conformal transformations,
respectively. Together the latter give an orthonormal frame field
on a $2n$-dimensional manifold. Because of the symmetry maintained
between the translations of the conformal group and the special conformal
transformations, these spaces are called \emph{biconformal}. Here
we explore the large class of torsion-free biconformal spaces, showing
that they reduce to general relativity on Lagrangian submanifolds.
It is the consequences of the action, Eq.(\ref{Action}) that will
occupy our present inquiry. 

\subsubsection{The biconformal gauging}

The construction of a biconformal space begins with a flat, $n$-dimensional
space with $SO\left(p,q\right)$ invariant metric, $p+q=n$. Compactifying
this space by adding an appropriate point or null cones at infinity
(See Appendix A) allows us to define its conformal group, $SO\left(p+1,q+1\right)$.
The biconformal quotient \cite{Ivanov,NCG} is then $SO\left(p+1,q+1\right)/SO(p,q)\times SO\left(1,1\right)$,
where $SO\left(1,1\right)$ transformations represent dilatations
and the full subgroup $\mathcal{W}\equiv SO(p,q)\times SO\left(1,1\right)$
is the homogeneous Weyl group. The quotient gives rise to a principal
fiber bundle with $2n$ dimensional base manifold and homogeneous
Weyl fibers. The connection of this \emph{flat biconformal space }is
then generalized, giving rise to conformal Lie algebra-valued $2$-form
curvatures. These are required to be horizontal and the resulting
Cartan equations integrable. The $2n$-dimensional curved base manifolds
are \emph{biconformal spaces} while local $SO\left(p,q\right)$ and
dilatational invariance remain, together comprising the \emph{biconformal
bundle.}

\emph{Biconformal gravity} is the gravity theory following from variation
of the action, Eq.(\ref{Action}), with respect to each of the conformal
gauge fields, together with the Cartan structure equations to define
the curvatures in terms of the connection, and the generalized Bianchi
identities arising as integrability conditions. The construction of
these models is described in full detail in \cite{Hazboun Wheeler}.
See also \cite{Ivanov,NCG,WW,Hazboun dissertation}.

\subsection{Relationship to double field theory}

Biconformal spaces share many features in common with double field
theories.

Double field theory is a means of making the $O\left(d,d\right)$
symmetry of $T$-duality manifest. By introducing scalars to produce
an additional $d$ dimension, Duff \cite{Duff} doubled the $X(\sigma,\tau)$
string variables to make this $O\left(d,d\right)$ symmetry manifest.
Siegel brought the idea to full fruition by deriving results from
superstring theory \cite{Siegel1,Siegel2,Siegel3}. Allowing fields
to depend on all $2d$ coordinates, Siegel introduced generalized
Lie brackets, gauge transformations, covariant derivatives, and a
section condition on the full doubled space, thereby introducing torsions
and curvatures in addition to the manifest T-duality.

There has been substantial subsequent development. Much of this development
is reviewed in \cite{Pedagogical double field theory}; the introduction
to \cite{Berman Blair and Perry} gives a concise summary. Briefly,
double field theory arises by making $T$-duality manifest in string
theory. When we compactify $n$ dimensions on a torus, the windings
of string about the torus can be interpreted as momenta. $T$-duality
is a mapping between the original spatial directions and these momenta.
Double field theory arises when these two $n$-spaces are kept present
simultaneously, making $T$-duality manifest and leading to an overall
$O\left(n,n\right)$ symmetry. The $T$-duality is identified with
the Weyl group of $O\left(n,n\right)$, consisting of permutations
of the distinct circles of the maximum torus and interchange of phases.

\subsubsection{Invariant tensors in double field theory and biconformal spaces}

In double field theory, doubled coordinates are introduced, extending
the spacetime coordinate $x^{\alpha}$ by an equal number of momenta,
\[
x^{M}=\left(\begin{array}{c}
x^{\alpha}\\
y_{\beta}
\end{array}\right)
\]
where $M,N,\cdots=1,\cdots,2n$ and $\alpha,\beta,\cdots=1,\cdots,n$
are coordinate indices. There are at least two important invariant
tensors identified in \cite{Berman Blair and Perry}. Defining the
$O\left(n,n\right)$ symmetry, there is the $2n\times2n$ quadratic
form
\[
K_{AB}=\left(\begin{array}{cc}
0 & \delta_{b}^{a}\\
\delta_{a}^{b} & 0
\end{array}\right)
\]
with $A,B,\cdots,L=1,\cdots,2n$ and lower case Latin indices, $a,b,\cdots=1,\cdots,n$
orthonormal. The second invariant object is the spacetime/dual generalized
metric, $M_{ab}$, built from the spacetime metric and the Kalb-Ramond
potential,  which takes the orthonormal form
\[
M_{AB}=\left(\begin{array}{cc}
\eta_{ab} & 0\\
0 & \eta^{ab}
\end{array}\right)
\]
where $\eta_{ab}$ is either Euclidean or Lorentzian, depending on
the model considered.

These are only half the invariant structures in biconformal spaces,
all of which arise from natural invariances of the conformal group.
Again letting $\eta_{ab}$ be Euclidean or Lorentzian (or in our main
development, any $\left(p,q\right)$ metric), we make use of the Killing
form of the conformal group of a compactified $\left(p,q\right)$
space,
\[
K_{\Sigma\Delta}=\left(\begin{array}{cccc}
\delta_{b}^{a}\delta_{d}^{c}-\eta^{ac}\eta_{bd} & 0 & 0 & 0\\
0 & 0 & \delta_{b}^{a} & 0\\
0 & \delta_{a}^{b} & 0 & 0\\
0 & 0 & 0 & 1
\end{array}\right)
\]
where the upper left block is the norm on Lorentz or Euclidean transformations,
the next $n$ rows and columns arise from tranlations and the next
$n$ from special conformal transformations. The final $1$ in the
lower right gives the Killing norm on dilatations. Upper case Greek
indices run over the dimension of the conformal group. When $K_{\Sigma\Delta}$
is restricted to the biconformal manifold, we have only the translations
and special conformal portion, and this is precisely the $O\left(n,n\right)$
metric,
\[
K_{AB}=\left(\begin{array}{cc}
0 & \delta_{b}^{a}\\
\delta_{a}^{b} & 0
\end{array}\right)
\]
Use of the Killing form as metric was first mentioned in \cite{NCG}
with explicit use in biconformal spaces in \cite{AWQM,Spencer Wheeler,Hazboun Wheeler}
where the orthonormal basis $\left(\mathbf{e}^{a},\mathbf{f}_{b}\right)$
is taken to satisfy
\begin{eqnarray}
\left\langle \mathbf{e}^{a},\mathbf{e}^{b}\right\rangle  & = & 0\label{Inner product ee}\\
\left\langle \mathbf{e}^{a},\mathbf{f}_{b}\right\rangle  & = & \delta_{b}^{a}\label{Inner product ef}\\
\left\langle \mathbf{f}_{a},\mathbf{f}_{b}\right\rangle  & = & 0\label{Inner product ff}
\end{eqnarray}
General linear changes of the original $\left(\mathbf{e}^{a},\mathbf{f}_{b}\right)$
basis are allowed,
\begin{eqnarray}
\boldsymbol{\chi}^{a} & = & A_{\;\;\;b}^{a}\mathbf{e}^{b}+B^{ab}\mathbf{f}_{b}\label{Change of basis solder}\\
\boldsymbol{\psi}_{a} & = & C_{ab}\mathbf{e}^{b}+D_{a}^{\;\;\;b}\mathbf{f}_{b}\label{Change of basis cosolder}
\end{eqnarray}
These become $O\left(n,n\right)$ transformations when they are required
to preserve the inner product given in Eqs.(\ref{Inner product ee})
- (\ref{Inner product ff}). These basis forms $\left(\mathbf{\chi}^{a},\boldsymbol{\psi}_{b}\right)$
are local, but may be defined globally when the structure equations
and field equations provide an appropriate involution. Such alternative
choices of basis have been explored in \cite{Spencer Wheeler,Hazboun Wheeler,LoveladyWheeler}.

There are further objects, discussed in detail in \cite{Spencer Wheeler}
and more comprehensively in \cite{Hazboun Wheeler}, where it is shown
that there exists a Kähler structure on biconformal space. The complex
structure arises from the symmetry of the conformal Lie algebra given
by interchanging translation and special conformal transformation
generators and changing the sign of the dilatation generator. This
is essentially an inversion, and when carried through to its effect
as a linear operation on the basis forms, may be written as
\begin{eqnarray*}
J_{\;\;\;B}^{A} & = & \left(\begin{array}{cc}
0 & -\eta^{ab}\\
\eta_{ab} & 0
\end{array}\right)
\end{eqnarray*}
Further, it has long been recognized that the Maurer-Cartan equation
of dilatations and, generically, its Cartan generalization describe
a symplectic form,
\begin{eqnarray*}
\mathbf{d}\boldsymbol{\omega} & = & \mathbf{e}^{a}\wedge\mathbf{f}_{a}
\end{eqnarray*}
The symplectic character is manifest since the left side shows the
$2$-form to be closed while the right shows it to be non-degenerate.
As a matrix in this basis, the symplectic form is
\[
S_{AB}=\left(\begin{array}{cc}
0 & \delta_{b}^{a}\\
-\delta_{a}^{b} & 0
\end{array}\right)
\]
The two of these may be used to define a Kähler metric via
\begin{eqnarray*}
M_{AB} & \equiv & S_{AC}J_{\;\;\;B}^{C}\\
 & = & \left(\begin{array}{cc}
0 & \delta_{a}^{c}\\
-\delta_{c}^{a} & 0
\end{array}\right)\left(\begin{array}{cc}
0 & -\eta^{cb}\\
\eta_{cb} & 0
\end{array}\right)\\
 & = & \left(\begin{array}{cc}
\eta_{ab} & 0\\
0 & \eta^{ab}
\end{array}\right)
\end{eqnarray*}
which is exactly the $M_{AB}$ of double field theory. All of these
objects arise from properties of the conformal group. We note that
the Killing metric $K_{AB}$ is \emph{not} the metric defined by the
almost Kähler structure.

The change of basis of Eq.(\ref{Change of basis solder}) and Eq.(\ref{Change of basis cosolder})
will be restricted further depending on which of these objects the
change is required to preserve. For example, the time theorem of \cite{Spencer Wheeler}
requires invariance of the inner product, Eqs.(\ref{Inner product ee})
- (\ref{Inner product ff}), and preservation of the symplectic form,
$S_{AB}$, reducing the allowed change of basis to the form
\[
\left(\begin{array}{cc}
A & B\\
C & D
\end{array}\right)=\left(\begin{array}{cc}
A & 0\\
0 & \left(A^{t}\right)^{-1}
\end{array}\right)
\]
This is simply an instance of spontaneous symmetry breaking. Solutions
typically do not preserve the full symmetry of a system of equations.

\subsubsection{Connection and action}

The principal differences between the usual treatment of double field
theories and biconformal gravity lie in the connection, the action,
and the means by which the doubled dimension is reduced back to an
$n$-dimensional spacetime.

There have been multiple proposals for a connection in dual field
theory \cite{Berman Blair and Perry}, including the Weitzenböch connection,
$\Gamma_{\;\;\;BC}^{A}=e_{M}^{\;\;\;A}\partial_{B}e_{C}^{\;\;\;M}$.
While this is compatible with the double field theory structures,
it leads to vanishing curvature and nonvanishing torsion. Even its
generalization has vanishing curvature. Constructing an action becomes
problematic. One proposal, given in \cite{Berman Blair and Perry},
is an action on the full doubled space given by
\begin{eqnarray*}
S & = & \int dxdye^{-2d}L
\end{eqnarray*}
where $d=\Phi-\frac{1}{2}\ln\left(\left|\deg g_{ij}\right|\right)$
generalizes the dilaton $\Phi$ and $L$ is given by 
\begin{eqnarray*}
L & = & \frac{1}{8}M^{AB}\partial_{A}M^{CD}\partial_{B}M_{CD}-\frac{1}{2}M^{AB}\partial_{A}M^{CD}\partial_{C}M_{BD}+4M^{AB}\partial_{A}\partial_{B}d-2\partial_{A}\partial_{B}M^{AB}\\
 &  & -4M^{AB}\partial_{A}d\partial_{B}d+4\partial_{A}M^{AB}\partial_{B}d+\frac{1}{2}\eta^{AB}\eta^{CD}\partial_{A}e_{C}^{\;\;\;M}\partial_{B}e_{DM}
\end{eqnarray*}

There are also multiple proposals in double field theory for finding
a condition that will reduce the full space back to an $n$-dimensional
spacetime. One proposal \cite{Scherk and Schwarz}, due to Scherk
and Schwarz, proposes requiring the functional dependence of fields
to be of the form 
\[
V_{\;\;\;B}^{A}\left(X,Y\right)=\left(W^{-1}\right)_{\;\;\;\hat{A}}^{A}\left(Y\right)W_{\;\;\;B}^{\hat{B}}\left(Y\right)\left[\hat{V}\left(x\right)\right]_{\;\;\;\hat{B}}^{\hat{A}}
\]
with the scalar field additively separable, $d\left(X,Y\right)=\hat{d}\left(X\right)+\lambda\left(Y\right)$.
Here the hatted indices, $\hat{A}$, refer to the gauged double field
theory while unhatted $A,B$ are associated to the double field theory
before applying the Scherk-Schwarz reduction. Alternatively, Berman
et al. \cite{Berman Blair and Perry} propose additive separability
with the ``section condition'' $\eta^{AB}\partial_{A}\partial_{B}=0$
acting on all fields.

Faced with these divergent approaches, the authors of \cite{Berman Blair and Perry}
summarize a set of desirable properties for a connection on double
field theory. We quote (replacing their notation with ours and numbering
the points for convenient reference below):
\begin{quotation}
``. . . we might want it to
\end{quotation}
\begin{enumerate}
\item define a covariant derivative that maps generalised tensors into generalised
tensors,
\item be compatible with the generalised metric $M_{AB}$,
\item be compatible with the $O\left(n,n\right)$ structure $K_{AB}$,
\item be completely determined in terms of the physical fields, in particular
the vielbein and its derivatives,
\item be torsion-free,
\item lead to a curvature that may be contracted with the metric to give
the scalar which appears in the action.''
\end{enumerate}
Their proposal satisfies conditions $1-4$.

The situation is quite different in biconformal geometry because it
has been developed first as a gravity theory, and all the relevant
structures are present from the Cartan construction. In particular,
the connection is automatically given by the $SO\left(p,q\right)$
spin connection and the Weyl vector, and these are compatible with
not only the generalized metric (the biconformal Kähler metric) $M_{AB}$
and the $O\left(n,n\right)$ structure $K_{AB}$ present as the restricted
conformal Killing form, but also the almost complex structure $J_{\;\;\;B}^{A}$
and symplectic form $S_{AB}$. This satisfies points $1,2$ and $3$.

While an $O\left(n\right)$ rather than an $O\left(n,n\right)$ connection
may seem restrictive, $O\left(n,n\right)$ transformations of the
orthonormal basis still retain the larger symmetry. Moreover, the
spin connection and Weyl vector start as general $1$-forms on a $2n$-dimensional
space, 
\begin{eqnarray}
\boldsymbol{\omega}_{\;\;\;b}^{a} & = & \omega_{\;\;\;bc}^{a}\left(x^{\alpha},y_{\beta}\right)\mathbf{e}^{c}\left(x^{\alpha},y_{\beta}\right)+\omega_{\;\;\;b}^{a\quad c}\left(x^{\alpha},y_{\beta}\right)\mathbf{f}_{c}\left(x^{\alpha},y_{\beta}\right)\label{Fully expanded spin connection}\\
\boldsymbol{\omega} & = & W_{c}\left(x^{\alpha},y_{\beta}\right)\mathbf{e}^{c}\left(x^{\alpha},y_{\beta}\right)+W^{c}\left(x^{\alpha},y_{\beta}\right)\mathbf{f}_{c}\left(x^{\alpha},y_{\beta}\right)\label{Fully expanded Weyl form}
\end{eqnarray}
It is because the spin connection performs the same $O\left(n\right)$
rotation simultaneously on each subspace that it is able to preserve
the multiple structures. In fact, Eq.(\ref{Fully expanded spin connection})
displays far more generality than we ultimately want: we would like
for all fields to be determined purely the the spacetime solder form,
$\mathbf{e}^{c}\left(x^{\alpha}\right)$, and this will require reduction
of both components (e.g., $\left(\omega_{\;\;\;bc}^{a},\omega_{\;\;\;b}^{a\quad c}\right)\rightarrow\omega_{\;\;\;bc}^{a}$)
and of independent variables ($\left(x^{\alpha},y_{\beta}\right)\rightarrow x^{\alpha}$).
Accomplishing this will satisfy point $4$ above, and this is the
central accomplishment of the current presentation.

The only assumption we make, beyond the Cartan biconformal construction
and the the field equations following from the action (\ref{Action}),
is vanishing torsion. This is a natural constraint for a spacetime
gravity theory, is consistent with existing measurements in general
relativity, and satisfies point $5$.

Point $6$ is satisfied by the action, (\ref{Action}), which despite
retaining scale invariance, is linear in the biconformal curvatures.
Notice that the $\alpha$ term in Eq.(\ref{Action}) is completely
analogous to the Einstein-Hilbert action written in similar language,
i.e., $S_{EH}=\int\mathbf{R}^{ab}\land\mathbf{e}^{c}\land\ldots\land\mathbf{e}^{d}\varepsilon_{abc\cdots d}$.

We therefore claim that the reduction presented here satisfies all
six desired conditions. 

\subsection{Additional potential advantages}

There are further potential advantages of biconformal models.

The biconformal theory developed here also overlaps strongly with
calculations in twistor space. Twistor space, in arbitrary dimension,
is the space of spinors of the conformal group, up to projection by
an overall factor. Witten showed in \cite{Witten} (see also \cite{Berkovits,CachazoSvrcekTwistorString,Mason})
that when a topological string theory is formulated in twistor space
it is equivalent to $N=4$ supersymmetric Yang-Mills theory. Most
notably, twistor string theory provided a string/gauge theory equivalence
that allowed remarkably efficient calculations of scattering amplitudes
for gauge theory, reducing months of supercomputer calculations to
fewer than two dozen integrals. The effort slowed considerably when
it was thought that it would necessarily lead to fourth order Weyl
gravity instead of general relativity. While a few alternative formulations
were found, Mason was led to conclude \cite{Mason}:
\begin{quote}
Clearly, more work is required to discover what other twistor\textendash string
theories can be constructed. In particular, one would like to have
twistor\textendash string theories that give rise to Poincaré supergravities,
or to pure super-Yang\textendash Mills, or that incorporate other
representations of the gauge and Lorentz groups.
\end{quote}
Biconformal gravity might be an ideal ground state for twistor string
since it arises from conformal symmetry, maintains scale invariance,
and reduces to general relativity. It is therefore of interest to
formulate twistor string theory in a biconformal space. These will
naturally use the spinor representation, as in the supergravity extension
of biconformal space \cite{AW}.

\medskip{}

Biconformal spaces also seem well suited to string compactification.
In the present work, we show that these $2n$-dimensional gravity
theories reduce via their field equations to $n$-dimensional general
relativity. As a result, a string theory written in a 10-dimensional
biconformal background will require only two dimensional compactification
to describe $4$-dimensional general relativity. There are only a
countable number of $2$-dimensional topologies, compared to the truly
huge number of $6$-dimensionsal compact spaces available when going
from $10$ directly to $4$-dimensions. 

The situation is even more restrictive than all $2$-dimensional compact
spaces because the compactification is required to go between two
biconformal spaces. It therefore must include one basis direction
of each conformal weight. This necessarily restricts the compactification
to a $2$-torus or possibly a $2$-sphere.

\subsection{Organization}

The organization of the paper begins with the basic equations of biconformal
gravity, including some notational conventions and ending with the
field equations. In Section \ref{sec:Reducing-the-curvatures}, we
show the effects of vanishing torsion on the remaining curvatures,
using the Bianchi identites and field equations to reduce the number
of components. From the effect of vanishing torsion on the curvtures,
it is immediate to see by symmetry that if the co-torsion were also
to be set to zero, the additional constraints would force the solution
to be trivial. We begin solving for the connection in Section \ref{Structure equation reduction}
by making use of the Frobenius theorem on the involution of the solder
form. This clarifies the meaning of the doubled dimension, showing
that the biconformal space is foliated by an $n$-dimensional Lie
group. This foliation may be interpreted as the translation group
of the co-tangent bundle, the torus of double field theory, or as
a new, nonabelian internal symmetry.

In Section \ref{The full space}, we extend the partial solution for
the connection from the involution back to the full biconformal space
and substitute into each structure equation to find the resulting
form of the curvatures, then use these results to reduce the field
equations. From this point on, the solution divides into two cases
depending on whether the Lie group of the foliation is abelian or
non-abelian. Each of these cases merits a Section (\ref{sec:Abelian-case:},\ref{sec:Non-abelian-case:}).
Finally, we summarize our results in Section \ref{sec:Conclusions}.

\section{The field equations of biconformal gravity }

The first construction of the biconformal quotient was carried out
by Ivanov and Niederle \cite{Ivanov}, who used it to describe a gravity
theory using a curvature-quadratic action. Subsequently, the geometry
was revived \cite{NCG} and a curvature-linear action was introduced
\cite{WW} to give biconformal gravity. The details of the construction
are given in \cite{Hazboun Wheeler}, along with a demonstration of
the signature-changing properties derived in \cite{Spencer Wheeler}.
Here, we rely on the specifics given in \cite{Hazboun Wheeler}, providing
only a basic description and introducing some convenient nomenclature,
then moving quickly to the Cartan structure equations, Bianchi identities,
and the linear action.

From the action, we find the field equations and study their consequences
with only the assumption of vanishing torsion. Throughout, we work
in arbitrary dimension with arbitrary signature for the conformal
metric class. 

\subsection{Building the structure equations}

Consider a space of dimension $n=p+q$, with an $SO\left(p,q\right)$-symmetric
orthonormal metric $\eta$. We compactify with appropriate null cones
at infinity, to permit the inversions that give the space a well-defined
conformal symmetry, $\mathcal{C}=SO\left(p+1,q+1\right)$. The homogeneous
Weyl subgroup $\mathcal{W}=SO\left(p,q\right)\times SO\left(1,1\right)\subset\mathcal{C}$
consists of the pseudo-rotations and dilatations. The quotient $\mathcal{C}/\mathcal{W}$
is a $2n$-dimensional homogeneous manifold from which we immediately
have a principal fiber bundle with fiber symmetry $\mathcal{W}$.
We take the local structure of this bundle as a model for a curved
space à la Cartan, modifying the manifold (if desired) and altering
the connection subject to two conditions:
\begin{enumerate}
\item The resulting curvature $2$-forms must be horizontal.
\item The resulting Cartan structure equations satisfy their integrability
conditions (generalized Bianchi identities).
\end{enumerate}
Let the connection forms dual to the generators of the Lie algebra
be written as $\boldsymbol{\omega}_{\;\;b}^{a}$ ($SO\left(p,q\right)$
transformations), $\mathbf{e}^{a}$ (translations), $\mathbf{f}_{a}$
(special conformal transformations, called co-translations in the
context of these biconformal geometries), and $\boldsymbol{\omega}$
(dilatations). Then the Cartan structure equations are:
\begin{eqnarray}
\mathbf{d}\boldsymbol{\omega}_{\;b}^{a} & = & \boldsymbol{\omega}_{\;b}^{c}\wedge\boldsymbol{\omega}_{\;c}^{a}+2\Delta_{cb}^{ad}\mathbf{f}_{d}\wedge\mathbf{e}^{c}+\boldsymbol{\Omega}_{\;\;b}^{a}\label{Curvature structure equation}\\
\mathbf{d}\mathbf{e}^{a} & = & \mathbf{e}^{b}\wedge\boldsymbol{\omega}_{\;\;b}^{a}+\boldsymbol{\omega}\wedge\mathbf{e}^{b}+\mathbf{T}^{a}\label{Torsion structure equation}\\
\mathbf{d}\mathbf{f}_{a} & = & \boldsymbol{\omega}_{\;\;a}^{b}\wedge\mathbf{f}_{b}+\mathbf{f}_{a}\wedge\boldsymbol{\omega}+\mathbf{S}_{a}\label{Co torsion structure equation}\\
\mathbf{d}\boldsymbol{\omega} & = & \mathbf{e}^{a}\wedge\mathbf{f}_{a}+\boldsymbol{\Omega}\label{Dilatation structure equation}
\end{eqnarray}
Horizontality requires the curvature to be expanded in the $\left(\mathbf{e}^{a},\mathbf{f}_{b}\right)$
basis, giving each of the components $\left(\boldsymbol{\Omega}_{\;\;b}^{a},\mathbf{T}^{a},\mathbf{S}_{a},\boldsymbol{\Omega}\right)$
the general form
\begin{equation}
\boldsymbol{\Omega}^{A}=\frac{1}{2}\Omega_{\;\;\;cd}^{A}\,\mathbf{e}^{c}\wedge\mathbf{e}^{d}+\Omega_{\quad d}^{Ac}\,\mathbf{f}_{c}\wedge\mathbf{e}^{d}+\frac{1}{2}\Omega^{Acd}\,\mathbf{f}_{c}\wedge\mathbf{f}_{d}\label{Expanded curvature}
\end{equation}
and integrability follows from the Poincaré lemma, $\mathbf{d}^{2}\equiv0$.

The $\frac{\left(n-1\right)\left(n+2\right)}{2}$ curvature components
$\left(\boldsymbol{\Omega}_{\;\;b}^{a},\mathbf{T}^{a},\mathbf{S}_{a},\boldsymbol{\Omega}\right)$
together comprise a single conformal curvature tensor. However, the
local symmetries of the homogeneous Weyl symmetry of the biconformal
bundle do not mix these four separate parts. Thereofore, we call the
$SO\left(p,q\right)$ part of the full conformal curvature $\boldsymbol{\Omega}_{\;\;b}^{a}$
the \emph{curvature}, the translational part of the curvature $\mathbf{T}^{a}$
the \emph{torsion}, the special conformal part of the curvature the
\emph{co-torsion}, $\mathbf{S}_{a}$, and the dilatational portion
$\boldsymbol{\Omega}$ the \emph{dilatational curvature} or simply
the \emph{dilatation}.

Each of the curvatures each has three distinguishable parts, as seen
in Eq.(\ref{Expanded curvature}). We call the $\mathbf{e}^{a}\wedge\mathbf{e}^{b}$
term the \emph{spacetime term}, the $\mathbf{f}_{a}\wedge\mathbf{e}^{b}$
term the \emph{cross term,} and the $\mathbf{f}_{a}\wedge\mathbf{f}_{b}$
term the \emph{momentum term}. While it may be somewhat abusive to
call a signature $\left(p,q\right)$ space ``spacetime'', for the
gravitational applications we consider the name is ultimately appropriate.
In the cases where the co-solder forms generate a nonabelian Lie group,
the name ``momentum'' is not appropriate, and we will speak of the
relevant group manifold.

To avoid introducing too many symbols, the symbols for the three parts
of curvatures are distinguished purely by index position. Thus, $\Omega_{\;\;\;b\quad d}^{a\quad c}$
denotes the cross-term of the $SO\left(p,q\right)$ curvature and
$\Omega_{\;\;\;bcd}^{a}$ the spacetime term of the $SO\left(p,q\right)$
curvature. \emph{These are independent functions}. We therefore do
not raise or lower indices unless, on some submanifold, there is no
chance for ambiguity. Note also that the raised and lowered index
positions indicate the conformal weights, $+1$ and $-1$ respectively,
of all definite weight objects. Therefore, the torsion cross-term
$T_{\quad c}^{ab}$ has net conformal weight $+1$, the spacetime
term of the co-torsion $S_{abc}$ has conformal weight $-3$, and
the full torsion 2-form $\mathbf{T}^{a}$ has conformal weight $+1$.

Note the similarity between Eqs.(\ref{Torsion structure equation})
and (\ref{Co torsion structure equation}). This occurs because, by
taking the quotient of the conformal group by its homogeneous Weyl
subgroup instead of the more common inhomogeneous Weyl group, symmetry
is maintained between the translations and the special conformal transformations.
Indeed, in their action on the defining compactified $\left(p,q\right)$
space, the special conformal transformations are simply translations
in inverse coordinates, $y_{\mu}=\frac{x_{\mu}}{x^{2}}$. As a result,
they behave near infinity exactly as translations do at the origin;
correspondingly, the effect of a simple translation expressed in inverse
coordinates is the same as that of a special conformal transformation
at the origin. In the biconformal space, the resulting gauge field
of translations, $\mathbf{e}^{a}$, and the gauge field of special
conformal transformations, $\mathbf{f}_{a}$, form a cotangent basis.
Each locally spans an $n$-dimensional subspace of the full biconformal
cotangent space, which we ultimately show to be submanifolds. In parallel
to calling $\mathbf{e}^{a}$ the solder form, we call $\mathbf{f}_{a}$
the \emph{co-solder form}. Similarly, just as the field strength of
the solder form is called the torsion, $\mathbf{T}^{a}$, we refer
to the field strength of the co-solder form as the \emph{co-torsion},
$\mathbf{S}_{a}$. 

\subsection{Bianchi identities}

The generalized Bianchi identities are the integrability conditions
for the Cartan equations. They are found by applying the Poincaré
lemma, $\mathbf{d}^{2}\equiv0$, to each structure equation, then
using the structure equations again to eliminate all but curvature
terms. They always give covariant expressions \textendash{} we are
guaranteed that all purely connection terms must cancel because when
all curvatures vanish the Cartan equations reduce to the Maurer-Cartan
equations, for which the integrability conditions are the Jacobi identities,
and therefore are automatically satisfied. 

Knowing that all connection terms must cancel when we replace exterior
derivatives with the corresponding curvatures makes it easier to derive
the identities. Furthermore, every exterior derivative of a curvature
becomes a covariant derivative. Using this knowledge, we may quickly
find the identities. Thus, for the $SO\left(p,q\right)$ curvature,
we take the exterior derivative of Eq.(\ref{Curvature structure equation}),
\begin{eqnarray*}
0 & \equiv & \mathbf{d}^{2}\boldsymbol{\omega}_{\;\;b}^{a}\\
 & = & \mathbf{d}\boldsymbol{\omega}_{\;\;b}^{c}\wedge\boldsymbol{\omega}_{\;\;c}^{a}-\boldsymbol{\omega}_{\;\;b}^{c}\wedge\mathbf{d}\boldsymbol{\omega}_{\;\;c}^{a}+2\Delta_{db}^{ac}\mathbf{d}\mathbf{f}_{c}\wedge\mathbf{e}^{d}-2\Delta_{db}^{ac}\mathbf{f}_{c}\wedge\mathbf{d}\mathbf{e}^{d}+\mathbf{d}\boldsymbol{\Omega}_{\;\;b}^{a}\\
0 & = & \boldsymbol{\Omega}_{\;\;b}^{c}\wedge\boldsymbol{\omega}_{\;\;c}^{a}-\boldsymbol{\omega}_{\;\;b}^{c}\wedge\boldsymbol{\Omega}_{\;\;c}^{a}+2\Delta_{db}^{ac}\mathbf{S}_{c}\wedge\mathbf{e}^{d}-2\Delta_{db}^{ac}\mathbf{f}_{c}\wedge\mathbf{T}^{d}+\mathbf{d}\boldsymbol{\Omega}_{\;\;b}^{a}\\
0 & = & \mathbf{D}\boldsymbol{\Omega}_{\;\;b}^{a}+2\Delta_{db}^{ac}\mathbf{S}_{c}\wedge\mathbf{e}^{d}-2\Delta_{db}^{ac}\mathbf{f}_{c}\wedge\mathbf{T}^{d}
\end{eqnarray*}
where we have identified the covariant exterior derivative, $\mathbf{D}\boldsymbol{\Omega}_{\;\;b}^{a}=\mathbf{d}\boldsymbol{\Omega}_{\;\;b}^{a}+\boldsymbol{\Omega}_{\;\;b}^{c}\wedge\boldsymbol{\omega}_{\;\;c}^{a}-\boldsymbol{\omega}_{\;\;b}^{c}\wedge\boldsymbol{\Omega}_{\;\;c}^{a}$.
Proceeding through Eqs.(\ref{Curvature structure equation}) - (\ref{Dilatation structure equation}),
we find the full set of integrability conditions,
\begin{eqnarray}
\mathbf{D}\boldsymbol{\Omega}_{\;\;b}^{a}+2\Delta_{cb}^{ad}\left(\mathbf{S}_{d}\wedge\mathbf{e}^{c}-\mathbf{f}_{d}\wedge\mathbf{T}^{c}\right) & = & 0\label{Curvature Bianchi}\\
\mathbf{D}\mathbf{T}^{a}-\mathbf{e}^{b}\wedge\boldsymbol{\Omega}_{\;b}^{a}+\boldsymbol{\Omega}\wedge\mathbf{e}^{a} & = & 0\label{Torsion Bianchi}\\
\mathbf{D}\mathbf{S}_{a}+\boldsymbol{\Omega}_{\;a}^{b}\wedge\mathbf{f}_{b}-\mathbf{f}_{a}\wedge\boldsymbol{\Omega} & = & 0\label{Co-torsion Bianchi}\\
\mathbf{D}\boldsymbol{\Omega}+\mathbf{T}^{a}\wedge\mathbf{f}_{a}-\mathbf{e}^{a}\wedge\mathbf{S}_{a} & = & 0\label{Dilatation Bianchi}
\end{eqnarray}
where the covariant derivatives are given by
\begin{eqnarray}
\mathbf{D}\boldsymbol{\Omega}_{\;\;b}^{a} & = & \mathbf{d}\boldsymbol{\Omega}_{\;\;b}^{a}+\boldsymbol{\Omega}_{\;\;b}^{c}\wedge\boldsymbol{\omega}_{\;c}^{a}-\boldsymbol{\omega}_{\;b}^{c}\wedge\boldsymbol{\Omega}_{\;\;c}^{a}\nonumber \\
\mathbf{D}\mathbf{T}^{a} & = & \mathbf{d}\mathbf{T}^{a}+\mathbf{T}^{b}\wedge\boldsymbol{\omega}_{\;b}^{a}-\boldsymbol{\omega}\wedge\mathbf{T}^{a}\nonumber \\
\mathbf{D}\mathbf{S}_{a} & = & \mathbf{d}\mathbf{S}_{a}-\boldsymbol{\omega}_{\;a}^{b}\wedge\mathbf{S}_{b}+\mathbf{S}_{a}\wedge\boldsymbol{\omega}\nonumber \\
\mathbf{D}\boldsymbol{\Omega} & = & \mathbf{d}\boldsymbol{\Omega}\label{Covariant derivtives}
\end{eqnarray}

Since each Bianchi identity contains the covariant derivative of a
curvature, it is typically difficult to use them to help find solutions
to the field equations. They are simply the conditions on the curvatures
that guarantee that a solution exists, and if we find a solution to
the field equations, the Bianchi identities are necessarily satisfied.
However, if one of the curvatures vanishes the relations become algebraic
and can be extremely helpful.

\subsection{The action functional}

\subsubsection{Notational conventions}

The metric is $\eta_{ab}$ with pseudo-rotational invariance under
$SO\left(p,q\right)$, $p+q=n$. Lower case Latin indices run $a,b\ldots=1,2\ldots,n$,
and refer to orthonormal frames $\left(\mathbf{e}^{a},\mathbf{f}_{a}\right)$.
When coordinates are introduced they are given Greek indices. Thus,
we may write
\[
\mathbf{e}^{a}=e_{\mu}^{\;\;a}\mathbf{d}x^{\mu}+e^{\mu a}\mathbf{d}y_{\mu}
\]
Until we have established appropriate submanifolds, we cannot use
the components of the solder form, $e_{\mu}^{\;\;\;a}$, to change
basis.

An antisymmetric projection operator on type $\left(\begin{array}{c}
0\\
2
\end{array}\right)$ tensors may be written as
\[
P_{fb}^{\quad ed}=\frac{1}{2}\left(\delta_{f}^{e}\delta_{b}^{d}-\delta_{f}^{d}\delta_{b}^{e}\right)
\]
If we raise the $f$ index and lower $e$, this becomes
\begin{eqnarray*}
\Delta_{db}^{ac} & \equiv & \eta_{de}\eta^{af}\frac{1}{2}\left(\delta_{f}^{e}\delta_{b}^{c}-\delta_{f}^{c}\delta_{b}^{e}\right)\\
 & = & \frac{1}{2}\left(\delta_{d}^{a}\delta_{b}^{c}-\eta^{ac}\eta_{bd}\right),
\end{eqnarray*}
the antisymmetric projection operator on type $\left(\begin{array}{c}
1\\
1
\end{array}\right)$ tensors. This symbol occurs frequently.

\subsubsection{The volume form}

The volume form is unusual, having two types of index. Since we can
distinguish the conformal weight $+1$ solder forms $\mathbf{e}^{a}$
from the conformal weight $-1$ co-solder forms, $\mathbf{f}_{a}$,
we can always partially re-arrange. Thus, while the $2n$-dim volume
form may be written as
\[
e_{\left(\begin{array}{c}
a\\
\cdot
\end{array}\right)\left(\begin{array}{c}
\cdot\\
d
\end{array}\right)\cdots\left(\begin{array}{c}
b\\
\cdot
\end{array}\right)\left(\begin{array}{c}
\cdot\\
e
\end{array}\right)\left(\begin{array}{c}
c\\
\cdot
\end{array}\right)\left(\begin{array}{c}
\cdot\\
f
\end{array}\right)}=e_{\left[\left(\begin{array}{c}
a\\
\cdot
\end{array}\right)\left(\begin{array}{c}
\cdot\\
d
\end{array}\right)\cdots\left(\begin{array}{c}
b\\
\cdot
\end{array}\right)\left(\begin{array}{c}
\cdot\\
e
\end{array}\right)\left(\begin{array}{c}
c\\
\cdot
\end{array}\right)\left(\begin{array}{c}
\cdot\\
f
\end{array}\right)\right]}
\]
where $\left(\begin{array}{c}
\cdot\\
a
\end{array}\right)$ represents an index that contracts with $\boldsymbol{\omega}^{a}$
and $\left(\begin{array}{c}
a\\
\cdot
\end{array}\right)$ represents an index that contracts with $\boldsymbol{\omega}_{a}$,
we can always insist that the weight $+1$ indices go first and the
weight $-1$ go last,
\[
e_{\left[\left(\begin{array}{c}
a\\
\cdot
\end{array}\right)\left(\begin{array}{c}
b\\
\cdot
\end{array}\right)\cdots\left(\begin{array}{c}
c\\
\cdot
\end{array}\right)\left(\begin{array}{c}
\cdot\\
d
\end{array}\right)\left(\begin{array}{c}
\cdot\\
e
\end{array}\right)\cdots\left(\begin{array}{c}
\cdot\\
f
\end{array}\right)\right]}\equiv e_{\qquad de\ldots f}^{ab\ldots c}
\]
thereby reducing the $\left(2n\right)!$ permutations to $n!n!$.
Locally (and globally once submanifolds are established), there exist
distinguishable subspaces on which we may write $e_{\qquad de\ldots f}^{ab\ldots c}$
as a pair of $n$-dim Levi-Civita tensors,
\[
e_{\qquad de\ldots f}^{ab\ldots c}=e^{ab\ldots c}e_{de\ldots f}
\]
This convention means that a contraction, $e_{\qquad ae\ldots f}^{ab\ldots c}$
is meaningful, when it would vanish immediately with the full antisymmetrization.
This is, nonetheless, correct since there exists an unambiguous local
separation by conformal weight, each with its own induced volume form.
Since variation of the action is local, we may use this to find the
field equations. A single contraction gives $e_{\qquad ae\ldots f}^{ab\ldots c}=\delta_{e\ldots f}^{b\ldots c}$
and in general, contracting all but $k$ pairs of indices,
\begin{equation}
\varepsilon^{c\cdots de\cdots f}\varepsilon_{a\cdots bg\cdots h}=\left(-1\right)^{q}\left(k-1\right)!\left(n-k+1\right)!\delta_{a\cdots b}^{c\cdots d}\label{Contracted Levi-Civita tensors}
\end{equation}
The presence of this conformal separation also allows the dilatational
curvature to be included as the $\beta$ term in the action. It may
be argued that this is not allowed if the subspaces are not integrable.
We find that the subspaces \emph{are} integrable, but have checked
that setting $\beta=0$ throughout does not alter any of our conclusions.

There exist conditions that guarantee that such a splitting into subspaces
is integrable across the full biconformal space. For example, the
$\mathbf{e}^{a}$ subspace is certainly integrable to a submanifold
if the basis structure equation,
\begin{eqnarray*}
\mathbf{d}\mathbf{e}^{a} & = & \mathbf{e}^{b}\wedge\boldsymbol{\omega}_{\;b}^{a}+\boldsymbol{\omega}\wedge\mathbf{e}^{a}+\mathbf{T}^{a}
\end{eqnarray*}
is in involution, and this is true if the torsion $\mathbf{T}^{a}$
is suitably restricted. Specifically, if the momentum term of the
torsion vanishes, $T^{acd}=0$, then the Eq.(\ref{Expanded curvature})
for the torsion reduces to
\begin{eqnarray}
\mathbf{T}^{a} & = & \frac{1}{2}T_{\;\;cd}^{a}\mathbf{e}^{c}\wedge\mathbf{e}^{d}+T_{\quad d}^{ac}\mathbf{f}_{c}\wedge\mathbf{e}^{d}\label{Expanded torsion}
\end{eqnarray}
and $\mathbf{e}^{a}$ is in involution. Similarly, the co-solder equation
will be involute provided $S_{acd}=0$. 

We make no assumptions about the torsion or co-torsion in deriving
the field equations. Though neither occurs explicitly in the curvature-linear
action, integrations by parts after variation nonetheless introduce
them into the field equations.

We define the Hodge dual of unity as a convenient volume form,
\begin{eqnarray}
\boldsymbol{\Phi} & \equiv & ^{*}1\nonumber \\
 & = & \frac{1}{n!n!}e_{\qquad de\ldots f}^{ab\ldots c}\mathbf{e}^{d}\wedge\mathbf{e}^{e}\wedge\cdots\wedge\mathbf{e}^{f}\wedge\mathbf{f}_{a}\wedge\mathbf{f}_{b}\wedge\cdots\wedge\mathbf{f}_{c}\label{Volume form}
\end{eqnarray}
It follows that
\begin{eqnarray}
\mathbf{e}^{d}\wedge\mathbf{e}^{e}\wedge\cdots\wedge\mathbf{e}^{f}\wedge\mathbf{f}_{a}\wedge\mathbf{f}_{b}\wedge\cdots\wedge\mathbf{f}_{c} & = & e_{\qquad ab\cdots c}^{de\cdots f}\boldsymbol{\Phi}\label{Volume form replacement}
\end{eqnarray}
Eq.(\ref{Volume form replacement}) is useful for finding the field
equations. Taking a second dual,
\begin{eqnarray*}
\,^{*}\boldsymbol{\Phi} & \equiv & \,^{**}1\\
 & = & \,^{*}\left(\frac{1}{n!n!}e_{\qquad de\ldots f}^{ab\ldots c}\mathbf{e}^{d}\wedge\mathbf{e}^{e}\wedge\cdots\wedge\mathbf{e}^{f}\wedge\mathbf{f}_{a}\wedge\mathbf{f}_{b}\wedge\cdots\wedge\mathbf{f}_{c}\right)\\
 & = & \frac{1}{n!n!}e_{\qquad de\ldots f}^{ab\ldots c}\eta_{aa'}\eta_{bb;}\ldots\eta_{cc'}\eta^{dd'}\eta^{ee'}\ldots\eta^{ff'}e_{\qquad d'e'\ldots f'}^{a'b'\ldots c'}\\
 & = & \left(-1\right)^{2q}\frac{1}{n!n!}n!n!\\
 & = & 1
\end{eqnarray*}
regardless of the dimension or signature.

\subsection{The action functional}

Eq.(\ref{Action}) is the most general action linear in biconformal
curvatures. It is defined on the $2n$-dimensional base manifold of
the bundle, spanned by $\left(\mathbf{e}^{a},\mathbf{f}_{a}\right)$.
The initial conformally symmetric space has metric $\eta_{ab}$ of
any dimension $n>2$ and any signature $\left(p,q\right)$.

We find the field equations by varying the full set of connection
$1$-forms, $\left\{ \boldsymbol{\omega}_{\;b}^{a},\mathbf{e}^{a},\mathbf{f}_{a},\boldsymbol{\omega}\right\} $.
Each variation has two parts when we expand in the $\left(\mathbf{e}^{a},\mathbf{f}_{b}\right)$
basis, for example,
\[
\delta\boldsymbol{\omega}_{\;\;b}^{a}=\delta A_{\;\;bc}^{a}\mathbf{e}^{c}+\delta B_{\;\;\;b}^{a\quad c}\mathbf{f}_{c}
\]
with $\delta A_{\;\;bc}^{a}$ and $\delta B_{\;\;\;b}^{a\quad c}$
independent, arbitrary variations. We therefore find eight sets of
field equations. To illustrate details of the variation technique,
the variation of the spin connection $\boldsymbol{\omega}_{\;\;b}^{a}$
is given in Appendix B.

Carrying out each of the connection variations, we arrive at the final
field equations:

\begin{eqnarray}
T_{\quad e}^{ae}-T_{\quad e}^{ea}-S_{e}^{\;\;ae} & = & 0\label{Field equation Weyl 1}\\
T_{\;\;ca}^{a}+S_{c\quad a}^{\;\;a}-S_{a\quad c}^{\;\;a} & = & 0\label{Field equation Weyl 2}\\
\alpha\Delta_{sb}^{ar}\left(T_{\quad a}^{mb}-\delta_{a}^{m}T_{\quad e}^{eb}-\delta_{a}^{m}S_{c}^{\;\;bc}\right) & = & 0\label{Field equation spin 1}\\
\alpha\Delta_{sb}^{ar}\left(\delta_{c}^{b}T_{\;\;ad}^{d}+S_{c\quad a}^{\;\;\;b}-\delta_{c}^{b}S_{d\quad a}^{\;\;\;d}\right) & = & 0\label{Field equation spin 2}\\
\alpha\left(\Omega_{\;\;e\quad b}^{a\quad e}-\Omega_{\;\;d\quad c}^{c\quad d}\delta_{\;\;b}^{a}\right)+\beta\left(\Omega_{\quad b}^{a}-\Omega_{\quad c}^{c}\delta_{\;\;b}^{a}\right)+\Lambda\delta_{\;\;b}^{a} & = & 0\label{Field equation solder 1}\\
\alpha\Omega_{\;\;acb}^{c}+\beta\Omega_{ab} & = & 0\label{Field equation solder 2}\\
\alpha\left(\Omega_{\;\;b\quad c}^{c\quad a}-\Omega_{\;\;e\quad c}^{c\quad e}\delta_{\;\;b}^{a}\right)+\beta\left(\Omega_{\quad b}^{a}-\Omega_{\quad c}^{c}\delta_{\;\;b}^{a}\right)+\Lambda\delta_{\;\;b}^{a} & = & 0\label{Field equation co solder 1}\\
\alpha\Omega_{\;\;c}^{a\quad cb}+\beta\Omega^{ab} & = & 0\label{Field equation co solder 2}
\end{eqnarray}
where the constant $\Lambda$ is defined to be $\Lambda\equiv\left(\left(n-1\right)\alpha-\beta+n^{2}\gamma\right)$.
Ultimately, all our results depend on a single parameter, $\chi=\frac{1}{n-1}\frac{\Lambda}{\left(n-1\right)\alpha-\beta}$.

\subsection{Biconformal spaces}

The system we wish to study consists of the structure equations Eqs.(\ref{Curvature structure equation})-(\ref{Dilatation structure equation}),
their associated Bianchi identities Eqs.(\ref{Curvature Bianchi})-(\ref{Dilatation Bianchi}),
and the field equations Eqs.(\ref{Field equation Weyl 1})-(\ref{Field equation co solder 2}).
These have been written above with no additional conditions, and they
apply to the biconformal geometry constructed from the conformal group
in any dimension $n$ and any signature $\left(p,q\right)$.

Our goal is to show how the full set of biconformal curvatures in
$2n$-dimensions reduces to only those required to describe $n$-dimensional
general relativity. Assuming only vanishing torsion and the field
equations, we show in the next Section that the curvatures, each initially
in the general form given in Eq. (\ref{Expanded curvature}), reduce
to 
\begin{eqnarray*}
\boldsymbol{\Omega}_{\;\;b}^{a} & = & \frac{1}{2}\Omega_{\;\;bcd}^{a}\mathbf{e}^{c}\wedge\mathbf{e}^{d}+2\chi\Delta_{db}^{ac}\mathbf{f}_{c}\wedge\mathbf{e}^{d}\\
\mathbf{T}^{a} & = & 0\\
\mathbf{S}_{a} & = & \frac{1}{2}S_{acd}\mathbf{e}^{c}\wedge\mathbf{e}^{d}+S_{a\quad d}^{\;\;c}\mathbf{f}_{c}\wedge\mathbf{e}^{d}+S_{a}^{\;\;cd}\mathbf{f}_{c}\wedge\mathbf{f}_{d}\\
\boldsymbol{\Omega} & = & \chi\mathbf{e}^{c}\wedge\mathbf{f}_{c}
\end{eqnarray*}
In Sections (\ref{Structure equation reduction}) and (\ref{The full space}),
we use the structure equations to reduce the coordinate dependence
to $x^{\alpha}$ only, with the exception of a few explicit terms
linear in $y_{\alpha}$. There is also further reduction of the curvatures.

\section{Reducing the curvatures of torsion-free biconformal spaces \label{sec:Reducing-the-curvatures}}

We seek to reduce the field equations as far as possible. In particular,
we will show that scale-invariant general relativity emerges from
the vanishing torsion field equations. As in Riemannian geometry,
vanishing torsion is a natural constraint on the full generality of
a biconformal space. This has three definite consequences corresponding
to the three parts in the expansion given by Eq.(\ref{Expanded curvature}),
\[
\mathbf{T}^{a}=\frac{1}{2}T_{\;\;cd}^{a}\mathbf{e}^{c}\wedge\mathbf{e}^{d}+T_{\quad d}^{ac}\mathbf{f}_{c}\wedge\mathbf{e}^{d}+\frac{1}{2}T^{acd}\mathbf{f}_{c}\wedge\mathbf{f}_{d}
\]
We expect the first term in this expansion to give the spacetime torsion,
which is zero in general relativity. The cross term, $T_{\quad d}^{ac}\mathbf{f}_{c}\wedge\mathbf{e}^{d}$,
gives the extrinsic curvature of the spacetime submanifold in the
full space, while the final term measures the non-involution \textendash{}
the degree to which the solder form fails to be in involution \cite{Hazboun dissertation}.
Taking the full torsion to vanish therefore has clear geometric consequences:
it guarantees the existence of a spacetime submanifold with vanishing
spacetime torsion, embeded with no extrinsic curvature in the larger
biconformal space.

Note that it is important that we do \emph{not} constrain the co-torsion.
Indeed, we show at the end of this Section that setting both torsion
and co-torsion to zero is overly restrictive, forcing the full space
to have at most constant curvature and dilatation. Naturally, setting
the torsion but not the co-torsion to zero breaks some of the symmetry
between the solder form and the co-solder form. It would be equivalent
to break the symmetry the other way, setting the co-torsion to zero
and not the torsion.

We begin with the consequences of vanishing torsion in the Bianchi
identities. 

\subsection{Consequences of the Bianchi identities}

If the torsion vanishes, $\mathbf{T}^{a}=0$, then the second Bianchi
identity, eq.(\ref{Torsion Bianchi}), becomes an algebraic condition
on the curvature and dilatation:
\begin{eqnarray*}
\mathbf{e}^{b}\wedge\boldsymbol{\Omega}_{\;b}^{a} & = & \boldsymbol{\Omega}\wedge\mathbf{e}^{a}
\end{eqnarray*}
Expanding each of the curvatures in components,
\begin{eqnarray*}
\mathbf{e}^{b}\wedge\left(\frac{1}{2}\Omega_{\;\;bcd}^{a}\mathbf{e}^{c}\wedge\mathbf{e}^{d}+\Omega_{\;\;b\quad d}^{a\quad c}\mathbf{f}_{c}\wedge\mathbf{e}^{d}+\frac{1}{2}\Omega_{\;\;b}^{a\quad cd}\mathbf{f}_{c}\wedge\mathbf{f}_{d}\right) & = & \left(\frac{1}{2}\Omega_{cd}\mathbf{e}^{c}\wedge\mathbf{e}^{d}+\Omega_{\quad d}^{c}\mathbf{f}_{c}\wedge\mathbf{e}^{d}+\frac{1}{2}\Omega^{cd}\mathbf{f}_{c}\wedge\mathbf{f}_{d}\right)\wedge\mathbf{e}^{a}
\end{eqnarray*}
This breaks into three independent equations, with components related
by
\begin{eqnarray}
\Omega_{\;\;\left[bcd\right]}^{a} & = & \delta_{[b}^{a}\Omega_{cd]}\label{Vanishing torsion effect on curvature 1}\\
\Omega_{\;\;\left[b\quad d\right]}^{a\quad c} & = & \delta_{[b}^{a}\Omega_{\quad d]}^{c}\label{Vanishing torsion effect on curvature 2}\\
\Omega_{\;\;\;b}^{a\quad cd} & = & \delta_{b}^{a}\Omega^{cd}\label{Vanishing torsion effect on curvature 3}
\end{eqnarray}
Note in particular that since $\Omega_{\;\;\;b}^{a\quad cd}$ is antisymmetric
on $ab$, the trace gives $\Omega_{\;\;\;a}^{a\quad cd}=0$, and the
final condition requires 
\begin{equation}
\Omega^{cd}=0\label{Vanishing momentum dilatation}
\end{equation}
and therefore, the momentum space component of the $SO\left(p,q\right)$
curvature vanishes, 
\begin{equation}
\Omega_{\;\;\;b}^{a\quad cd}=0\label{Vanishing momentum curvature}
\end{equation}

The cross-term Bianchi may be used to express the cross curvature
in terms of the cross dilatation. Expanding the antisymmetry in Eq.(\ref{Vanishing torsion effect on curvature 2}),
\begin{eqnarray*}
\Omega_{\;\;\;b\quad d}^{a\quad c}-\Omega_{\;\;\;d\quad b}^{a\quad c} & = & \delta_{\;\;b}^{a}\Omega_{\quad d}^{c}-\delta_{\;\;d}^{a}\Omega_{\quad b}^{c}
\end{eqnarray*}
we formally lower the $a$ index to $e$,
\[
0=\eta_{ea}\Omega_{\;\;b\quad d}^{a\quad c}-\eta_{ea}\Omega_{\;\;d\quad b}^{a\quad c}-\eta_{eb}\Omega_{\quad d}^{c}+\eta_{ed}\Omega_{\quad b}^{c}
\]
and cycle the $b,e,d$ indices. Then adding the first two permutations
and subtracting the third, we get
\begin{eqnarray*}
0 & = & \eta_{ea}\Omega_{\;\;b\quad d}^{a\quad c}-\eta_{ea}\Omega_{\;\;d\quad b}^{a\quad c}+\eta_{ba}\Omega_{\;\;d\quad e}^{a\quad c}-\eta_{ba}\Omega_{\;\;e\quad d}^{a\quad c}-\eta_{da}\Omega_{\;\;e\quad b}^{a\quad c}+\eta_{da}\Omega_{\;\;b\quad e}^{a\quad c}\\
 &  & -\eta_{eb}\Omega_{\quad d}^{c}+\eta_{ed}\Omega_{\quad b}^{c}-\eta_{bd}\Omega_{\quad e}^{c}+\eta_{be}\Omega_{\quad d}^{c}+\eta_{de}\Omega_{\quad b}^{c}-\eta_{db}\Omega_{\quad e}^{c}
\end{eqnarray*}
Now use the antisymmetry $\eta_{ea}\Omega_{\;\;b\quad d}^{a\quad c}=-\eta_{ba}\Omega_{\;\;e\quad d}^{a\quad c}$
to solve,
\begin{eqnarray}
\Omega_{\;\;b\quad d}^{a\quad c} & = & -2\Delta_{db}^{ae}\Omega_{\quad e}^{c}\label{Cross curvature in terms of dilatation}
\end{eqnarray}

Vanishing torsion also affects the remaining Bianchi identities, but
the effects are most pronounced when those are combined with the field
equations. Therefore, we turn next to the simplification of the field
equations.

\subsection{Simplifications of the torsion and co-torsion equations\label{subsec:Vanishing-torsion}}

Of the torsion and co-torsion equations, the first two relate various
traces. Equations (\ref{Field equation Weyl 1}) and (\ref{Field equation Weyl 2})
identify two relationships between these traces. When the torsion
vanishes, these become
\begin{eqnarray}
S_{a}^{\;\;ca} & = & 0\label{Momentum co torsion field equation}\\
S_{c\quad a}^{\;\;a} & = & S_{a\quad c}^{\;\;a}\label{Cross co-torsion field equation 1}
\end{eqnarray}
Using these in the next pair, Eqs.(\ref{Field equation spin 1}) is
now identically satisfied while (\ref{Field equation spin 2}) determines
the antisymmetric part of the cross-terms of the co-torsion, in terms
of its trace,
\begin{eqnarray}
\Delta_{sb}^{ar}S_{c\quad a}^{\;\;\;b} & = & \Delta_{sc}^{ar}S_{a\quad e}^{\;\;e}\label{Cross co-torsion equation 2}
\end{eqnarray}
The $rc$ trace fixes the remaining possible contraction,
\begin{eqnarray*}
\eta^{ac}S_{c\quad a}^{\;\;\;b} & = & -\left(n-2\right)\eta^{bc}S_{c\quad e}^{\;\;e}
\end{eqnarray*}
There is only one independent contraction of the cross-term, and it
determines the antisymmetric part of the full cross-term.

\subsection{Simplifications of the curvature and dilatation equations}

Now consider the remaining four equations for the curvature and dilatation,
Eqs.(\ref{Field equation solder 1}) - (\ref{Field equation co solder 2}).
Eq.(\ref{Field equation co solder 2}) is already satisfied by the
consequences of the torsion Bianchi identity, Eqs.(\ref{Vanishing momentum dilatation})
and (\ref{Vanishing momentum curvature}). The difference of the two
cross curvature equations, Eq.(\ref{Field equation solder 1}) and
Eq.(\ref{Field equation co solder 1}), shows the equality of the
traces,
\begin{eqnarray*}
\Omega_{\;\;e\quad b}^{a\quad e} & = & \Omega_{\;\;b\quad c}^{c\quad a}
\end{eqnarray*}
Eq.(\ref{Field equation solder 1}) together with Eq.(\ref{Cross curvature in terms of dilatation})
allows us to completely determine the cross terms of the curvature
and dilatation. Starting with the trace of Eq.(\ref{Field equation solder 1}),
\begin{align*}
\alpha\Omega_{\;\;\;d\quad c}^{c\quad d}+\beta\Omega_{\quad c}^{c} & =\frac{n}{n-1}\Lambda
\end{align*}
and substituting this back into Eq.(\ref{Field equation solder 1}),
we find
\begin{eqnarray}
\alpha\Omega_{\;\;e\quad b}^{a\quad e}+\beta\Omega_{\quad b}^{a} & = & \frac{1}{n-1}\Lambda\delta_{\;\;b}^{a}\label{Reduced cross curvature Bianchi}
\end{eqnarray}
Now, using the $\left(ad\right)$ trace of Eq.(\ref{Cross curvature in terms of dilatation})
\begin{eqnarray*}
\Omega_{\;\;b\quad a}^{a\quad c} & = & -\left(n-1\right)\Omega_{\quad b}^{c}
\end{eqnarray*}
the equality of the cross curvature traces allows us to substitute
into Eq.(\ref{Reduced cross curvature Bianchi})
\[
-\left(n-1\right)\alpha\Omega_{\quad b}^{a}+\beta\Omega_{\quad b}^{a}=\frac{1}{n-1}\Lambda\delta_{\;\;b}^{a}
\]
to show that
\begin{equation}
\Omega_{\quad b}^{a}=-\chi\delta_{\;\;b}^{a}\label{Solution for the cross dilatation}
\end{equation}
where we define
\begin{eqnarray*}
\chi & \equiv & \frac{1}{n-1}\frac{1}{\left(\left(n-1\right)\alpha-\beta\right)}\Lambda
\end{eqnarray*}
The cross-term of the cuvature is now given by Eq.(\ref{Cross curvature in terms of dilatation}),
\begin{eqnarray}
\Omega_{\;\;b\quad d}^{a\quad c} & = & 2\chi\Delta_{db}^{ac}\label{Solution for the cross curvature}
\end{eqnarray}

Next, we examine the remaining vanishing torsion Bianchi identity,
Eq.(\ref{Vanishing torsion effect on curvature 1}). Expanding the
antisymmetry and taking the $ad$ trace,
\begin{eqnarray*}
\Omega_{\;\;bcd}^{a}+\Omega_{\;\;cdb}^{a}+\Omega_{\;\;dbc}^{a} & = & \delta_{b}^{a}\Omega_{cd}+\delta_{c}^{a}\Omega_{db}+\delta_{d}^{a}\Omega_{bc}\\
\Omega_{\;\;cab}^{a}-\Omega_{\;\;bac}^{a} & = & \left(n-2\right)\Omega_{bc}
\end{eqnarray*}
Combining this with the field equation, Eq.(\ref{Field equation solder 2}),
for the corresponding components, $\Omega_{\;\;bac}^{a}=-\frac{\beta}{\alpha}\Omega_{bc}$,
we have
\begin{eqnarray*}
\left(\left(n-2\right)\alpha-2\beta\right)\Omega_{bc} & = & 0
\end{eqnarray*}
so the spacetime dilatation generically vanishes. The field equation
then implies
\begin{eqnarray}
\Omega_{ab} & = & 0\label{Spacetime dilatation}\\
\Omega_{\;\;acb}^{c} & = & 0\label{Spacetime curvature from torsion Bianchi}
\end{eqnarray}
The special case when $\left(\left(n-2\right)\alpha-2\beta\right)=0$
allows a non-integrable Weyl geometry and, likely being unphysical,
will not concern us further.

Because of the constant form of the components of the dilatation,
Eq.(\ref{Solution for the cross dilatation}), the dilatation Bianchi
identity gives constraints on the co-torsion. Starting with Eq.(\ref{Dilatation Bianchi})
with vanishing torsion and the complete dilatation now given by $\boldsymbol{\Omega}=\chi\mathbf{e}^{a}\mathbf{f}_{a}$,
Eq.(\ref{Dilatation Bianchi}) gives
\begin{eqnarray*}
0 & = & \mathbf{D}\left(\chi\mathbf{e}^{a}\wedge\mathbf{f}_{a}\right)-\mathbf{e}^{a}\wedge\mathbf{S}_{a}\\
 & = & -\left(1+\chi\right)\mathbf{e}^{a}\wedge\mathbf{S}_{a}
\end{eqnarray*}
with components
\begin{eqnarray}
\left(1+\chi\right)S_{\left[abc\right]} & = & 0\nonumber \\
\left(1+\chi\right)\left(S_{a\quad c}^{\;\;\;b}-S_{c\quad a}^{\;\;\;b}\right) & = & 0\nonumber \\
\left(1+\chi\right)S_{a}^{\;\;bc} & = & 0\label{Co-torsion constraints}
\end{eqnarray}
For generic constants in the action we may cancel the $1+\chi$ factor,
but the $\chi=-1$ case permits the presence of a non-abelian internal
symmetry.

\subsection{A theorem: Vanishing torsion \emph{and} co-torsion\label{subsec:TheoremVanishingSandT}}

We digress briefly to prove a useful result. From our results so far,
we can easily prove the following theorem. We start with the definition
of a \emph{flat }and\emph{ trivial biconformal space}. Because of
the ``cosmological constant'' term $\Lambda$ in Eqs.(\ref{Field equation solder 1})
and (\ref{Field equation co solder 1}), we cannot, in general, set
all curvatures to zero unless $\Lambda=0$ as well. We therefore define
a \emph{flat biconformal space} \cite{NCG} to have vanishing curvatures
and $\Lambda=0$, and a \emph{trivial biconformal space} to have vanishing
curvatures except for constant curvature and dilatation cross-terms,
which then have the $\Lambda$-dependent forms given in Eqs.(\ref{Solution for the cross dilatation})
and (\ref{Solution for the cross curvature}). That these constant
values of the curvatures yield solutions to the field equations follows
as a special case of the generic torsion free solution below. 
\begin{description}
\item [{Triviality$\:$Theorem}] : Biconformal spaces in which both the
torsion and the co-torsion vanish are trivial biconformal spaces.
\item [{Proof:}] With vanishing torsion, we have already seen that the
momentum curvature and dilatation vanish. By the symmetry of biconformal
spaces, zero co-torsion requires the spacetime curvature and dilatation
to vanish as well. Since, by assumption we have both $\mathbf{T}^{a}=0$
and $\mathbf{S}_{a}=0$, the only nonvanishing curvature components
are the dilatation and curvature cross-terms, shown above to necessarily
have the forms given in Eqs.(\ref{Solution for the cross dilatation})
and (\ref{Solution for the cross curvature}),
\begin{eqnarray*}
\Omega_{\;\;b}^{a} & = & -\chi\delta_{\;\;b}^{a}\\
\Omega_{\;\;b\quad d}^{a\quad c} & = & 2\chi\Delta_{db}^{ac}
\end{eqnarray*}
vanishing if and only if $\chi\equiv\frac{1}{n-1}\frac{1}{\alpha\left(n-1\right)-\beta}\Lambda$
is zero. The biconformal space is therefore trivial.
\end{description}
There are interesting properties to trivial biconformal spaces. These
homogeneous manifolds have been shown to be Kähler \cite{Hazboun Wheeler},
and allow time to emerge as part of the solution from the properties
of the underlying conformal group \cite{Spencer Wheeler,Hazboun Wheeler}. 

Still, there can be no spacetime or momentum space curvature if both
the torsion and the co-torsion vanish completely, and therefore no
local gravity. To achieve a meaningful gravity theory it is necessary
that at least part of either the torsion or the co-torsion remains
nonzero. 

\subsection{Summary of curvatures and remaining field equations}

Initially, the four curvatures (``curvature'', torsion, co-torsion,
and dilatation) have the three independent terms displayed in eq.(\ref{Expanded curvature}).
Using the assumption of vanishing torsion, we have now reduced these
to
\begin{eqnarray}
\boldsymbol{\Omega}_{\;\;b}^{a} & = & \frac{1}{2}\Omega_{\;\;bcd}^{a}\mathbf{e}^{c}\wedge\mathbf{e}^{d}+2\chi\Delta_{db}^{ac}\mathbf{f}_{c}\wedge\mathbf{e}^{d}\nonumber \\
\mathbf{T}^{a} & = & 0\nonumber \\
\mathbf{S}_{a} & = & \frac{1}{2}S_{acd}\mathbf{e}^{c}\wedge\mathbf{e}^{d}+S_{a\quad d}^{\;\;c}\mathbf{f}_{c}\wedge\mathbf{e}^{d}+S_{a}^{\;\;cd}\mathbf{f}_{c}\wedge\mathbf{f}_{d}\nonumber \\
\boldsymbol{\Omega} & = & \chi\mathbf{e}^{c}\wedge\mathbf{f}_{c}\label{Reduced curvatures}
\end{eqnarray}
together with the remaining field equations
\begin{eqnarray*}
s_{c}\;\;\equiv\;\;S_{c\quad a}^{\;\;a} & = & S_{a\quad c}^{\;\;a}\;\;\\
\Delta_{sb}^{ar}S_{c\quad a}^{\;\;\;b} & = & \Delta_{sc}^{ar}s_{a}\\
S_{c}^{\;\;ac} & = & 0\\
\Omega_{\;\;acb}^{c} & = & 0
\end{eqnarray*}
 and remaining Bianchi conditions,
\begin{eqnarray*}
\left(1+\chi\right)S_{\left[abc\right]} & = & 0\\
\left(1+\chi\right)\left(S_{a\quad c}^{\;\;\;b}-S_{c\quad a}^{\;\;\;b}\right) & = & 0\\
\left(1+\chi\right)S_{a}^{\;\;bc} & = & 0\\
\Omega_{\;\;\left[bcd\right]}^{a} & = & 0
\end{eqnarray*}

Even when $1+\chi\neq0$, the equations involving the co-torsion cross-term
do not determine the co-torsion further; we must turn to the structure
equations to proceed.

While the severe restrictions evident in Eqs.(\ref{Reduced curvatures})
reduce the space considerably toward an $n$-dimensional theory, the
remaining fields are still functions of all $2n$ coordinates. It
is only by using the structure equations that we fully reduce the
theory to $n$-dimensional scale-covariant general relativity.

\section{The meaning of the doubled dimension \label{Structure equation reduction}}

With $\mathbf{T}^{a}=0$, the torsion Eq.(\ref{Torsion structure equation})
is in involution. This lets us first solve the structure equations
on a submanifold and results in a substantial restriction of the connection
forms. Extending back to the full space, we then work through the
full structure equations to determine the final form of each connection
form. 

\subsection{The involution}

The involution of the solder form,
\[
\mathbf{d}\mathbf{e}^{a}=\mathbf{e}^{b}\land\boldsymbol{\omega}_{\;b}^{a}+\boldsymbol{\omega}\land\mathbf{e}^{a}
\]
allows us to apply the Frobenius theorem, which tells us that there
exist $n$ functions on the manifold, $x^{\mu}$, such that
\[
\mathbf{e}^{a}=e_{\mu}^{\;\;\;a}\mathbf{d}x^{\mu}
\]
Furthermore, holding those functions constant, $x^{\mu}=x_{0}^{\mu}$,
so that $\mathbf{d}x^{\mu}=0$ and $\mathbf{e}^{a}=0$, the remaining
structure equations describe submanifolds of a foliation of the full
space. These remaining equations are
\begin{eqnarray}
\mathbf{d}\tilde{\boldsymbol{\omega}}_{\;b}^{a} & = & \tilde{\boldsymbol{\omega}}_{\;b}^{c}\land\tilde{\boldsymbol{\omega}}_{\;c}^{a}+\tilde{\boldsymbol{\Omega}}_{\;\;b}^{a}\nonumber \\
\mathbf{d}\tilde{\mathbf{f}}_{a} & = & \tilde{\boldsymbol{\omega}}_{\;a}^{b}\land\tilde{\mathbf{f}}_{b}+\tilde{\mathbf{f}}_{a}\land\tilde{\boldsymbol{\omega}}+\tilde{\mathbf{S}}_{a}\nonumber \\
\mathbf{d}\tilde{\boldsymbol{\omega}}\, & = & \tilde{\boldsymbol{\Omega}}\label{Submanifold structure equations}
\end{eqnarray}
where the tilde indicates the restriction to vanishing solder form,
e.g.,
\[
\tilde{\boldsymbol{\omega}}_{\;b}^{a}\equiv\left.\boldsymbol{\omega}_{\;b}^{a}\right|_{x^{\mu}=x_{0}^{\mu}}
\]
We will also examine the restriction of the integrability (i.e., the
Bianchi identity) of the co-solder equation,
\begin{eqnarray}
0 & = & \mathbf{d}^{2}\tilde{\mathbf{f}}_{a}\nonumber \\
 & = & \tilde{\boldsymbol{\Omega}}_{\;\;b}^{a}\land\tilde{\mathbf{f}}_{b}-\tilde{\boldsymbol{\omega}}_{\;a}^{b}\land\tilde{\mathbf{S}}_{b}+\tilde{\mathbf{S}}_{a}\land\tilde{\boldsymbol{\omega}}-\tilde{\mathbf{f}}_{a}\land\tilde{\boldsymbol{\Omega}}+\mathbf{d}\tilde{\mathbf{S}}_{a}\nonumber \\
\tilde{\mathbf{D}}\tilde{\mathbf{S}}_{a} & = & \tilde{\mathbf{f}}_{a}\land\tilde{\boldsymbol{\Omega}}-\tilde{\boldsymbol{\Omega}}_{\;\;b}^{a}\land\tilde{\mathbf{f}}_{b}\label{Submanifold cotorsion Bianchi}
\end{eqnarray}

When $\mathbf{e}^{a}=0$, the curvature, co-torsion, and dilatation
simplify to $\tilde{\boldsymbol{\Omega}}_{\;\;b}^{a}=\frac{1}{2}\Omega_{\;\;b}^{a\quad cd}\tilde{\mathbf{f}}_{c}\land\tilde{\mathbf{f}}_{d},\:\tilde{\mathbf{S}}_{a}=\frac{1}{2}S_{a}^{\;\;cd}\tilde{\mathbf{f}}_{c}\land\tilde{\mathbf{f}}_{d}$,
and $\tilde{\boldsymbol{\Omega}}=\frac{1}{2}\Omega^{cd}\tilde{\mathbf{f}}_{c}\land\tilde{\mathbf{f}}_{d}$.
In the previous section we showed that these components of the curvature
and dilatation, $\Omega_{\;\;b}^{a\quad cd}$ and $\Omega^{cd}$,
vanish. Therefore, the structure equations and basis integrability
reduce to
\begin{eqnarray*}
\mathbf{d}\tilde{\boldsymbol{\omega}}_{\;b}^{a} & = & \tilde{\boldsymbol{\omega}}_{\;b}^{c}\land\tilde{\boldsymbol{\omega}}_{\;c}^{a}\\
\mathbf{d}\tilde{\boldsymbol{\omega}}\, & = & 0\\
\mathbf{d}\tilde{\mathbf{f}}_{a} & = & \tilde{\boldsymbol{\omega}}_{\;a}^{b}\land\tilde{\mathbf{f}}_{b}+\tilde{\mathbf{f}}_{a}\land\tilde{\boldsymbol{\omega}}+\frac{1}{2}S_{a}^{\;\;cd}\tilde{\mathbf{f}}_{c}\land\tilde{\mathbf{f}}_{d}\\
\tilde{\mathbf{D}}\tilde{\mathbf{S}}_{a} & = & 0
\end{eqnarray*}
Let this submanifold be spanned by coordinates $y_{\mu}$. The first
two equations show that the spin connection, $\tilde{\boldsymbol{\omega}}_{\;b}^{a}$,
and Weyl vector, $\tilde{\boldsymbol{\omega}}$, are pure gauge on
the submanifold, 
\begin{eqnarray*}
\tilde{\boldsymbol{\omega}}_{\;\;b}^{a}\left(x_{0},y\right) & = & -\bar{F}_{\;\;c}^{a}\left(x_{0}^{\mu},y_{\nu}\right)\mathbf{d}F_{\;\;b}^{c}\left(x_{0}^{\mu},y_{\nu}\right)\\
\tilde{\boldsymbol{\omega}}\left(x_{0},y\right) & = & \mathbf{d}f\left(x_{0}^{\mu},y_{\nu}\right)
\end{eqnarray*}
where at each $x_{0}^{\mu}$ we are free to choose a local $SO\left(p,q\right)$
transformation $\Lambda_{\;\;c}^{a}\left(y\right)$ and a local dilatation
$\phi\left(y\right)$. This allows us to gauge both $\tilde{\boldsymbol{\omega}}_{\;\;b}^{a}\left(x_{0},y\right)$
and $\tilde{\boldsymbol{\omega}}\left(x_{0},y\right)$ to zero if
desired. It proves convenient to rename the restriction of the basis,
$\mathbf{h}_{a}\equiv\tilde{\mathbf{f}}_{a}$ and the restriction
of the spin connection as $\boldsymbol{\xi}_{\:\:\:b}^{a}\equiv\tilde{\boldsymbol{\omega}}_{\;\;\;b}^{a}\left(x_{0},y\right)$,
while gauging the Weyl vector to zero. The basis $\mathbf{h}_{a}$
must span the co-tangent space to the submanifold, so it must be nondegenerate.
The submanifold is then described by
\begin{eqnarray}
\mathbf{d}\boldsymbol{\xi}_{\;\;\;b}^{a} & = & \boldsymbol{\xi}_{\;\;\;b}^{c}\land\boldsymbol{\xi}_{\;\;\;c}^{a}\label{Spin connection on momentum submanifold}\\
\mathbf{d}\mathbf{h}_{a} & = & \boldsymbol{\xi}_{\;\;\;a}^{b}\land\mathbf{h}_{b}+\frac{1}{2}S_{a}^{\;\;cd}\mathbf{h}_{c}\land\mathbf{h}_{d}\label{Co solder form on submanifold}\\
\tilde{\mathbf{D}}\tilde{\mathbf{S}}_{a} & = & 0\label{Bianchi on submanifold}
\end{eqnarray}
To continue, we examine a manifold with these conditions, Eqs.(\ref{Spin connection on momentum submanifold}-\ref{Bianchi on submanifold}).
Notice that Eqs.(\ref{Spin connection on momentum submanifold})-(\ref{Bianchi on submanifold})
describe a differentiable manifold with flat connection for which
the momentum part of the co-torsion $\frac{1}{2}S_{a}^{\;\;cd}\mathbf{h}_{c}\land\mathbf{h}_{d}$
is the torsion of the submanifold. This submanifold torsion is constrained
by Eq.(\ref{Co-torsion constraints}),
\begin{equation}
\left(1+\chi\right)S_{a}^{\;\;\;bc}=0\label{Momentum co-torsion constraint}
\end{equation}
The importance of these properties will be seen in this and the following
Sections. 

\subsection{Foliation by a Lie group}

We quote a well-known theorem due to Auslander and Markus \cite{AuslanderMarkus}:
\begin{quotation}
THEOREM 5. Let M be a differentiable manifold with complete, flat,
affine connection $\Gamma$ and holonomy group H($\mathcal{M}$; $\Gamma$)
= 0. Then $\mathcal{M}$ is a complete Riemann space with Christoffel
connection $\Gamma$ and $\mathcal{M}$ is diferentiably isometric
with a torus space.
\end{quotation}
F. W. Kamber and Ph. Tondeur generalize this theorem \cite{KamberTondeur},
introducing their proof with the following:
\begin{quotation}
Consider a linear connection on a smooth manifold. The connection
is flat, if the curvature tensor $R$ is zero. If the torsion tensor
$T$ has vanishing covariant derivative, the torsion is said to be
parallel. A linear connection is complete, if every geodesic can be
defined for any real value of the affine parameter. In this note the
following structure theorem for smooth manifolds admitting a complete
flat connection with parallel torsion is proved: \emph{Any such manifold
is the orbit space of a simply connected Lie group $\mathcal{G}$
under a properly discontinuous and fixed-point free action of a subgroup
of the affine group of $\mathcal{G}$.} This Theorem includes the
classical cases of flat Riemannian manifolds and flat affine manifolds
(Auslander and Markus), where the torsion is assumed to be zero and
$\mathcal{G}$ turns out to be $\mathbb{R}^{n}$, and also generalizes
a theorem of Hicks {[}Theorem 6{]} for complete connections with trivial
holonomy group and parallel torsion tensor, stating that a manifold
with such a connection is homogeneous. We consider the case where
the curvature vanishes, without requiring the holonomy group to be
trivial.
\end{quotation}
As noted above, our equations, Eqs.(\ref{Spin connection on momentum submanifold}-\ref{Bianchi on submanifold}),
exactly describe a manifold with flat connection $\boldsymbol{\xi}_{\;\;\;b}^{a}$
but with torsion satisfying only $\mathbf{D}\tilde{\mathbf{S}}_{a}=0$.
This torsion conditions is weaker than those of the theorems above.
Moreover, the additional conditions may or may not hold. Spacetime,
and the more general $SO\left(p,q\right)$ spaces we consider may
be pseudo-Riemannian rather than Riemannian. Further, we know that
spacetimes are generically incomplete \cite{GerochKronheimerPenrose,HawkingEllis,Wald}
and that our physical spacetime contains black hole singularities
and initial time incompleteness; the corresponding properties of the
momentum subspace depend on the manifold chosen during the quotient
construction. Finally, with our general considerations we cannot be
certain of the remaining specifications regarding holonomy present
in both theorems. Therefore, we do not attempt to apply Auslander-Markus
or Kamber-Tondeur theorems, but derive our results directly, making
our assumptions explicit.

We consider the two possible solutions to Eq.(\ref{Momentum co-torsion constraint}):
\begin{description}
\item [{Case$\:$1:$\:$$S_{a}^{\;\;\;bc}=0.$}] In Sec.(\ref{sec:Abelian-case:})
below, we show that with no further assumptions, generic biconformal
spaces (i.e., those with $\chi\neq-1$) are foliated by an abelian
Lie group. They therefore describe either the co-tangent bundle or
torus space foliations over $SO\left(p,q\right)$ spaces. Generically,
therefore, the conclusion of Theorem 5 of Auslander and Markus holds
for the momentum submanifolds of biconformal space.
\item [{Case$\:$2:$\:$$\chi=-1.$}] In Sec.(\ref{sec:Non-abelian-case})
below, we show that the subclass of biconformal spaces with $1+\chi=0$
allows the possibility of foliation by a nonabelian Lie group. The
result is consistent with the claim of Kamber and Tondeur. To make
further progress, we too assume vanishing covariant derivative of
the torsion rather than vanishing covariant exterior derivative.
\end{description}
In the remainder of this Section and in Sec.(\ref{sec:The full space}),
we show results that hold for either Case 1 or Case 2 by assuming
$S_{a}^{\;\;\;bc}$ constant and placing no condition on $\chi$.
This is sufficient to show foliation by a Lie group; we leave detailed
topological discussion to subsequent studies. In Sec.(\ref{sec:The full space})
we extend back to the full biconformal space, substituting the form
of the connection into the structure equations to continue the reduction
of the system toward general relativity. The program is completed
in two different ways for Case 1 and Case 2, in Sections (\ref{sec:Abelian-case:})
and (\ref{sec:Non-abelian-case}) respectively.

\subsubsection{Co-torsion Bianchi}

We have seen that the vanishing torsion, $\mathbf{T}^{a}=0$, combined
with the dilatation Bianchi identity gives Eqs.(\ref{Momentum co-torsion constraint}).
For the remainder of this Section, we will place a weaker constraint
on the momentum co-torsion and $\chi$ consistent with both Cases
above. Thus, the conclusions of this Section for the discussions of
both Sec.(\ref{sec:Abelian-case:}) and Sec.(\ref{sec:Non-abelian-case}).

The integrability condition for the submanifold co-torsion, Eq.(\ref{Bianchi on submanifold})
is
\begin{eqnarray}
0 & = & \tilde{\mathbf{D}}\tilde{\mathbf{S}}_{a}\nonumber \\
 & = & \frac{1}{2}S_{a}^{\;\;\left[\alpha\beta;\mu\right]}\mathbf{d}y_{\alpha}\land\mathbf{d}y_{\beta}\land\mathbf{d}y_{\mu}\label{Submanifold Bianchi}
\end{eqnarray}
so the covariant exterior $y$-derivative of $\frac{1}{2}S_{a}^{\;\;cd}\mathbf{h}_{c}\wedge\mathbf{h}_{d}$
vanishes. Choosing the $y_{\alpha}$-dependent part of the gauge so
that the submanifold spin connection and Weyl vector vanish, the covariant
derivative reduces to a partial derivative,
\begin{eqnarray*}
0 & = & \tilde{\mathbf{d}}\tilde{\mathbf{S}}_{a}
\end{eqnarray*}
and therefore for some $1$-form, $\boldsymbol{\xi}_{a}$
\begin{eqnarray*}
\tilde{\mathbf{S}}_{a} & = & \tilde{\mathbf{d}\boldsymbol{\xi}_{a}}
\end{eqnarray*}
In coordinate components,
\[
S_{a}^{\;\;\alpha\beta}=\xi_{a}^{\;\;\alpha,\beta}-\xi_{a}^{\;\;\beta,\alpha}
\]

However, instead of such a general potential $\tilde{\boldsymbol{\xi}}_{a}$,
we assume
\begin{eqnarray}
\partial^{\mu}S_{a}^{\;\;\;\alpha\beta} & = & 0\label{Constancy of momentum co-torsion}
\end{eqnarray}
This is one of the assumptions of the Kamber-Tondeur Theorem.

With the momentum co-torsion independent of $y_{\mu}$ the structure
equation on the $\mathbf{e}^{a}=0$ submanifolds becomes
\begin{equation}
\mathbf{d}\mathbf{h}_{a}=\frac{1}{2}S_{a}^{\;\;cd}\left(x_{0}\right)\mathbf{h}_{c}\land\mathbf{h}_{d}\label{Structure equation for h}
\end{equation}
In terms of the basis $\mathbf{h}_{a}$ the integrability condition
for eq.(\ref{Structure equation for h}) is
\begin{eqnarray*}
0 & \equiv & \mathbf{d}^{2}\mathbf{h}_{a}\\
 & = & \frac{1}{2}S_{a}^{\;\;cd}\left(x_{0}\right)\mathbf{h}_{c}\land\mathbf{h}_{d}\\
 & = & S_{a}^{\;\;cd}\left(x_{0}\right)\mathbf{d}\mathbf{h}_{c}\land\mathbf{h}_{d}\\
 & = & \frac{1}{2}S_{a}^{\;\;cd}\left(x_{0}\right)S_{c}^{\;\;ef}\left(x_{0}\right)\mathbf{h}_{e}\land\mathbf{h}_{f}\land\mathbf{h}_{d}
\end{eqnarray*}
and therefore,
\[
S_{a}^{\;\;c[d}S_{c}^{\;\;ef]}=0
\]
With $S_{a}^{\;\;bc}\left(x_{0}^{\mu}\right)$ constant, we set $c_{a}^{\quad bc}\equiv-S_{a}^{\;\;bc}$,
and observe that the pair
\begin{eqnarray}
\mathbf{d}\mathbf{h}_{a} & = & -\frac{1}{2}c_{a}^{\;\;bc}\mathbf{h}_{b}\land\mathbf{h}_{c}\label{Maurer-Cartan for h}\\
c_{a}^{\;\;c[d}c_{c}^{\;\;ef]} & = & 0\label{Jacobi identity}
\end{eqnarray}
form the Maurer-Cartan equations and the Jacobi identity (in the adjoint
representation) for an $n$-dimensional Lie algebra. The field equation
for the momentum co-torsion, Eq.(\ref{Momentum co torsion field equation})
shows that the adjoint generators are traceless, so when the adjoint
representation is faithful the Lie group elements will have unit determinant.
With the observation that $\mathbf{h}_{a}$ has an $n$-dimensional
$SO\left(p,q\right)$ or $Spin\left(p,q\right)$ index (depending
on which representation we have chosen for the beginning group), we
have therefore proved the following theorem:\medskip{}

\begin{description}
\item [{Theorem:}] \emph{In any $2n$-dimensional, torsion-free biconformal
spaces with $\partial^{\mu}S_{a}^{\;\;\;\alpha\beta}=0$, there exists
an n-dimensional foliation by a Lie group. If the adjoint representation
is faithful, the group is special.}
\end{description}
\medskip{}

We conjecture that the theorem holds for all torsion free biconformal
spaces.

This is one of our most important new results, giving a definitive
interpretation to the doubled dimension of biconformal spaces.

Introducing vector fields $G^{a}$ dual to the one forms $\mathbf{h}_{a}$,
we have
\begin{equation}
\left[G^{a},G^{b}\right]=c_{c}^{\;\;ab}G^{c}\label{G Lie algebra}
\end{equation}
and
\begin{eqnarray}
\left[G^{a},\left[G^{b},G^{c}\right]\right]+\left[G^{b},\left[G^{c},G^{a}\right]\right]+\left[G^{c},\left[G^{a},G^{b}\right]\right] & = & 0\label{G Jacobi identity}
\end{eqnarray}
Let $\mathcal{G}$ be the Lie group generated by the $G^{a}$. Then
the $y$-submanifold at each $x_{0}$ is the group manifold.

Since $\mathbf{h}_{a}$ transforms as a vector under $SO\left(p,q\right)$,
there may be constraints between $SO\left(p,q\right)$ and $\mathcal{G}$.

Acting with $\Lambda_{\;\;b}^{a}\in SO\left(p,q\right)$ on the structure
equation of $\mathbf{h}_{a}$,
\[
\tilde{\mathbf{h}}_{a}=\mathbf{h}_{a}\Lambda_{\;\;b}^{a}
\]
invariance of the structure equation requires
\begin{eqnarray*}
\mathbf{d}\tilde{\mathbf{h}}_{a} & = & -\frac{1}{2}\tilde{c}_{a}^{\;\;bc}\tilde{\mathbf{h}}_{b}\land\tilde{\mathbf{h}}_{c}\\
\mathbf{d}\mathbf{h}_{a}\Lambda_{\;\;b}^{a} & = & -\frac{1}{2}\tilde{c}_{a}^{\;\;bc}\mathbf{h}_{d}\Lambda_{\;\;b}^{d}\land\mathbf{h}_{e}\Lambda_{\;\;c}^{e}\\
-\frac{1}{2}c_{a}^{\;\;de}\mathbf{h}_{d}\land\mathbf{h}_{e}\Lambda_{\;\;b}^{a} & = & -\frac{1}{2}\tilde{c}_{a}^{\;\;bc}\Lambda_{\;\;b}^{d}\Lambda_{\;\;c}^{e}\mathbf{h}_{d}\land\mathbf{h}_{e}\\
c_{f}^{\;\;de} & = & \bar{\Lambda}_{\;\;f}^{a}\tilde{c}_{a}^{\;\;bc}\Lambda_{\;\;b}^{d}\Lambda_{\;\;c}^{e}
\end{eqnarray*}
so the structure constants must transform as a $\left(\begin{array}{c}
2\\
1
\end{array}\right)$ tensor, consistent with $\mathbf{S}_{a}$ being a tensor. Since $SO\left(p,q\right)$
acts on itself, any subgroup of $SO\left(p,q\right)$ will be allowed,
but it is clear that there are additional possibilities. For example,
 the vanishing structure constants of an abelian group will be preserved,
as will partly abelian combinations. We develop a concrete example.

Starting with a $3$-dim representation of $SO\left(3\right)$, we
require a $3$-dimensional Lie group with structure constants that
transform as a tensor under $SO\left(3\right)$. Consider $ISO\left(2\right)$,
the two translations and one rotation of the plane. The Lie algebra
is
\begin{eqnarray*}
\left[R,T_{1}\right] & = & -2T_{2}\\
\left[R,T_{2}\right] & = & 2T_{1}
\end{eqnarray*}
where we may think of $R$ as the generator of rotations about the
$z$-axis and $T_{k}$ as translations in the $xy$-plane. The nonvanishing
structure constants (using the conventional index positions) are then
$c_{\;\;32}^{1}\;\;=\;\;c_{\;\;13}^{1}=-c_{\;\;23}^{1}\;\;=\;\;-c_{\;\;31}^{2}\;\;=\;\;2$.

While this 3-dim picture of the group is clearly rotationally invariant,
we may make the proof explicit by defining three unit vectors
\begin{eqnarray*}
n_{\left(a\right)}^{i} & \equiv & \delta_{a}^{i}
\end{eqnarray*}
Letting the generators in an arbitrary basis form a 3-vector $\mathbf{G}=\left(G_{1},G_{2},G_{3}\right)$
we set
\begin{eqnarray*}
R & = & \mathbf{n}_{\left(3\right)}\cdot\mathbf{G}\\
T_{1} & = & \mathbf{n}_{\left(1\right)}\cdot\mathbf{G}\\
T_{2} & = & \mathbf{n}_{\left(2\right)}\cdot\mathbf{G}
\end{eqnarray*}
Then the structure constants may be written in terms of the unit vectors
as
\[
c_{\;\;jk}^{i}=2n_{\left(1\right)}^{i}\left(n_{\left(3\right)j}n_{\left(2\right)k}-n_{\left(2\right)j}n_{\left(3\right)k}\right)-2n_{\left(2\right)}^{i}\left(n_{\left(3\right)j}n_{\left(1\right)k}-n_{\left(1\right)j}n_{\left(3\right)k}\right)
\]
which now manifestly transforms as a $\left(\begin{array}{c}
1\\
2
\end{array}\right)$ tensor under rotations.

For $n=4$, the electroweak group, $SU\left(2\right)\times U\left(1\right)$,
naturally springs to mind. This, and other particular cases will be
explored explicitly in subsequent work.

\medskip{}

Now consider the effect on $S_{a}^{\;\;bc}$ of allowing $x^{\mu}$
to vary. At each value of $x^{\mu}$, $S_{a}^{\;\;cd}\left(x\right)$,
comprises the structure constants of a Lie group. However, while the
structure constants depend on the choice of the basis of group generator,
we are limited to differentiable changes. Since Lie algebras are classified
by a discrete set of possible root diagrams, a continuous transformation
as we vary $x^{\mu}$ cannot change to structure constants with a
different root diagram. Moreover, by choosing the basis of dual $1$-forms
appropriately at each point, we may bring the structure constants
to a given standard form $c_{a}^{\quad bc}$. With $S_{a}^{\;\;bc}\left(x\right)=-c_{a}^{\quad bc}$
at each value of $x^{\mu}$, there is no $x$-dependence. With any
such choice of basis, $\mathbf{d}S_{a}^{\quad bc}=0$, and we may
set
\begin{eqnarray*}
S_{a}^{\quad bc}\left(x,y\right) & = & -c_{a}^{\quad bc}
\end{eqnarray*}
across the entire biconformal manifold.

If the group $\mathcal{G}$ is abelian, the structure constants are
zero and the momentum co-torsion vanishes. In this case, $\mathbf{d}\mathbf{h}_{a}=0$
and we have
\[
\mathbf{h}_{a}=f_{a}^{\;\;\mu}\left(x\right)\mathbf{d}y_{\mu}
\]
where the $y$-dependence of the coefficients must now vanish, though
in this case it is useful to allow the dependence on $x^{\mu}$. This
describes an exact, orthonormal frame and therefore a flat space.
Since we evaluate at fixed $x^{\alpha}=x_{0}^{\alpha}$, the coefficients
$f_{a}^{\;\;\mu}\left(x_{0}\right)$ are constants on the submanifold,
but may be functions of $x^{\mu}$ when we extend back to the full
biconformal space. This is equivalent to the abelian Lie algebra of
$n$ translations, and the $\mathcal{G}$-foliation of the biconformal
space may be identified as the co-tangent bundle of the remaining
$SO\left(p,q\right)$ space. Alternatively, the abelian group may
be taken as a compactification on a torus.

\subsubsection{Parameterization of the group elements as coordinates}

The integral of the structure equations,
\begin{eqnarray*}
\mathbf{d}\mathbf{h}_{a} & = & -\frac{1}{2}c_{a}^{\;\;bc}\mathbf{h}_{b}\land\mathbf{h}_{c}\\
S_{a}^{\;\;c[d}S_{c}^{\;\;ef]} & = & 0
\end{eqnarray*}
gives the group manifold, which is most easily coordinatized by the
group elements. We may find these by exponentiating the Lie algebra,
$V=\left\{ y_{a}G^{a}|y_{a}\in R^{n}\right\} $ where $G^{a}$ satisfy
the Lie algebra relations in Eqs.(\ref{G Lie algebra}) and (\ref{G Jacobi identity}).
The group elements may be parameterized by coordinates $y_{a}$, by
exponentiating the Lie algebra, $g\left(y\right)=e^{y_{a}G^{a}}\in\mathcal{G}$.

The basis forms, $\mathbf{h}_{a}$ may be explicitly turned into Lie
algebra valued one forms using any desired linear representation of
the generators, $\boldsymbol{\xi}_{\;\;B}^{C}\equiv-\mathbf{h}_{a}\left[G^{a}\right]_{B}^{\;\;C}$.
For example, using the adjoint representation we contract a copy of
the structure constants with the Maurer-Cartan equations, eq.(\ref{Maurer-Cartan for h}),
and define $\boldsymbol{\xi}_{\;\;b}^{c}\equiv-\mathbf{h}_{a}\left[G^{a}\right]_{b}^{\;\;c}=-c_{b}^{\;\;ac}\mathbf{h}_{a}$.
Then, using the Jacobi identity,
\begin{eqnarray*}
\mathbf{d}\left(c_{b}^{\;\;ac}\mathbf{h}_{a}\right) & = & -\frac{1}{2}c_{b}^{\;\;ac}c_{a}^{\;\;de}\mathbf{h}_{d}\land\mathbf{h}_{e}\\
-\mathbf{d}\boldsymbol{\xi}_{\;\;b}^{c} & = & \frac{1}{2}\left(c_{b}^{\;\;ad}c_{a}^{\;\;ec}+c_{b}^{\;\;ae}c_{a}^{\;\;cd}\right)\mathbf{h}_{d}\land\mathbf{h}_{e}\\
 & = & \frac{1}{2}\left(-c_{b}^{\;\;da}\mathbf{h}_{d}\land c_{a}^{\;\;ec}\mathbf{h}_{e}-c_{b}^{\;\;ea}\mathbf{h}_{e}\land c_{a}^{\;\;dc}\mathbf{h}_{d}\right)\\
 & = & \frac{1}{2}\left(-\boldsymbol{\xi}_{\;\;b}^{a}\land\boldsymbol{\xi}_{\;\;a}^{c}-\boldsymbol{\xi}_{\;\;b}^{a}\land\boldsymbol{\xi}_{\;\;a}^{c}\right)\\
 & = & -\boldsymbol{\xi}_{\;\;b}^{a}\land\boldsymbol{\xi}_{\;\;a}^{c}
\end{eqnarray*}
or
\begin{eqnarray*}
\mathbf{d}\boldsymbol{\xi}_{\;\;b}^{c} & = & \boldsymbol{\xi}_{\;\;b}^{a}\land\boldsymbol{\xi}_{\;\;a}^{c}
\end{eqnarray*}
This is the structure equation of a connection on a flat manifold,
so we may write $\boldsymbol{\xi}_{\;\;b}^{a}$ as a pure gauge connection,
\begin{eqnarray*}
\boldsymbol{\xi}_{\;\;b}^{a} & = & -\bar{g}_{\;b}^{c}\mathbf{d}g_{\;\;c}^{a}
\end{eqnarray*}
where, in the adjoint representation, $g_{\;\;b}^{a}=\exp\left(y_{c}\left[G^{c}\right]_{\;\;b}^{a}\right)=\exp\left(y_{c}c_{b}^{\;\;ca}\right)$.
We check that this solves the structure equation,
\begin{eqnarray*}
\mathbf{d}\boldsymbol{\xi}_{\;\;b}^{c} & = & \boldsymbol{\xi}_{\;\;b}^{a}\land\boldsymbol{\xi}_{\;\;a}^{c}\\
\mathbf{d}\left(-\bar{g}_{\;b}^{e}\mathbf{d}g_{\;\;e}^{c}\right) & = & \left(-\bar{g}_{\;b}^{e}\mathbf{d}g_{\;\;e}^{a}\right)\land\left(-\bar{g}_{\;a}^{f}\mathbf{d}g_{\;\;f}^{c}\right)\\
-\mathbf{d}\bar{g}_{\;b}^{e}\land\mathbf{d}g_{\;\;e}^{c} & = & -\mathbf{d}\bar{g}_{\;b}^{e}g_{\;\;e}^{a}\land\bar{g}_{\;a}^{f}\mathbf{d}g_{\;\;f}^{c}\\
 & = & -\mathbf{d}\bar{g}_{\;b}^{e}\land\mathbf{d}g_{\;\;e}^{c}
\end{eqnarray*}
as required. We note that while this construction gives an explicit
form for $\mathbf{h}_{a}$, this is not the usual connection, $\boldsymbol{\omega}_{\;\;b}^{c}\equiv-\frac{1}{2}\mathbf{h}_{a}\left[G^{a}\right]_{b}^{\;\;c}$,
which gives constant curvature \cite{Milnor}.

\section{Returning to the full space \label{sec:The full space}}

We have established a geometric breakdown of the $2n$-dim biconformal
space into an $n$-dimensional foliation with a Lie group for leaves.
However, the connection forms still retain their dependence on the
full set of coordinates $\left(x^{\alpha},y_{\beta}\right)$. In this
Section, and for the generic $\chi\neq-1$ case in the next Section,
we show that the structure equations further restrict this dependence
so that except for certain explicit linear $y_{\alpha}$ dependence,
all fields depend only on the $x^{\alpha}$, up to coordinate choices
and gauge transformations. To this end, we turn to the full space
and the Cartan structure equations.

When we restore the solder form, letting $x^{\mu}$ vary again, the
connection forms must be given by their $\mathbf{e}^{a}=0$ parts,
Eqs.(\ref{Spin connection on momentum submanifold}, \ref{Maurer-Cartan for h},
and the vanishing Weyl vector) plus additional parts proportional
to the solder form. Therefore,
\begin{eqnarray}
\boldsymbol{\omega}_{\;\;b}^{a} & = & \omega_{\;\;bc}^{a}\left(x,y\right)\mathbf{e}^{c}\label{Reduced spin connection}\\
\mathbf{e}^{a} & = & e_{\mu}^{\;\;\;a}\left(x,y\right)\mathbf{d}x^{\mu}\label{Reduced solder form}\\
\mathbf{f}_{a} & = & h_{a}^{\;\;\mu}\left(x,y\right)\mathbf{d}y_{\mu}+c_{ab}\left(x,y\right)\mathbf{e}^{b}\nonumber \\
 & \equiv & \mathbf{h}_{a}+c_{ab}\mathbf{e}^{b}\nonumber \\
 & \equiv & \mathbf{h}_{a}+\mathbf{c}_{a}\label{Reduced co solder form}\\
\boldsymbol{\omega} & = & W_{a}\left(x,y\right)\mathbf{e}^{a}\label{Reduced Weyl vector}
\end{eqnarray}
Eqs. (\ref{Reduced spin connection}) - (\ref{Reduced Weyl vector})
hold as long as we perform only $x$-dependent fiber transformations.
While this form is convenient for recognizing the content of the geometry,
the biconformal space is unchanged by more general transformations.
General $\left(x^{\alpha},y_{\beta}\right)$-dependent transformations
on biconformal space act similarly to canonical transformations on
phase spaces. They do not change the underlying physics.

We substitute these forms into the structure equations, with the reduced
curvatures as given in Eq.(\ref{Reduced curvatures}), 
\begin{eqnarray}
\mathbf{d}\boldsymbol{\omega}_{\;b}^{a} & = & \boldsymbol{\omega}_{\;b}^{c}\wedge\boldsymbol{\omega}_{\;c}^{a}+2\left(1+\chi\right)\Delta_{cb}^{ad}\mathbf{f}_{d}\wedge\mathbf{e}^{c}+\frac{1}{2}\Omega_{\;\;bcd}^{a}\mathbf{e}^{c}\wedge\mathbf{e}^{d}\label{Reduced curvature}\\
\mathbf{d}\mathbf{e}^{a} & = & \mathbf{e}^{b}\wedge\boldsymbol{\omega}_{\;\;b}^{a}+\boldsymbol{\omega}\wedge\mathbf{e}^{a}\label{Reduced torsion}\\
\mathbf{d}\mathbf{f}_{a} & = & \boldsymbol{\omega}_{\;a}^{b}\wedge\mathbf{f}_{b}+\mathbf{f}_{a}\wedge\boldsymbol{\omega}+\frac{1}{2}S_{acd}\mathbf{e}^{c}\wedge\mathbf{e}^{d}+S_{a\quad d}^{\;\;c}\mathbf{f}_{c}\wedge\mathbf{e}^{d}-\frac{1}{2}c_{a}^{\;\;cd}\mathbf{f}_{c}\wedge\mathbf{f}_{d}\label{Reduced co torsion}\\
\mathbf{d}\boldsymbol{\omega}\, & = & \left(1+\chi\right)\mathbf{e}^{c}\wedge\mathbf{f}_{c}\label{Reduced dilatation}
\end{eqnarray}
where $\chi\equiv\frac{1}{n-1}\frac{1}{\left(n-1\right)\alpha-\beta}\Lambda$
and $\left(1+\chi\right)$ factors appear where we have combined the
dilatation and curvature cross terms with matching pieces of the connection.

\subsection{The basis structure equations \label{subsec:The-basis-structure-equation}}

First consider the solder form equation, Eq.(\ref{Reduced torsion}).
Substituting Eqs.(\ref{Reduced spin connection}), (\ref{Reduced solder form})
and (\ref{Reduced Weyl vector}) for the current form of the connection,
\begin{eqnarray*}
\mathbf{d}x^{\mu}\wedge\partial_{\mu}\mathbf{e}^{a}+\mathbf{d}y_{\mu}\wedge\partial^{\mu}\mathbf{e}^{a} & = & \boldsymbol{\omega}_{\;\;bc}^{a}\mathbf{e}^{b}\wedge\mathbf{e}^{c}+W_{b}\mathbf{e}^{b}\wedge\mathbf{e}^{a}
\end{eqnarray*}
The sole mixed term must vanish, $\mathbf{d}y_{\mu}\wedge\partial^{\mu}\mathbf{e}^{a}=0$,
and this requires the solder form to be independent of $y_{\alpha}$.
Therefore,
\begin{equation}
e_{\mu}^{\;\;a}\left(x,y\right)=e_{\mu}^{\;\;a}\left(x\right)\label{x dependence of solder form}
\end{equation}

\subsubsection{Solving the solder form equation for the spin connection}

The next step is to solve the solder form equation for the spin connection.
In the remaining $\mathbf{e}^{a}\wedge\mathbf{e}^{b}$ part of the
reduced solder form equation, Eq.(\ref{Reduced solder form}) we may
separate the connection into the familiar metric compatible piece,
and a Weyl vector piece. Let $\boldsymbol{\alpha}_{\;\;b}^{a}$ be
chosen as the $\mathbf{e}^{a}$-compatible connection, so that
\begin{equation}
\mathbf{d}\mathbf{e}^{a}=\mathbf{e}^{b}\wedge\boldsymbol{\alpha}_{\;b}^{a}\label{Compatible connection}
\end{equation}
We note that as a consequence of Eqs.(\ref{x dependence of solder form})
and (\ref{Compatible connection}), $\boldsymbol{\alpha}_{\;b}^{a}=\boldsymbol{\alpha}_{\;b}^{a}\left(x\right)$.
Then writing $\boldsymbol{\omega}_{\;b}^{a}=\boldsymbol{\alpha}_{\;b}^{a}+\boldsymbol{\beta}_{\;b}^{a}$
with antisymmetry on each piece, $\boldsymbol{\alpha}_{\;b}^{a}=-\eta^{ac}\eta_{bd}\boldsymbol{\alpha}_{\;c}^{d}$
and $\boldsymbol{\beta}_{\;b}^{a}=-\eta^{ac}\eta_{bd}\boldsymbol{\beta}_{\;c}^{d}$,
we must have
\[
0=\mathbf{e}^{b}\wedge\boldsymbol{\beta}_{\;b}^{a}+\boldsymbol{\omega}\wedge\mathbf{e}^{a}
\]
Since the solution is unique up to local Weyl transformations, we
need only find an expression that works. Using the antisymmetric $\left(\begin{array}{c}
1\\
1
\end{array}\right)$ projection operator $\Delta_{bd}^{ac}$ to impose antisymmetry, and
requiring linearity in the Weyl vector and the solder form, we guess
that
\begin{equation}
\boldsymbol{\beta}_{\;\;b}^{a}=-2\Delta_{db}^{ac}W_{c}\mathbf{e}^{d}\label{Beta}
\end{equation}
and check
\begin{eqnarray*}
\mathbf{e}^{b}\wedge\boldsymbol{\beta}_{\;b}^{a}+\boldsymbol{\omega}\wedge\mathbf{e}^{a} & = & -2\Delta_{db}^{ac}W_{c}\mathbf{e}^{b}\wedge\mathbf{e}^{d}+\boldsymbol{\omega}\wedge\mathbf{e}^{a}\\
 & = & -\left(\delta_{d}^{a}\delta_{b}^{c}-\eta^{ac}\eta_{bd}\right)W_{c}\mathbf{e}^{b}\wedge\mathbf{e}^{d}+\boldsymbol{\omega}\wedge\mathbf{e}^{a}\\
 & = & -W_{b}\mathbf{e}^{b}\wedge\mathbf{e}^{a}+\boldsymbol{\omega}\wedge\mathbf{e}^{a}\\
 & = & 0
\end{eqnarray*}
as required. Therefore, the spin connection is
\begin{equation}
\boldsymbol{\omega}_{\;b}^{a}=\boldsymbol{\alpha}_{\;b}^{a}+\boldsymbol{\beta}_{\;b}^{a}=\boldsymbol{\alpha}_{\;b}^{a}-2\Delta_{db}^{ac}W_{c}\mathbf{e}^{d}\label{Spin connection}
\end{equation}
where $\boldsymbol{\alpha}_{\;b}^{a}$ is the connection compatible
with $\mathbf{e}^{a}$. Any $y$-dependence must come from the Weyl
vector.

\subsubsection{Coordinate form of the connection}

We can also find the coordinate form of the connection. Starting from
the solder form equation we expand,
\begin{eqnarray*}
0 & = & \mathbf{d}\mathbf{e}^{a}-\mathbf{e}^{b}\wedge\boldsymbol{\omega}_{\;b}^{a}-\boldsymbol{\omega}\wedge\mathbf{e}^{a}\\
 & = & \left(\partial_{\mu}e_{\nu}^{\;\;a}+e_{\nu}^{\;\;b}\omega_{\;b\mu}^{a}-W_{\mu}e_{\nu}^{\;\;a}\right)\mathbf{d}x^{\mu}\wedge\mathbf{d}x^{\nu}
\end{eqnarray*}
As the antisymmetric part of the coefficient expression in parentheses
must vanish, it must equal a symmetric object, i.e.,
\[
\Sigma_{\;\;\nu\mu}^{a}\equiv\partial_{\mu}e_{\nu}^{\;\;a}+e_{\nu}^{\;\;b}\omega_{\;b\mu}^{a}-W_{\mu}e_{\nu}^{\;\;a}
\]
where $\Sigma_{\;\;\nu\mu}^{a}=\Sigma_{\;\;\mu\nu}^{a}$. Writing
$\Sigma_{\;\;\nu\mu}^{a}=e_{\alpha}^{\;\;a}\Sigma_{\;\;\nu\mu}^{\alpha}$
the equation takes the form of a vanishing covariant derivative,
\begin{eqnarray*}
D_{\mu}e_{\nu}^{\;\;a} & = & \partial_{\mu}e_{\nu}^{\;\;a}-e_{\alpha}^{\;\;a}\Sigma_{\;\;\nu\mu}^{\alpha}+e_{\nu}^{\;\;b}\omega_{\;b\mu}^{a}-W_{\mu}e_{\nu}^{\;\;a}\;\;=\;\;0
\end{eqnarray*}
We easily check that $\Sigma_{\;\;\nu\mu}^{\alpha}$ is indeed the
expected connection. First, contract with $\eta_{ab}e_{\beta}^{\;\;b}$,
\begin{eqnarray*}
0 & = & \eta_{ab}e_{\beta}^{\;\;b}\partial_{\mu}e_{\nu}^{\;\;a}-\eta_{ab}e_{\beta}^{\;\;b}e_{\alpha}^{\;\;a}\Sigma_{\;\;\nu\mu}^{\alpha}+\eta_{ab}e_{\beta}^{\;\;b}e_{\nu}^{\;\;c}\omega_{\;c\mu}^{a}-\eta_{ab}e_{\beta}^{\;\;b}W_{\mu}e_{\nu}^{\;\;a}
\end{eqnarray*}
Now symmetrize on $\beta\nu$ and use $g_{\alpha\beta}=\eta_{ab}e_{\alpha}^{\;\;a}e_{\beta}^{\;\;b}$,
\begin{eqnarray*}
0 & = & \eta_{ab}e_{\beta}^{\;\;b}\partial_{\mu}e_{\nu}^{\;\;a}-g_{\alpha\beta}\Sigma_{\;\;\nu\mu}^{\alpha}+e_{\beta}^{\;\;b}e_{\nu}^{\;\;c}\eta_{ab}\omega_{\;c\mu}^{a}-\eta_{ab}e_{\beta}^{\;\;b}e_{\nu}^{\;\;a}W_{\mu}\\
 &  & +\eta_{ab}e_{\nu}^{\;\;b}\partial_{\mu}e_{\beta}^{\;\;a}-g_{\alpha\nu}\Sigma_{\;\;\beta\mu}^{\alpha}+e_{\nu}^{\;\;b}e_{\beta}^{\;\;c}\eta_{ab}\omega_{\;c\mu}^{a}-\eta_{ab}e_{\nu}^{\;\;b}e_{\beta}^{\;\;a}W_{\mu}\\
 & = & \partial_{\mu}g_{\nu\beta}-g_{\alpha\beta}\Sigma_{\;\;\nu\mu}^{\alpha}-g_{\nu\alpha}\Sigma_{\;\;\beta\mu}^{\alpha}-2g_{\nu\beta}W_{\mu}
\end{eqnarray*}
This is precisely the conformal metric compatibility of $g_{\nu\beta}$
with $\Sigma_{\;\;\beta\mu}^{\alpha}$ in a Weyl geometry. Solving
for the connection by the usual cyclic permution of $\mu\nu\beta$,
adding the first two permutations and subtracting the third, we recover
the explicit form of the compatible connection of a Weyl geometry
\cite{WeylGeom}:
\begin{eqnarray}
\Sigma_{\nu\beta\mu} & = & \frac{1}{2}\left(\partial_{\mu}g_{\nu\beta}+\partial_{\beta}g_{\mu\nu}-\partial_{\nu}g_{\beta\mu}\right)-\left(g_{\nu\beta}W_{\mu}+g_{\mu\nu}W_{\beta}-g_{\beta\mu}W_{\nu}\right)\nonumber \\
 & = & \Gamma_{\nu\beta\mu}-\left(g_{\nu\beta}W_{\mu}+g_{\mu\nu}W_{\beta}-g_{\beta\mu}W_{\nu}\right)\label{Weyl connection}
\end{eqnarray}
where $\Gamma_{\nu\beta\mu}$ is the Christoffel connection. This
connection is compatible with the conformal class of metrics, $\left\{ \left.g_{\alpha\beta}e^{2\phi}\right|all\:\phi\left(x,y\right)\right\} $.

\subsubsection{The covariant derivative of the solder form}

The Weyl covariant derivative is compatible with the component matrix
of the solder form,
\begin{eqnarray}
D_{\mu}e_{\nu}^{\;\;a} & = & \partial_{\mu}e_{\nu}^{\;\;a}-e_{\alpha}^{\;\;a}\Sigma_{\;\;\nu\mu}^{\alpha}+e_{\nu}^{\;\;b}\omega_{\;b\mu}^{a}-W_{\mu}e_{\nu}^{\;\;a}\;\;=\;\;0\label{Covariant derivative of solder form}
\end{eqnarray}
For the inverse component matrix, we contract with $e_{a}^{\;\;\beta}$
and use the product rule to express the result as $e_{\nu}^{\;\;a}D_{\mu}e_{a}^{\;\;\beta}$,
recognizing the covariant derivative of the inverse,
\begin{eqnarray}
D_{\mu}e_{a}^{\;\;\beta} & = & \partial_{\mu}e_{a}^{\;\;\beta}+e_{a}^{\;\;\alpha}\Sigma_{\;\;\alpha\mu}^{\beta}-e_{b}^{\;\;\beta}\omega_{\;a\mu}^{b}+e_{a}^{\;\;\beta}W_{\mu}\;\;=\;\;0\label{Covariant derivative of inverse solder form}
\end{eqnarray}

Knowing the coordinate form of the covariant derivative lets us compute
the covariant derivative of $y_{a}$. Notice that multiplying by $e_{a}^{\;\;\mu}$
changes the conformal weight.
\begin{eqnarray*}
\mathbf{D}y_{a} & = & \mathbf{d}y_{a}-y_{b}\boldsymbol{\omega}_{\;\;a}^{b}+y_{a}\boldsymbol{\omega}\\
 & = & \mathbf{d}e_{a}^{\;\;\beta}y_{\beta}+e_{a}^{\;\;\mu}\mathbf{d}y_{\mu}-y_{b}\boldsymbol{\omega}_{\;\;a}^{b}+y_{a}\boldsymbol{\omega}\\
 & = & \left(-e_{a}^{\;\;\alpha}\Sigma_{\;\;\alpha\mu}^{\beta}\mathbf{d}x^{\mu}+e_{b}^{\;\;\beta}\omega_{\;a\mu}^{b}\mathbf{d}x^{\mu}-e_{a}^{\;\;\beta}W_{\mu}\mathbf{d}x^{\mu}\right)y_{\beta}+e_{a}^{\;\;\mu}\mathbf{d}y_{\mu}-y_{b}\boldsymbol{\omega}_{\;\;a}^{b}+y_{a}\boldsymbol{\omega}\\
 & = & y_{b}\boldsymbol{\omega}_{\;a}^{b}-e_{a}^{\;\;\alpha}y_{\beta}\boldsymbol{\Sigma}_{\;\;\alpha}^{\beta}-y_{a}\boldsymbol{\omega}+e_{a}^{\;\;\mu}\mathbf{d}y_{\mu}-y_{b}\boldsymbol{\omega}_{\;\;a}^{b}+y_{a}\boldsymbol{\omega}\\
 & = & e_{a}^{\;\;\alpha}\left(\mathbf{d}y_{\mu}-y_{\beta}\boldsymbol{\Sigma}_{\;\;\alpha\mu}^{\beta}\mathbf{d}x^{\mu}\right)
\end{eqnarray*}
This will be of use shortly.

\subsection{Curvature equation}

We next study the curvature equation, Eq.(\ref{Reduced curvature}).
We begin with its integrability condition, which places strong constraints
on the co-torsion. Then, substituting the connection forms from Eqs.(\ref{Reduced spin connection})-(\ref{Reduced co solder form}),
we impose the curvature field equation, $\Omega_{\;\;acb}^{c}=0$.

\subsubsection{Curvature Bianchi}

Expanding the curvature Bianchi identity, Eq.(\ref{Curvature Bianchi}),
and substititing the reduced curvatures, it becomes
\begin{eqnarray}
0 & = & \mathbf{d}\left(\frac{1}{2}\Omega_{\;\;bcd}^{a}\mathbf{e}^{c}\land\mathbf{e}^{d}\right)+\frac{1}{2}\Omega_{\;\;bef}^{c}\mathbf{e}^{e}\land\mathbf{e}^{f}\land\boldsymbol{\omega}_{\;c}^{a}-\frac{1}{2}\Omega_{\;\;cef}^{a}\mathbf{e}^{e}\land\mathbf{e}^{f}\land\boldsymbol{\omega}_{\;b}^{c}\nonumber \\
 &  & +2\left(1+\chi\right)\Delta_{cb}^{ad}\left(\frac{1}{2}S_{def}\mathbf{e}^{e}\land\mathbf{e}^{f}\land\mathbf{e}^{c}+S_{d\quad f}^{\;\;e}\mathbf{f}_{e}\land\mathbf{e}^{f}\land\mathbf{e}^{c}-\frac{1}{2}c_{d}^{\;\;ef}\mathbf{f}_{e}\land\mathbf{f}_{f}\land\mathbf{e}^{c}\right)\label{Expanded reduced curvature}
\end{eqnarray}
To find the independent parts, we must break the exterior derivative
into $\mathbf{d}_{\left(x\right)}$ and $\mathbf{d}_{\left(y\right)}$.
This also requires separating the independent parts of the co-solder
form,
\begin{eqnarray*}
\mathbf{f}_{a} & = & h_{a}^{\;\;\mu}\left(x,y\right)\mathbf{d}y_{\mu}+c_{ab}\left(x,y\right)\mathbf{e}^{b}\\
 & = & \mathbf{h}_{a}+\mathbf{c}_{a}
\end{eqnarray*}
where the solder form already lies only in the $x$-sector, $\mathbf{e}^{b}=e_{\alpha}^{\;\;\;b}\mathbf{d}x^{\alpha}$.
Then the only term proportional to $\mathbf{h}_{a}\wedge\mathbf{h}_{b}\wedge\mathbf{e}^{c}$
is the final one, so $0=2\left(1+\chi\right)\Delta_{cb}^{ad}c_{d}^{\;\;ef}\mathbf{h}_{e}\land\mathbf{h}_{f}\land\mathbf{e}^{c}$.
Since the structure constants are already antisymmetric we may drop
the basis forms. Then the $ac$ contraction shows that
\begin{eqnarray*}
0 & = & \left(1+\chi\right)c_{a}^{\;\;bc}
\end{eqnarray*}
Dropping this factor from the remaining parts of Eq.(\ref{Expanded reduced curvature}),
we move to the $\mathbf{h}_{a}\wedge\mathbf{e}^{b}\wedge\mathbf{e}^{c}$
components, antisymmetrizing to remove the basis forms
\begin{eqnarray}
0 & = & \partial^{\alpha}\Omega_{\;\;bfc}^{a}+2\left(1+\chi\right)\Delta_{cb}^{ad}S_{d\quad f}^{\;\;e}h_{e}^{\;\;\;\alpha}-2\left(1+\chi\right)\Delta_{fb}^{ad}S_{d\quad c}^{\;\;e}h_{e}^{\;\;\;\alpha}\label{Mixed term of expanded reduced curvature}
\end{eqnarray}
Taking the $af$ trace and using the field equation, $\Omega_{\;\;\;bac}^{a}=0$,
\begin{eqnarray*}
0 & = & \left(1+\chi\right)\left(\left(n-2\right)S_{b\quad c}^{\;\;e}+\eta_{bc}\eta^{ad}S_{d\quad a}^{\;\;e}\right)
\end{eqnarray*}
A further contraction with $\eta^{bc}$ shows that $\left(1+\chi\right)\left(n-1\right)\eta^{ad}S_{d\quad a}^{\;\;e}=0$
and therefore, $0=\left(1+\chi\right)S_{b\quad c}^{\;\;a}$. Substitutinng
this back into Eq.(\ref{Mixed term of expanded reduced curvature})
shows that the spacetime curvature is independent of $y_{\alpha}$,
$\partial^{\alpha}\Omega_{\;\;bcd}^{a}=0$.

Finally, defining the $x$-covariant derivative of the spacetime component
of the curvature,
\begin{eqnarray*}
\mathbf{D}_{\left(\omega,x\right)}\left(\frac{1}{2}\Omega_{\;\;bcd}^{a}\mathbf{e}^{c}\land\mathbf{e}^{d}\right) & \equiv & \mathbf{d}_{\left(x\right)}\left(\frac{1}{2}\Omega_{\;\;bcd}^{a}\mathbf{e}^{c}\land\mathbf{e}^{d}\right)+\left(\frac{1}{2}\Omega_{\;\;bef}^{c}\mathbf{e}^{e}\land\mathbf{e}^{f}\right)\land\boldsymbol{\omega}_{\;c}^{a}-\left(\frac{1}{2}\Omega_{\;\;cef}^{a}\mathbf{e}^{e}\land\mathbf{e}^{f}\right)\land\boldsymbol{\omega}_{\;b}^{c}
\end{eqnarray*}
we conclude
\begin{eqnarray}
0 & = & D_{[e}^{\left(\omega,x\right)}\Omega_{\;\;\left|b\right|cd]}^{a}+2\left(1+\chi\right)\Delta_{\left[e\right|b}^{af}S_{f\left|cd\right]}\label{Curvature Bianchi eee}\\
0 & = & \left(1+\chi\right)S_{a\quad c}^{\;\;b}\label{Curvature Bianchi hee}\\
0 & = & \left(1+\chi\right)c_{a}^{\;\;bc}\label{Curvature Bianchi hhe}
\end{eqnarray}
along with the further consequence of Eq.(\ref{Curvature Bianchi hee}),
\begin{equation}
\partial^{\mu}\Omega_{\;\;bcd}^{a}=0\label{No y dependence of conformal curvature}
\end{equation}

\subsubsection{The curvature structure equation}

We next find the components of the curvature, $\Omega_{\;\;bcd}^{a}$,
in terms of the connection, and impose the field equation.

Substituting Eq.(\ref{Reduced co solder form}) for $\mathbf{f}_{a}$
and expanding the exterior derivatve $\mathbf{d}\boldsymbol{\omega}_{\;\;\;b}^{a}=\mathbf{d}_{\left(x\right)}\boldsymbol{\omega}_{\;\;\;b}^{a}-2\Delta_{db}^{ac}\mathbf{d}_{\left(y\right)}W_{c}\mathbf{e}^{d}$
in the structure equation, Eq.(\ref{Reduced curvature}), allows separation
of the $\mathbf{e}^{c}\wedge\mathbf{e}^{d}$ and $\mathbf{e}^{c}\wedge\mathbf{h}_{d}$
parts into two equations
\begin{eqnarray}
\mathbf{d}_{\left(x\right)}\boldsymbol{\omega}_{\;b}^{a} & = & \boldsymbol{\omega}_{\;b}^{c}\land\boldsymbol{\omega}_{\;c}^{a}+2\left(1+\chi\right)\Delta_{db}^{ac}c_{ce}\mathbf{e}^{e}\land\mathbf{e}^{d}+\frac{1}{2}\Omega_{\;\;bcd}^{a}\mathbf{e}^{c}\land\mathbf{e}^{d}\label{Spacetime part of curvature}\\
0 & = & 2\Delta_{db}^{ac}\left(\partial^{\mu}W_{c}+\left(1+\chi\right)h_{c}^{\;\;\mu}\right)\mathbf{d}y_{\mu}\land\mathbf{e}^{d}\label{Cross-term of curvature equation}
\end{eqnarray}
The $ad$ trace of Eq.(\ref{Cross-term of curvature equation}) requires
\[
0=\left(n-1\right)\left(\partial^{\mu}W_{b}+\left(1+\chi\right)h_{b}^{\;\;\mu}\right)
\]
which solves the full cross-term equation. This same condition is
also required by the dilatation equation below. For the $\mathbf{e}^{c}\mathbf{e}^{d}$
terms, notice that we may still have some $y_{\mu}$ dependence.

\subsubsection{The spacetime equation \label{The Weyl curvature emerges}}

It is convenient to define the curvature $2$-form of the connection
compatible with the solder form,
\begin{eqnarray}
\mathbf{R}_{\;\;b}^{a}\left(x\right) & \equiv & \mathbf{d}_{\left(x\right)}\boldsymbol{\alpha}_{\;b}^{a}-\boldsymbol{\alpha}_{\;b}^{c}\land\boldsymbol{\alpha}_{\;c}^{a}\label{Riemann curvature}
\end{eqnarray}
This is the Riemann curvature built from $\boldsymbol{\alpha}_{\;\;b}^{a}\left(x\right)$,
not the full scale-invariant curvature of the biconformal space. Writing
the spin connection as
\begin{eqnarray*}
\boldsymbol{\omega}_{\;b}^{a} & = & \boldsymbol{\alpha}_{\;b}^{a}+\boldsymbol{\beta}_{\;b}^{a}
\end{eqnarray*}
where $\boldsymbol{\beta}_{\;b}^{a}\left(x,y\right)=-2\Delta_{db}^{ac}W_{c}\mathbf{e}^{d}$,
we substitute into Eq.(\ref{Spacetime part of curvature}),
\begin{eqnarray*}
\mathbf{d}_{\left(x\right)}\boldsymbol{\alpha}_{\;b}^{a}+\mathbf{d}_{\left(x\right)}\boldsymbol{\beta}_{\;b}^{a} & = & \boldsymbol{\alpha}_{\;b}^{c}\land\boldsymbol{\alpha}_{\;c}^{a}+\boldsymbol{\alpha}_{\;b}^{c}\land\boldsymbol{\beta}_{\;c}^{a}+\boldsymbol{\beta}_{\;b}^{c}\land\boldsymbol{\alpha}_{\;c}^{a}+\boldsymbol{\beta}_{\;b}^{c}\land\boldsymbol{\beta}_{\;c}^{a}\\
 &  & +2\left(1+\chi\right)\Delta_{db}^{ac}c_{ce}\mathbf{e}^{e}\land\mathbf{e}^{d}+\frac{1}{2}\Omega_{\;\;bcd}^{a}\mathbf{e}^{c}\land\mathbf{e}^{d}
\end{eqnarray*}
Solving for $\frac{1}{2}\Omega_{\;\;bcd}^{a}\mathbf{e}^{c}\land\mathbf{e}^{d}$,
using Eq.(\ref{Riemann curvature}), and recognizing the $\boldsymbol{\alpha}$-covariant
derivative of $\boldsymbol{\beta}_{\;b}^{a}$ as
\begin{eqnarray*}
\mathbf{D}_{\left(\alpha,x\right)}\boldsymbol{\beta}_{\;b}^{a} & \equiv & \mathbf{d}_{\left(x\right)}\boldsymbol{\beta}_{\;b}^{a}+\boldsymbol{\beta}_{\;c}^{a}\land\boldsymbol{\alpha}_{\;b}^{c}-\boldsymbol{\beta}_{\;b}^{c}\land\boldsymbol{\alpha}_{\;c}^{a}
\end{eqnarray*}
this becomes
\begin{eqnarray*}
\frac{1}{2}\Omega_{\;\;bcd}^{a}\mathbf{e}^{c}\land\mathbf{e}^{d} & = & \mathbf{R}_{\;b}^{a}+\mathbf{D}_{\left(\alpha,x\right)}\boldsymbol{\beta}_{\;b}^{a}-\boldsymbol{\beta}_{\;b}^{c}\land\boldsymbol{\beta}_{\;c}^{a}-2\left(1+\chi\right)\Delta_{db}^{ac}c_{ce}\mathbf{e}^{e}\land\mathbf{e}^{d}
\end{eqnarray*}

Recalling that $\mathbf{D}_{\left(\alpha,x\right)}\mathbf{e}^{a}=0$
the covariant exterior derivative of $\boldsymbol{\beta}_{\;b}^{a}$
becomes
\begin{eqnarray*}
\mathbf{D}_{\left(\alpha,x\right)}\boldsymbol{\beta}_{\;b}^{a} & = & \mathbf{D}_{\left(\alpha,x\right)}\left(-2\Delta_{db}^{ac}W_{c}\mathbf{e}^{d}\right)\\
 & = & -2\Delta_{db}^{ac}\left(\mathbf{D}_{\left(\alpha,x\right)}W_{c}\right)\land\mathbf{e}^{d}
\end{eqnarray*}
The $\boldsymbol{\beta}_{\;b}^{c}\land\boldsymbol{\beta}_{\;c}^{a}$
term may be simplified considerably. With $W^{2}\equiv\eta^{ab}W_{a}W_{b}$,

\begin{eqnarray*}
\boldsymbol{\beta}_{\;b}^{c}\land\boldsymbol{\beta}_{\;c}^{a} & = & 2\Delta_{db}^{ce}2\Delta_{gc}^{af}W_{e}W_{f}\mathbf{e}^{d}\land\mathbf{e}^{g}\\
 & = & \left(\delta_{g}^{a}\delta_{d}^{f}\delta_{b}^{e}-\eta^{ef}\eta_{bd}\delta_{g}^{a}-\eta^{af}\eta_{gd}\delta_{b}^{e}+\delta_{g}^{e}\eta_{bd}\eta^{af}\right)W_{e}W_{f}\mathbf{e}^{d}\land\mathbf{e}^{g}\\
 & = & -\left(\delta_{c}^{a}W_{b}W_{d}+\eta_{bd}\eta^{af}W_{c}W_{f}-\eta_{bd}\delta_{c}^{a}\eta^{ef}W_{e}W_{f}\right)\mathbf{e}^{c}\land\mathbf{e}^{d}\\
 & = & -\left(\delta_{c}^{a}\delta_{b}^{e}W_{e}W_{d}-\frac{1}{2}\eta_{ed}\delta_{b}^{e}\delta_{c}^{a}W^{2}+\eta_{bd}\eta^{ae}W_{c}W_{e}-\frac{1}{2}\eta_{bd}\eta^{ae}\eta_{ce}W^{2}\right)\mathbf{e}^{c}\land\mathbf{e}^{d}\\
 & = & -\left(\delta_{c}^{a}\delta_{b}^{e}\left(W_{e}W_{d}-\frac{1}{2}\eta_{ed}W^{2}\right)-\eta_{bc}\eta^{ae}\left(W_{d}W_{e}-\frac{1}{2}\eta_{de}W^{2}\right)\right)\mathbf{e}^{c}\land\mathbf{e}^{d}\\
 & = & -2\Delta_{cb}^{ae}\left(W_{e}W_{d}-\frac{1}{2}\eta_{ed}W^{2}\right)\mathbf{e}^{c}\land\mathbf{e}^{d}
\end{eqnarray*}
Therefore,
\begin{eqnarray*}
\frac{1}{2}\Omega_{\;\;bcd}^{a}\mathbf{e}^{c}\land\mathbf{e}^{d} & = & \mathbf{R}_{\;\;b}^{a}-2\Delta_{db}^{ae}\left(D_{c}^{\left(\alpha,x\right)}W_{e}+W_{e}W_{c}-\frac{1}{2}\eta_{ec}W^{2}+\left(1+\chi\right)c_{ec}\right)\mathbf{e}^{c}\land\mathbf{e}^{d}
\end{eqnarray*}

It is convenient to define the curvature $2$-form of the full spin
connection as well,
\begin{eqnarray}
\boldsymbol{\mathscr{R}}_{\;\;b}^{a} & = & \mathbf{d}_{\left(x\right)}\boldsymbol{\omega}_{\;b}^{a}-\boldsymbol{\omega}_{\;b}^{c}\land\boldsymbol{\omega}_{\;c}^{a}\nonumber \\
 & = & \mathbf{R}_{\;\;b}^{a}-2\Delta_{db}^{ae}\left(D_{c}^{\left(\alpha,x\right)}W_{e}+W_{e}W_{c}-\frac{1}{2}\eta_{ec}W^{2}\right)\mathbf{e}^{c}\land\mathbf{e}^{d}\label{Riemann tensor of Weyl geometry}
\end{eqnarray}
where $\boldsymbol{\omega}_{\;b}^{a}=\boldsymbol{\alpha}_{\;b}^{a}-2\Delta_{db}^{ac}W_{c}\mathbf{e}^{d}$.
This may be recognized as the curvature tensor of an $n$-dim Weyl
geometry \cite{WeylGeom}.

We also identify the Schouten tensor 
\begin{eqnarray}
\boldsymbol{\mathcal{R}}_{a} & = & \mathcal{R}_{ab}\mathbf{e}^{b}\;\;\equiv\;\;\frac{1}{\left(n-2\right)}\left(R_{ab}-\frac{1}{2\left(n-1\right)}\eta_{ab}R\right)\mathbf{e}^{b}\label{Schouten tensor}
\end{eqnarray}
In any dimension greater than two, knowing the Schouten tensor is
equivalent to knowing the Ricci tensor, since we may always invert,
$R_{ab}=\left(n-2\right)\mathcal{R}_{ab}+\eta_{ab}\mathcal{R}$. In
terms of the Schouten tensor, the decomposition of the Riemann curvature
into the traceless Weyl conformal tensor, $\mathbf{C}_{\;\;b}^{a}$
and its Ricci parts, takes the simple form \cite{WeylGeom}, 
\begin{equation}
\mathbf{R}_{\;\;b}^{a}=\mathbf{C}_{\;\;b}^{a}-2\Delta_{db}^{ae}\boldsymbol{\mathcal{R}}_{e}\wedge\mathbf{e}^{d}\label{Partition of Riemann}
\end{equation}

Using this decomposition, the Ricci parts of the curvature combine
with the additional terms from the scale covariance, 
\begin{eqnarray}
\frac{1}{2}\Omega_{\;\;bcd}^{a}\mathbf{e}^{c}\land\mathbf{e}^{d} & = & \mathbf{C}_{\;\;b}^{a}-2\Delta_{db}^{ae}\left(\mathcal{R}_{ec}+D_{c}^{\left(\alpha,x\right)}W_{e}+W_{e}W_{c}-\frac{1}{2}\eta_{ec}W^{2}+\left(1+\chi\right)c_{ec}\right)\mathbf{e}^{c}\land\mathbf{e}^{d}\label{Expanded SO(p,q) spacetime curvature}
\end{eqnarray}
To impose the field equation, set $P_{ec}\equiv\mathcal{R}_{ec}+D_{c}^{\left(\alpha,x\right)}W_{e}+W_{e}W_{c}-\frac{1}{2}\eta_{ec}W^{2}+\left(1+\chi\right)c_{ec}$.
Then substituting $\Omega_{\;\;bcd}^{a}=C_{\;\;bcd}^{a}-2\Delta_{db}^{ae}P_{ec}+2\Delta_{cb}^{ae}P_{ed}$
into Eq.(\ref{Spacetime curvature from torsion Bianchi}), 
\begin{eqnarray*}
C_{\;\;bcd}^{c}-2\Delta_{db}^{ae}P_{ec}+2\Delta_{cb}^{ae}P_{ed} & = & 0
\end{eqnarray*}
Since $C_{\;\;bcd}^{c}=0$, this has the component form, for any $P_{ab}$,
$0=\Delta_{db}^{ce}P_{ec}-\Delta_{cb}^{ce}P_{ed}$, which, expanding
the projections and combining result with the further contraction
with $\eta^{bd}$, is seen to be true if and only if $P_{ab}=0$.

Applying this general result to the field equation by replacing $P_{ec}$,
we have
\begin{equation}
\mathcal{R}_{ec}+D_{c}^{\left(\alpha,x\right)}W_{e}+W_{e}W_{c}-\frac{1}{2}\eta_{ec}W^{2}+\left(1+\chi\right)c_{ec}=0\label{Curvature field equation for Ricci}
\end{equation}
This determines $c_{ab}$ unless $1+\chi=0$. The symmetric part of
the first four terms on the left side is the Weyl-Schouten tensor,
\[
\boldsymbol{\mathscr{R}}_{a}\equiv\boldsymbol{\mathcal{R}}_{a}+\left(W_{\left(a;b\right)}+W_{a}W_{b}-\frac{1}{2}W^{2}\eta_{ab}\right)\mathbf{e}^{b}
\]
and we see that there is an antisymmetric part to the trace of the
Riemann-Weyl tensor,
\begin{eqnarray*}
\mathscr{R}_{bd}\equiv\mathscr{R}_{\;\;bcd}^{c} & = & R_{bd}+\left(n-2\right)\left(D_{d}^{\left(\alpha,x\right)}W_{b}+W_{b}W_{d}-\frac{1}{2}\eta_{bd}W^{2}\right)-\eta_{bd}\eta^{ce}\left(D_{c}^{\left(\alpha,x\right)}W_{e}+W_{e}W_{c}-\frac{1}{2}\eta_{ec}W^{2}\right)\\
\mathscr{R}_{\left[bd\right]} & = & \left(n-2\right)W_{\left[b;d\right]}
\end{eqnarray*}
This agrees with the trace of the corresponding term of the torsion-free
Bianchi identity arising from Eq.(\ref{Reduced torsion}), and shows
that $c_{ab}$ may have both symmetric and antisymmetric parts.

Returning to the full spacetime curvature after satisfying the field
equation,
\begin{eqnarray}
\Omega_{\;\;bcd}^{a} & = & C_{\;bcd}^{a}\left(\alpha\right)\label{Solution for spacetime conformal curvature}
\end{eqnarray}
so the spacetime piece of the biconformal curvature reduces to the
Weyl (conformal) curvature of the metric compatible connection. This
part of the curvature is independent of $y_{\mu}$ as required by
Eq.(\ref{No y dependence of conformal curvature}), since the only
$y_{\mu}$-dependence of the connection must arise from the Weyl vector,
and as seen in Eq.(\ref{Expanded SO(p,q) spacetime curvature}) the
Weyl vector is only present in the trace terms of the curvature. The
full $SO\left(p,q\right)$ curvature may now be written as
\begin{eqnarray*}
\boldsymbol{\Omega}_{\;\;b}^{a} & = & \left(\frac{1}{2}C_{\;\;bcd}^{a}+2\chi\Delta_{db}^{ae}c_{ec}\right)\mathbf{e}^{c}\wedge\mathbf{e}^{d}+2\chi\Delta_{db}^{ac}\mathbf{h}_{c}\wedge\mathbf{e}^{d}\\
 & = & \mathbf{C}_{\;\;b}^{a}-\frac{2\chi}{1+\chi}\Delta_{db}^{ae}\boldsymbol{\mathscr{R}}_{e}\wedge\mathbf{e}^{d}+2\chi\Delta_{db}^{ac}\mathbf{h}_{c}\wedge\mathbf{e}^{d}
\end{eqnarray*}
with $\mathbf{h}_{a}=h_{a}^{\;\;\alpha}\mathbf{d}y_{\alpha}$. The
second expression holds only when $1+\chi\neq0$.

\subsubsection{Form of the spacetime Bianchi identity}

When we combine this solution for $\boldsymbol{\Omega}_{\;\;b}^{a}$
with the spacetime part of the curvature Bianchi identity, we have
Eq.(\ref{Curvature Bianchi eee})
\[
D_{[e}^{\left(\omega,x\right)}\Omega_{\;\;\left|b\right|cd]}^{a}+2\left(1+\chi\right)\Delta_{\left[e\right|b}^{af}S_{f\left|cd\right]}=0
\]
With $\frac{1}{2}\Omega_{\;\;bcd}^{a}\mathbf{e}^{c}\land\mathbf{e}^{d}=\frac{1}{2}C_{\;\;bcd}^{a}\mathbf{e}^{c}\land\mathbf{e}^{d}=\mathbf{C}_{\;\;b}^{a}$,
the covariant exterior derivative of the Weyl curvature is
\begin{eqnarray*}
\mathbf{D}_{\left(\omega,x\right)}\mathbf{C}_{\;\;b}^{a} & \equiv & \mathbf{d}_{\left(x\right)}\mathbf{C}_{\;\;b}^{a}+\mathbf{C}_{\;\;b}^{a}\land\boldsymbol{\omega}_{\;c}^{a}-\mathbf{C}_{\;\;b}^{a}\land\boldsymbol{\omega}_{\;b}^{c}\\
 & = & \mathbf{D}_{\left(\alpha,x\right)}\mathbf{C}_{\;\;b}^{a}+\mathbf{C}_{\;\;b}^{a}\land\boldsymbol{\beta}_{\;c}^{a}-\mathbf{C}_{\;\;c}^{a}\land\boldsymbol{\beta}_{\;b}^{c}
\end{eqnarray*}
where we have expanded $\boldsymbol{\omega}_{\;c}^{a}=\boldsymbol{\alpha}_{\;c}^{a}+\boldsymbol{\beta}_{\;c}^{a}$.
But $\mathbf{C}_{\;\;b}^{a}$ is the usual traceless part of the Riemann
curvature, which satisfies
\begin{eqnarray*}
0 & \equiv & \mathbf{D}_{\left(\alpha,x\right)}\mathbf{R}_{\;\;b}^{a}\\
 & = & \mathbf{D}_{\left(\alpha,x\right)}\mathbf{C}_{\;\;b}^{a}-2\Delta_{db}^{ae}\mathbf{D}_{\left(\alpha,x\right)}\boldsymbol{\mathcal{R}}_{e}\wedge\mathbf{e}^{d}
\end{eqnarray*}
and we may rewrite the covariant exterior derivative of the Weyl curvature
in terms of the derivative of the Schouten tensor. Making this replacement
and setting $\boldsymbol{\beta}_{\;c}^{a}=-2\Delta_{dc}^{ae}W_{e}\mathbf{e}^{d}$,
the Bianchi identity may be written as

\begin{eqnarray*}
0 & = & 2\Delta_{df}^{ge}\left(\delta_{b}^{f}\delta_{g}^{a}\mathbf{D}_{\left(\alpha,x\right)}\boldsymbol{\mathcal{R}}_{e}-\delta_{g}^{a}W_{e}\mathbf{C}_{\;\;b}^{f}+\delta_{g}^{c}\delta_{b}^{f}W_{e}\mathbf{C}_{\;\;c}^{a}+\left(1+\chi\right)\delta_{g}^{a}\delta_{b}^{f}\mathbf{S}_{e}^{\left(ee\right)}\right)\land\mathbf{e}^{d}
\end{eqnarray*}
where we set $\mathbf{S}_{b}^{\left(ee\right)}\equiv\frac{1}{2}S_{emn}\mathbf{e}^{m}\land\mathbf{e}^{n}$.
Now we expand the $\Delta$-projections, distribute and re-collect
terms, then use the first Riemannian Bianchi $\mathbf{C}_{db}\land\mathbf{e}^{d}=0$,
to show that
\begin{eqnarray}
0 & = & \Delta_{db}^{ac}\left(\mathbf{D}_{\left(\alpha,x\right)}\boldsymbol{\mathcal{R}}_{c}-W_{e}\mathbf{C}_{\;\;c}^{e}+2\left(1+\chi\right)\mathbf{S}_{c}^{\left(ee\right)}\right)\land\mathbf{e}^{d}\label{Spacetime curvature Bianchi}
\end{eqnarray}

Expanding both the $\Delta$-projection and the triple antisymmetrization,
we show that for all $n>3$, Eq.(\ref{Spacetime curvature Bianchi})
holds if and only if

\begin{eqnarray*}
2\left(1+\chi\right)S_{bfg} & = & D_{g}^{\left(\alpha,x\right)}\mathcal{R}_{bf}-D_{f}^{\left(\alpha,x\right)}\mathcal{R}_{bg}+W_{e}C_{\;\;bfg}^{e}
\end{eqnarray*}
Restoring two basis forms, we may write this as
\begin{eqnarray}
\mathbf{D}^{\left(\alpha,x\right)}\boldsymbol{\mathcal{R}}_{b}-W_{e}\mathbf{C}_{\;\;b}^{e}+2\left(1+\chi\right)\mathbf{S}_{b}^{\left(ee\right)} & = & 0\label{Reduced spacetime curvature Bianchi}
\end{eqnarray}
which solves the full Bianchi relation, Eq.(\ref{Spacetime curvature Bianchi}).
From Eq.(\ref{Reduced spacetime curvature Bianchi}) it follows that:
\begin{description}
\item [{Theorem:}] In any torsion-free biconformal space with integrable
Weyl vector, $W_{\alpha}=\partial_{\alpha}\phi$, and $1+\chi\neq0$,
the spacetime co-torsion is the obstruction to conformal Ricci flatness.
\end{description}
Complete details of the algebra leading to Eq.(\ref{Spacetime curvature Bianchi})
are given in Appendix C.

\subsection{The dilatation and co-torsion structure equations}

Expanding the dilatation equation, Eq.(\ref{Reduced dilatation}),
using Eqs.(\ref{Reduced solder form})-(\ref{Reduced Weyl vector}),
to display the independent parts
\begin{eqnarray*}
\mathbf{d}_{\left(x\right)}\boldsymbol{\omega}+\mathbf{d}_{\left(y\right)}\boldsymbol{\omega} & = & \left(1+\chi\right)\mathbf{e}^{c}\wedge\mathbf{h}_{c}+\left(1+\chi\right)\mathbf{e}^{c}\wedge\mathbf{c}_{c}
\end{eqnarray*}
we find two independent equations,
\begin{eqnarray}
\mathbf{d}_{\left(x\right)}\boldsymbol{\omega} & = & \left(1+\chi\right)\mathbf{c}\label{Dilatation structure equation ee}\\
\mathbf{d}_{\left(y\right)}\boldsymbol{\omega} & = & \left(1+\chi\right)\mathbf{e}^{c}\wedge\mathbf{h}_{c}\label{Dilatation structure equation eh}
\end{eqnarray}
where we have set $\mathbf{c}\equiv\mathbf{e}^{c}\wedge\mathbf{c}_{c}$.
The Bianchi identity reduces to $0\equiv\mathbf{d}^{2}\boldsymbol{\omega}=\left(1+\chi\right)\mathbf{d}\left(\mathbf{e}^{a}\wedge\mathbf{f}_{a}\right)=\left(1+\chi\right)\mathbf{D}\left(\mathbf{e}^{a}\wedge\mathbf{f}_{a}\right)=-\left(1+\chi\right)\mathbf{e}^{a}\wedge\mathbf{S}_{a}$,
which reproduces Eqs.(\ref{Curvature Bianchi hee}) and (\ref{Curvature Bianchi hhe})
and shows that
\begin{eqnarray*}
\left(1+\chi\right)S_{\left[abc\right]} & = & 0
\end{eqnarray*}

The dilatation structure equations may be integrated exactly, but
the result depends crucially on whether or not $\left(1+\chi\right)=0$.
The two cases will be handled separately in the next two Sections.

The co-torsion structure equation also depends on the case considered.
In addition to the structure equations (\ref{Reduced co torsion}),
we still have the field equations
\begin{eqnarray}
s_{c}\;\;\equiv\;\;S_{c\quad a}^{\;\;a} & = & S_{a\quad c}^{\;\;a}\nonumber \\
\Delta_{sb}^{ar}S_{c\quad a}^{\;\;\;b} & = & \Delta_{sc}^{ar}s_{a}\nonumber \\
S_{c}^{\;\;ac} & = & 0\label{Field equations for co-torsion}
\end{eqnarray}
and constraints from the curvature and dilatation Bianchi identities,
\begin{eqnarray}
\left(1+\chi\right)S_{\left[abc\right]} & = & 0\nonumber \\
D_{[e}^{\left(\omega,x\right)}\Omega_{\;\;\left|b\right|cd]}^{a}+2\left(1+\chi\right)\Delta_{\left[e\right|b}^{af}S_{f\left|cd\right]} & = & 0\nonumber \\
\left(1+\chi\right)S_{a\quad c}^{\;\;b} & = & 0\nonumber \\
\left(1+\chi\right)S_{a}^{\;\;bc} & = & 0\label{Bianchi identities for co-torsion}
\end{eqnarray}

To complete the reduction of the biconformal space, we turn to the
$1+\chi\neq0$ and $1+\chi=0$ cases.  

\section{Generic case: $1+\chi\protect\neq0$\label{sec:Abelian-case:}}

In this Section, we consider the final reduction to spacetime for
generic values of the constants $\alpha,\beta,\gamma$ in the original
action, assuming 
\begin{equation}
1+\chi\neq0\label{Generic condition}
\end{equation}
It follows that we must have $S_{a}^{\;\;bc}=0$ and therefore on
the $\mathbf{e}^{a}=0$ submanifold,
\begin{eqnarray*}
\mathbf{d}_{\left(y\right)}\mathbf{h}_{a} & = & 0\\
\mathbf{h}_{a} & = & \mathbf{d}_{\left(y\right)}y_{a}
\end{eqnarray*}
where the functions $y_{a}$ may be written as coordinates with an
$x_{0}$-dependent linear transformation,
\[
h_{a}\left(x_{0}^{\mu},y_{\mu}\right)=h_{a}^{\;\;\;\mu}\left(x_{0}^{\mu}\right)\left(y_{\mu}+\beta_{\mu}\left(x_{0}^{\mu}\right)\right)
\]
When we return to the full biconformal space, the linear coefficients
$h_{a}^{\;\;\;\mu}\left(x\right)$ and $\beta_{\mu}\left(x\right)$
remain as arbitrary coordinate choices. The full co-solder form, Eq.(\ref{Reduced co solder form}),
then satisfies
\begin{eqnarray*}
\mathbf{f}_{a} & = & h_{a}^{\;\;\mu}\left(x\right)\mathbf{d}y_{\mu}+c_{ab}\left(x,y\right)\mathbf{e}^{b}\\
 & = & h_{a}^{\;\;\mu}\left(x\right)\mathbf{d}y_{\mu}+h_{a}^{\;\;\mu}\left(x\right)\mathbf{d}\beta_{\mu}\left(x\right)+c_{ab}\left(x,y\right)\mathbf{e}^{b}
\end{eqnarray*}
Thus, the coordinate choice of the origin for $y_{\mu}$ at each $x^{\alpha}$
changes $c_{ab}$. We continue to define the co-basis $\mathbf{h}_{a}$
as only the $\mathbf{d}y_{\mu}$ part,
\begin{eqnarray}
\mathbf{h}_{a} & \equiv & h_{a}^{\;\;\mu}\left(x\right)\mathbf{d}y_{\mu}\label{Cobasis}
\end{eqnarray}
The momentum space is therefore foliated by abelian group manifolds.
The foliation may be identified as $R^{n}$ or a toroidal compactification
of all or part of $R^{n}$. Being principally interested in the underlying
presence of general relativity, we take it to be $R^{n}$ and eventually
identify it with the cotangent space at each $x^{\alpha}$. However,
it may also be taken as the torus $T^{d}$ of double field theory
for other applications to string theory.

Given the form Eq.(\ref{Cobasis}) of the co-basis $\mathbf{h}_{a}$,
it is useful to begin with the natural inner product arising from
the conformal Killing form together with our freedom to choose the
$x$-coordinates. The coordinate freedom allows us to conveniently
choose the functions $h_{\mu}^{\;\;a}$ to be the inverse solder form,
enabling us to integrate the dilatation equation for the Weyl vector.

\subsection{The Killing metric}

The field $h_{a}^{\;\;\mu}\left(x\right)$ lets us choose a convenient
orthonormal basis for the $y_{\alpha}$ space at each point of the
$x^{\alpha}$ space. Taking the restriction of the conformal Killing
form to the biconformal manifold, $\left(\begin{array}{cc}
0 & \delta_{b}^{a}\\
\delta_{d}^{d} & 0
\end{array}\right)$, as the biconformal metric lets us usefully control this coordinate
freedom. The Killing metric gives the orthonormal inner product of
$\left(\mathbf{e}^{a},\mathbf{f}_{b}\right)$ basis, 
\begin{eqnarray*}
\left\langle \mathbf{e}^{a},\mathbf{e}^{b}\right\rangle  & = & 0\\
\left\langle \mathbf{e}^{a},\mathbf{f}_{b}\right\rangle  & = & \delta_{b}^{a}\\
\left\langle \mathbf{f}_{a},\mathbf{f}_{b}\right\rangle  & = & 0
\end{eqnarray*}
Choosing arbitrary coordinates, $\tilde{x}^{\mu}$ as the complement
to $y_{\mu}$, the first of these three relations, $\left\langle \mathbf{e}^{a},\mathbf{e}^{b}\right\rangle =0$,
shows that $\left\langle \mathbf{d}\tilde{x}^{\mu},\mathbf{d}\tilde{x}^{\nu}\right\rangle =0.$
Substituting $\mathbf{f}_{a}=\mathbf{h}_{a}+c_{ab}\mathbf{e}^{b}$
into $\left\langle \mathbf{e}^{a},\mathbf{f}_{b}\right\rangle $ and
expanding in coordinates,
\begin{eqnarray}
\delta_{b}^{a} & = & \left\langle \mathbf{e}^{a},\mathbf{h}_{b}+c_{bc}\mathbf{e}^{c}\right\rangle \nonumber \\
 & = & \left\langle \mathbf{e}^{a},\mathbf{h}_{b}\right\rangle \nonumber \\
 & = & h_{b}^{\;\;\mu}\left(\tilde{x}\right)e_{\nu}^{\;\;a}\left(\tilde{x}\right)\left\langle \mathbf{d}\tilde{x}^{\nu},\mathbf{d}y_{\mu}\right\rangle \nonumber \\
e_{b}^{\;\;\nu}\left(\tilde{x}\right) & = & h_{b}^{\;\;\mu}\left(\tilde{x}\right)\left\langle \mathbf{d}\tilde{x}^{\nu},\mathbf{d}y_{\mu}\right\rangle \label{dx dy inner product}
\end{eqnarray}
Using $\left\langle \mathbf{e}^{a},\mathbf{h}_{b}\right\rangle =\delta_{b}^{a}$
in the $\left\langle \mathbf{f}_{a},\mathbf{f}_{b}\right\rangle $
inner product, we find $\left\langle \mathbf{h}_{a},\mathbf{h}_{b}\right\rangle $,
\begin{eqnarray}
0 & = & \left\langle \mathbf{f}_{a},\mathbf{f}_{b}\right\rangle \nonumber \\
 & = & \left\langle \mathbf{h}_{a}+c_{ac}\mathbf{e}^{c},\mathbf{h}_{b}+c_{bd}\mathbf{e}^{d}\right\rangle \nonumber \\
\left\langle \mathbf{h}_{a},\mathbf{h}_{b}\right\rangle  & = & -\left(c_{ab}+c_{ba}\right)=-2c_{\left(ab\right)}\label{hh inner product}
\end{eqnarray}

We see from Eq.(\ref{dx dy inner product}) that the inner product
of $\mathbf{d}y_{\mu}$ with $\mathbf{d}\tilde{x}^{\nu}$ cannot depend
on $y_{\mu}$, 
\[
\left\langle \mathbf{d}\tilde{x}^{\nu},\mathbf{d}y_{\mu}\right\rangle =k_{\;\;\mu}^{\nu}\left(\tilde{x}\right)
\]
Moreover, like $e_{b}^{\;\;\nu}\left(\tilde{x}\right)$ and $h_{b}^{\;\;\mu}\left(\tilde{x}\right)$,
$k_{\;\;\mu}^{\nu}\left(\tilde{x}\right)$ must be invertible. Let
$x^{\alpha}=x^{\alpha}\left(\tilde{x}\right)$ be any coordinate transformation
of $\tilde{x}^{\alpha}$. Then in the new $x$-coordinates the inner
product becomes
\begin{eqnarray*}
\left\langle \mathbf{d}x^{\alpha},\mathbf{d}y_{\mu}\right\rangle  & = & \left\langle \frac{\partial x^{\nu}}{\partial\tilde{x}^{\alpha}}\mathbf{d}\tilde{x}^{\alpha},\mathbf{d}y_{\mu}\right\rangle \\
 & = & \frac{\partial x^{\nu}}{\partial\tilde{x}^{\alpha}}k_{\;\;\mu}^{\alpha}\left(\tilde{x}\right)
\end{eqnarray*}
Since $\frac{\partial x^{\nu}}{\partial\tilde{x}^{\alpha}}$ is an
arbitrary general linear transformation at each point and $k_{\;\;\mu}^{\nu}$
is invertible, we may choose $\frac{\partial x^{\nu}}{\partial\tilde{x}^{\alpha}}$
to be its inverse. Then
\[
\left\langle \mathbf{d}x^{\nu},\mathbf{d}y_{\mu}\right\rangle =\delta_{\mu}^{\nu}
\]
Writing eq.(\ref{dx dy inner product}) in these new coordinates,
we have
\[
h_{b}^{\;\;\mu}\left(x\right)=e_{b}^{\;\;\mu}\left(x\right)
\]
showing that in these coordinates $h_{b}^{\;\;\mu}\left(x\right)$
is just the inverse matrix to $e_{\mu}^{\;\;a}\left(x\right)$. This
fixes
\begin{equation}
\mathbf{h}_{a}=e_{b}^{\;\;\mu}\left(x\right)\mathbf{d}y_{\mu}\label{Momentum submanifold basis form}
\end{equation}

\subsection{The dilatation equation}

With the change of $x$-coordinate, the basis forms are now given
by 
\begin{eqnarray}
\mathbf{e}^{a} & = & e_{\mu}^{\;\;a}\left(x\right)\mathbf{d}x^{\mu}\label{Final solder form}\\
\mathbf{f}_{a} & = & e_{a}^{\;\;\alpha}\left(\mathbf{d}y_{\alpha}+\mathbf{d}\beta_{\alpha}+c_{\alpha\nu}\mathbf{d}x^{\nu}\right)\label{Near final cosolder form}
\end{eqnarray}
with the spin connection and Weyl vector given by Eq.(\ref{Reduced spin connection})
and Eq.(\ref{Reduced Weyl vector}). The $x$-dependent translation
$\beta_{\mu}$ remains an arbitrary coordinate choice.

Using the coefficients $e_{\mu}^{\;\;a}$ to change basis in the usual
way to convert between coordinate and orthonormal indices, we expand
$\mathbf{c}_{a}=e_{a}^{\;\;\;\alpha}c_{\alpha\nu}\mathbf{d}x^{\nu}$
and the Weyl vector $\boldsymbol{\omega}=W_{a}\mathbf{e}^{a}=W_{\mu}\mathbf{d}x^{\mu}$
in Eq.(\ref{Reduced dilatation}) in coordinates.
\begin{eqnarray*}
\mathbf{d}x^{\mu}\wedge\partial_{\mu}\left(W_{\nu}\mathbf{d}x^{\nu}\right)+\mathbf{d}y_{\mu}\wedge\partial^{\mu}\left(W_{\nu}\mathbf{d}x^{\nu}\right) & = & \left(1+\chi\right)\left(e_{\mu}^{\;\;a}\mathbf{d}x^{\mu}\right)\wedge\left(e_{a}^{\;\;\alpha}\left(\mathbf{d}y_{\alpha}+\beta_{\alpha,\nu}\mathbf{d}x^{\nu}+c_{\alpha\nu}\mathbf{d}x^{\nu}\right)\right)
\end{eqnarray*}
Equating independent parts,
\begin{eqnarray}
W_{\left[\mu,\nu\right]} & = & -\left(1+\chi\right)\left(\beta_{\left[\mu,\nu\right]}+c_{\left[\mu\nu\right]}\right)\label{Dilatation ee part}\\
\partial^{\mu}W_{\nu} & = & -\left(1+\chi\right)\delta_{\nu}^{\mu}\label{Dilatation eh part}
\end{eqnarray}
Eq.(\ref{Dilatation eh part}) is integrated immediately,
\begin{equation}
W_{\mu}=\left(1+\chi\right)\left(-y_{\mu}+\alpha_{\mu}\left(x\right)\right)\label{Integrated Weyl vector}
\end{equation}
This must satisfy both equations, so substituting into Eq.(\ref{Dilatation ee part}),
\begin{eqnarray}
c_{\left[\mu\nu\right]} & = & -\left(\alpha_{\left[\mu,\nu\right]}+\beta_{\left[\mu,\nu\right]}\right)\label{Potential for antisymmetric part of c}
\end{eqnarray}

Before making the obvious coordinate choice, $\beta_{\mu}=-\alpha_{\mu}$,
it is suggestive to comment on the form of Eq.(\ref{Integrated Weyl vector}).
The integration ``constant'' $\alpha_{\mu}\left(x\right)$ is a
potential for the antisymmetric part of $c_{\mu\nu}$ and the antisymmetric
part is independent of $y_{\mu}$. Since an $x$-dependent rescaling
does not affect the vanishing of the $\mathbf{f}_{a}$ component of
the Weyl vector, we may perform a dilatation to modify $\alpha_{\mu}\left(x\right)$.
This is precisely the form of the gauge transformation of the electromagnetic
potential, but as with the failed Weyl theory of electromagnetism,
it may lead to unphysical size changes since the dilatational curvature,
$\Omega_{\mu\nu}$ does not necessarily vanish. However, notice that
biconformal space has a symplectic form. Eq.(\ref{Reduced dilatation})
describes a manifestly nondegenerate 2-form, $\mathbf{e}^{a}\wedge\mathbf{f}_{a}$,
which is exact and therefore closed. This means we may interpret the
full biconformal space as a relativistic particle phase space with
canonical coordinates $\left(x^{\alpha},y_{\beta}\right)$. In this
view, $y_{\mu}$ is a momentum and the Weyl vector (\ref{Integrated Weyl vector})
has exactly the form and gauge properties of the electromagnetic conjugate
momentum if $\alpha_{\mu}$ is taken proportional to the vector potential.
Moreover, the previous well-known conflict with observation is avoided.
The transformation by $\beta_{\mu}$ to remove $\alpha_{\mu}$ is
then the cannonical transformation between the conjugate electromagnetic
momentum, $\pi_{\mu}=p_{\mu}-eA_{\mu}$ and the simple particle momentum
$p_{\mu}$. 

The original ill-fated attempt by Weyl to identify the Weyl vector
of a Weyl geometry as the vector potential of electromagnetism, $W_{\mu}=eA_{\mu}$,
leads to nonvanishing dilatation in the presence of electromagnetic
fields, $\Omega_{\mu\nu}=eF_{\mu\nu}$. Einstein immediately observed
that this conflicts with experiment, and it is easy to show, for example,
that two hydrogen atoms moving to produce a closed path that encloses
some electromagnetic flux would emerge with different sizes, and therefore
very different spectra. The precision of atomic spectra therefore
disproves the simplest version of the theory. The situation is completely
different in the biconformal setting. Because of the extra $\mathbf{e}^{a}\mathbf{f}_{a}$
term in the dilatation equation, it is possible to have vanishing
dilatational curvature and retain the interpretation of $\alpha_{\mu}$
(rather than $W_{\mu}$) as the vector potential. The idea has been
explored to some extent in \cite{NCG}. Here, the form of the dilatation
is given by

\begin{eqnarray*}
\boldsymbol{\Omega} & = & \Omega_{\;\;b}^{a}\mathbf{f}_{a}\wedge\mathbf{e}^{a}\\
 & = & \chi\mathbf{e}^{a}\wedge\mathbf{f}_{a}\\
 & = & \chi\delta_{\nu}^{\mu}e_{a}^{\;\;\nu}e_{\mu}^{\;\;a}\mathbf{d}x^{\mu}\wedge\mathbf{d}y_{\nu}+\chi c_{\mu\nu}e_{\mu}^{\;\;a}\mathbf{d}x^{\mu}\wedge\mathbf{d}x^{\nu}
\end{eqnarray*}
or since we may also write $\boldsymbol{\Omega}=\frac{1}{2}\Omega_{\mu\nu}\mathbf{d}x^{\mu}\mathbf{d}x^{\nu}+\Omega_{\;\;\nu}^{\mu}\mathbf{d}y_{\mu}\mathbf{d}x^{\nu}+\Omega^{\mu\nu}\mathbf{d}y_{\mu}\mathbf{d}y_{\nu}$
we have coordinate components
\begin{eqnarray*}
\Omega_{\mu\nu} & = & \chi\left(c_{\mu\nu}-c_{\nu\mu}\right)\\
\Omega_{\;\;\nu}^{\mu} & = & -\chi\delta_{\nu}^{\mu}\\
\Omega^{\mu\nu} & = & 0
\end{eqnarray*}
Therefore, while the space spanned by $\mathbf{f}_{a}=0$ shows no
unphysical size changes, $\Omega_{ab}=0$, the space defined by setting
$y_{\mu}=0$ has $\Omega_{\mu\nu}$ given by the antisymmetric part
of $c_{\mu\nu}$.

It is possible to avoid dilatational curvature altogether by setting
$\chi=0$. In this case, the full dilatational curvature is identically
zero. There is still a symplectic form in this subclass of theories,
since we still have $\mathbf{d}\boldsymbol{\omega}=\mathbf{e}^{a}\wedge\mathbf{f}_{a}$.
This permits the consistent interpretation of the Weyl vector as the
conjugate electromagnetic momentum according to Eq.(\ref{Integrated Weyl vector}).

\medskip{}

Notice that setting $\chi=0$ is inconsistent with the $1+\chi=0$
cases to be studied in the next Section. The possibility of a geometric
graviweak theory with $1+\chi=0$ is more appealing than this $\chi=0$
case, since the success of the standard model strongly suggests that
the electromagnetic and weak interactions should arise together. We
continue with the generic picture, but eventually choose the $y_{a}$
coordinate to be offset by $\beta_{\mu}\left(x\right)=-\alpha_{\mu}\left(x\right)$.
This makes $\Omega_{\mu\nu}=\chi c_{\left[\mu\nu\right]}$ vanish
without restricting the action, while it leaves the cross-dilatation
nonzero and $c_{\mu\nu}$ symmetric. There is no effect of this on
spacetime, but, identifying $y_{\mu}=\frac{i}{\hbar}p_{\mu}$ as argued
in \cite{WheelerQMComplex,WheelerQMandG} it leads to a non-integrability
in phase space of the form $\frac{i}{\hbar}\oint p_{\mu}dx^{\mu}\neq0$
arising from the interesting conjunction of the dilatational curvature
with the symplectic form. The result might be consistent with a quantum
interpretation. This idea has been explored in \cite{WheelerQMandG,WheelerQMComplex,AWQM}.

\medskip{}

Without further conjecture on the interpretation of the geometry,
we continue with the generic case of the reduction toward general
relativity. Without loss of generality, we choose the $y_{\mu}$ coordinate
so that $\alpha_{\mu}=\tilde{\alpha}\left(x\right)+\beta_{\mu}\left(x\right)=0$,
but this is merely a convenient coordinate choice. The solution retains
full coordinate covariance.

Collecting the forms for the connection and basis established in Eqs.(\ref{Spin connection})
and (\ref{Final solder form}), and writing the gauged form of the
Weyl vector co-solder form, we now have
\begin{eqnarray}
\boldsymbol{\omega}_{\;b}^{a} & = & \boldsymbol{\alpha}_{\;b}^{a}\left(x\right)-2\Delta_{db}^{ac}W_{c}\mathbf{e}^{d}\label{Spin connection-1}\\
W_{\alpha} & = & -\left(1+\chi\right)y_{\alpha}\label{Simplified Weyl vector}\\
\mathbf{e}^{a} & = & e_{\alpha}^{\;\;a}\left(x\right)\mathbf{d}x^{\alpha}\label{Simplified solder form}\\
\mathbf{f}_{a} & = & e_{a}^{\;\;\alpha}\mathbf{d}y_{\alpha}+c_{ab}\left(x,y\right)\mathbf{e}^{b}\;\;=\;\;\mathbf{h}_{a}+\mathbf{c}_{a}\label{Simplified co solder form}
\end{eqnarray}
where
\begin{eqnarray}
\mathbf{c}_{a} & = & -\frac{1}{1+\chi}\left(\boldsymbol{\mathcal{R}}_{a}+\mathbf{D}_{\left(\alpha,x\right)}W_{a}+W_{a}\boldsymbol{\omega}-\frac{1}{2}\eta_{ab}W^{2}\mathbf{e}^{b}\right)\label{Solution for ca}\\
\mathbf{c} & = & -\mathbf{d}\boldsymbol{\alpha}\;\;=\;\;0\nonumber 
\end{eqnarray}
The dilatation may now be written as 
\begin{eqnarray}
\boldsymbol{\Omega} & = & \chi\mathbf{e}^{a}\wedge\mathbf{f}_{a}\;\;=\;\;\chi\mathbf{e}^{a}\wedge\mathbf{h}_{a}\;\;=\;\;\chi\mathbf{d}x^{\mu}\wedge\mathbf{d}y_{\mu}\label{Final dilatation}
\end{eqnarray}

\subsection{The co-solder equation}

Now consider the co-solder equation, Eq.(\ref{Reduced co torsion}),
with the co-torsion constrained by the Bianchi identities, Eqs.(\ref{Bianchi identities for co-torsion})
\begin{eqnarray}
\mathbf{d}\mathbf{f}_{a} & = & \boldsymbol{\omega}_{\;a}^{b}\land\mathbf{f}_{b}+\mathbf{f}_{a}\land\boldsymbol{\omega}+\frac{1}{2}S_{acd}\mathbf{e}^{c}\land\mathbf{e}^{d}\label{Co solder structure equation reduced}
\end{eqnarray}
First, note that the the Bianchi identity $S_{\left[abc\right]}=0$
is identically satisfied, since contraction with the solder form vanishes
identically:
\begin{eqnarray*}
\frac{1}{2}S_{\left[acd\right]}\mathbf{e}^{a}\land\mathbf{e}^{c}\land\mathbf{e}^{d} & = & \mathbf{e}^{a}\land\mathbf{d}\mathbf{f}_{a}-\mathbf{e}^{a}\land\boldsymbol{\omega}_{\;a}^{b}\land\mathbf{f}_{b}-\mathbf{e}^{a}\land\mathbf{f}_{a}\land\boldsymbol{\omega}\\
 & = & -\mathbf{d}\left(\mathbf{e}^{a}\land\mathbf{f}_{a}\right)+\left(\mathbf{d}\mathbf{e}^{b}-\mathbf{e}^{a}\land\boldsymbol{\omega}_{\;a}^{b}-\boldsymbol{\omega}\land\mathbf{e}^{b}\right)\land\mathbf{f}_{b}\\
 & = & 0
\end{eqnarray*}

Now, solving Eq.(\ref{Co solder structure equation reduced}) for
the co-torsion and substituting for the connection forms
\begin{eqnarray*}
\mathbf{S}_{a} & = & \mathbf{d}\mathbf{f}_{a}-\boldsymbol{\omega}_{\;a}^{b}\wedge\mathbf{f}_{b}-\mathbf{f}_{a}\wedge\boldsymbol{\omega}\\
 & = & \mathbf{D}_{\left(\omega\right)}\mathbf{f}_{a}\\
 & = & \mathbf{D}_{\left(\omega\right)}\left(e_{a}^{\;\;\alpha}\mathbf{d}y_{\alpha}+c_{ab}\left(x,y\right)\mathbf{e}^{b}\right)\\
 & = & e_{a}^{\;\;\alpha}\mathbf{D}_{\left(\omega\right)}\left(\mathbf{d}y_{\alpha}\right)+\mathbf{D}_{\left(\omega\right)}\mathbf{c}_{a}\\
 & = & e_{a}^{\;\;\alpha}\left(\mathbf{d}\left(\mathbf{d}y_{\alpha}\right)+\mathbf{d}y_{\beta}\wedge\Sigma_{\;\;\alpha\mu}^{\beta}\mathbf{d}x^{\mu}\right)+\left(\mathbf{d}\mathbf{c}_{a}+\mathbf{c}_{b}\wedge\boldsymbol{\omega}_{\;\;a}^{b}-\boldsymbol{\omega}\wedge\mathbf{c}_{a}\right)\\
 & = & e_{a}^{\;\;\alpha}\Sigma_{\;\;\alpha\mu}^{\beta}\mathbf{d}y_{\beta}\wedge\mathbf{d}x^{\mu}+\left(\mathbf{d}\mathbf{c}_{a}+\mathbf{c}_{b}\wedge\boldsymbol{\omega}_{\;\;a}^{b}+\boldsymbol{\omega}\wedge\mathbf{c}_{a}\right)
\end{eqnarray*}
We first need the $\mathbf{d}y_{\beta}$ dependent pieces of $\mathbf{d}\mathbf{c}_{a}$,
where $\mathbf{c}_{a}$ is given by Eq.(\ref{Solution for ca}). Since
the only $y$-dependence is in the Weyl vector,
\begin{eqnarray*}
\mathbf{d}\mathbf{c}_{a} & = & \mathbf{d}_{\left(x\right)}\mathbf{c}_{a}-\frac{1}{1+\chi}\mathbf{d}_{\left(y\right)}\left(\mathbf{D}_{\left(\alpha,x\right)}W_{a}+W_{a}\boldsymbol{\omega}-\frac{1}{2}\eta_{ab}W^{2}\mathbf{e}^{b}\right)
\end{eqnarray*}
Expanding the $x$-dependent, $\alpha$-covariant derivative of the
Weyl vector,
\begin{eqnarray*}
\mathbf{D}_{\left(\alpha,x\right)}W_{b} & = & \mathbf{e}^{a}e_{a}^{\;\;\mu}D_{\mu}^{\left(\alpha,x\right)}\left(e_{b}^{\;\;\mu}W_{\mu}\right)\\
 & = & \left(1+\chi\right)\mathbf{e}^{a}e_{a}^{\;\;\mu}e_{b}^{\;\;\nu}D_{\mu}^{\left(\alpha,x\right)}\left(-y_{\nu}\right)\\
 & = & \left(1+\chi\right)\mathbf{e}^{a}e_{a}^{\;\;\mu}e_{b}^{\;\;\nu}\left(y_{\alpha}\Gamma_{\;\;\nu\mu}^{\alpha}\left(x\right)\right)
\end{eqnarray*}
the $y$-derivatives become
\begin{eqnarray*}
\mathbf{d}_{\left(y\right)}\mathbf{c}_{a} & = & -\frac{1}{1+\chi}\mathbf{d}_{\left(y\right)}\left(\mathbf{D}_{\left(\alpha,x\right)}W_{a}+W_{a}\boldsymbol{\omega}-\frac{1}{2}\eta_{ab}W^{2}\mathbf{e}^{b}\right)\\
 & = & -\frac{1}{1+\chi}e_{b}^{\;\;\mu}e_{a}^{\;\;\nu}\mathbf{d}_{\left(y\right)}\left(\left(1+\chi\right)y_{\alpha}\Gamma_{\;\;\nu\mu}^{\alpha}+W_{\mu}W_{\nu}-\frac{1}{2}g_{\mu\nu}g^{\alpha\beta}W_{\alpha}W_{\beta}\right)\wedge\mathbf{e}^{b}\\
 & = & -e_{b}^{\;\;\mu}e_{a}^{\;\;\nu}\left(\mathbf{d}y_{\alpha}\Gamma_{\;\;\nu\mu}^{\alpha}-\mathbf{d}y_{\mu}W_{\nu}-W_{\mu}\mathbf{d}y_{\nu}+g_{\mu\nu}g^{\alpha\beta}W_{\alpha}\mathbf{d}y_{\beta}\right)\wedge\mathbf{e}^{b}\\
 & = & -e_{b}^{\;\;\mu}e_{a}^{\;\;\nu}\left(\Gamma_{\;\;\nu\mu}^{\beta}-\delta_{\mu}^{\beta}W_{\nu}-\delta_{\nu}^{\beta}W_{\mu}+g_{\mu\nu}g^{\alpha\beta}W_{\alpha}\right)\mathbf{d}y_{\beta}\wedge\mathbf{e}^{b}\\
 & = & -e_{a}^{\;\;\nu}\Sigma_{\;\;\nu\mu}^{\beta}\mathbf{d}y_{\beta}\wedge\mathbf{d}x^{\mu}
\end{eqnarray*}
Substituting, the $\mathbf{d}y_{\alpha}$ terms cancel identically,
leaving
\begin{eqnarray*}
\mathbf{S}_{a} & = & e_{a}^{\;\;\alpha}\Sigma_{\;\;\alpha\mu}^{\beta}\mathbf{d}y_{\beta}\wedge\mathbf{d}x^{\mu}+\left(\mathbf{d}_{\left(x\right)}\mathbf{c}_{a}+\mathbf{c}_{b}\wedge\boldsymbol{\omega}_{\;\;a}^{b}+\boldsymbol{\omega}\wedge\mathbf{c}_{a}\right)-e_{a}^{\;\;\nu}\Sigma_{\;\;\nu\mu}^{\beta}\mathbf{d}y_{\beta}\wedge\mathbf{d}x^{\mu}\\
 & = & \mathbf{d}_{\left(x\right)}\mathbf{c}_{a}+\mathbf{c}_{b}\wedge\boldsymbol{\omega}_{\;\;a}^{b}+\boldsymbol{\omega}\wedge\mathbf{c}_{a}
\end{eqnarray*}
This shows once again that the cross-term of the co-torsion vanishes,
$S_{a\quad b}^{\;\;\;c}=0$. Now we expand the spin connection, rewriting
all of the derivatives as $x$-dependent, $\alpha$-covariant, $\mathbf{D}_{\left(\alpha,x\right)}$.
\begin{eqnarray*}
\mathbf{S}_{a} & = & \mathbf{d}_{\left(x\right)}\mathbf{c}_{a}+\mathbf{c}_{b}\wedge\left(\boldsymbol{\alpha}_{\;\;a}^{b}+\boldsymbol{\beta}_{\;\;a}^{b}\right)+\boldsymbol{\omega}\wedge\mathbf{c}_{a}\\
 & = & \mathbf{D}_{\left(\alpha,x\right)}\mathbf{c}_{a}+\mathbf{c}_{b}\wedge\boldsymbol{\beta}_{\;\;a}^{b}+\boldsymbol{\omega}\wedge\mathbf{c}_{a}\\
 & = & -\frac{1}{1+\chi}\mathbf{D}_{\left(\alpha,x\right)}\left(\boldsymbol{\mathcal{R}}_{a}+\mathbf{D}_{\left(\alpha,x\right)}W_{a}+W_{a}\boldsymbol{\omega}-\frac{1}{2}\eta_{ab}W^{2}\mathbf{e}^{b}\right)\\
 &  & -\frac{1}{1+\chi}\left(\boldsymbol{\mathcal{R}}_{b}+\mathbf{D}_{\left(\alpha,x\right)}W_{b}+W_{b}\boldsymbol{\omega}-\frac{1}{2}\eta_{bc}W^{2}\mathbf{e}^{c}\right)\wedge\boldsymbol{\beta}_{\;\;a}^{b}\\
 &  & -\frac{1}{1+\chi}\left(\boldsymbol{\omega}\wedge\boldsymbol{\mathcal{R}}_{a}+\boldsymbol{\omega}\wedge\mathbf{D}_{\left(\alpha,x\right)}W_{a}-\frac{1}{2}\eta_{ab}W^{2}\boldsymbol{\omega}\wedge\mathbf{e}^{b}\right)
\end{eqnarray*}
After distributing the covariant derivative and expanding $\boldsymbol{\beta}_{\;\;\;a}^{b}$,
we separate curvature terms and simplify,
\begin{eqnarray*}
\mathbf{S}_{a} & = & -\frac{1}{1+\chi}\left(\mathbf{D}_{\left(\alpha,x\right)}\boldsymbol{\mathcal{R}}_{a}+\mathbf{D}_{\left(\alpha,x\right)}\wedge\mathbf{D}_{\left(\alpha,x\right)}W_{a}-W_{b}\boldsymbol{\mathcal{R}}_{a}\wedge\mathbf{e}^{b}+\eta^{bd}\eta_{ea}W_{d}\boldsymbol{\mathcal{R}}_{b}\wedge\mathbf{e}^{e}\right)\\
 &  & -\frac{1}{1+\chi}\left(\mathbf{D}_{\left(\alpha,x\right)}W_{a}\wedge\boldsymbol{\omega}+\boldsymbol{\omega}\mathbf{D}_{\left(\alpha,x\right)}W_{a}+\eta^{bd}\eta_{ea}W_{d}\left(\mathbf{D}_{\left(\alpha,x\right)}W_{b}\right)\wedge\mathbf{e}^{e}-\eta_{ab}\eta^{cd}W_{c}\left(\mathbf{D}_{\left(\alpha,x\right)}W_{d}\right)\wedge\mathbf{e}^{b}\right)\\
 &  & -\frac{1}{1+\chi}\left(\eta_{ea}W^{2}\boldsymbol{\omega}\wedge\mathbf{e}^{e}-\frac{1}{2}\eta_{ea}W^{2}\boldsymbol{\omega}\wedge\mathbf{e}^{e}-\frac{1}{2}\eta_{ab}W^{2}\boldsymbol{\omega}\wedge\mathbf{e}^{b}\right)\\
 & = & -\frac{1}{1+\chi}\left(\mathbf{D}_{\left(\alpha,x\right)}\boldsymbol{\mathcal{R}}_{a}+\mathbf{D}_{\left(\alpha,x\right)}\wedge\mathbf{D}_{\left(\alpha,x\right)}W_{a}-W_{b}\left(\delta_{a}^{d}\delta_{e}^{b}-\eta^{bd}\eta_{ea}\right)\boldsymbol{\mathcal{R}}_{d}\wedge\mathbf{e}^{e}\right)
\end{eqnarray*}
Using the partition of the Riemann tensor, Eq.(\ref{Partition of Riemann}),
and the Ricci identity,
\begin{eqnarray*}
\mathbf{D}_{\left(\alpha,x\right)}\wedge\mathbf{D}_{\left(\alpha,x\right)}W_{a} & = & -W_{b}\mathbf{R}{}_{\;\;a}^{b}
\end{eqnarray*}
we see that $\mathbf{S}_{a}$ is
\begin{eqnarray}
\mathbf{S}_{a} & = & -\frac{1}{1+\chi}\left(\mathbf{D}_{\left(\alpha,x\right)}\boldsymbol{\mathcal{R}}_{a}-W_{b}\mathbf{C}_{\;\;a}^{b}\right)\label{Co-torsion}
\end{eqnarray}
If $W_{a}$ were the gradient of a function of $x^{\mu}$, then Eq.(\ref{Co-torsion})
would be the condition for the spacetime to be conformal to a Ricci-flat
spacetime. Since
\begin{eqnarray*}
\mathbf{d}_{\left(x\right)}\boldsymbol{\omega} & = & \mathbf{d}_{\left(x\right)}\left(-\left(1+\chi\right)y_{\alpha}\mathbf{d}x^{\alpha}\right)\\
 & = & 0
\end{eqnarray*}
this is the case, but only at constant $y_{\alpha}$.
\[
\mathbf{S}_{\mu\alpha\beta}=-\frac{1}{1+\chi}D_{\alpha}^{\left(\alpha,x\right)}\boldsymbol{\mathcal{R}}_{a}+y_{b}\mathbf{C}_{\;\;a}^{b}
\]
This is in agreement with our conclusion, Eq.(\ref{Reduced spacetime curvature Bianchi}),
from the spacetime Bianchi equation combined with the usual Riemannian
Bianchi identity. The result means that for vanishing co-torsion and
\emph{constant} $y_{\alpha}$ there exists an $x$-dependent gauge
transformation to a Ricci flat spacetime. This form is not unfamiliar,
the same expression having been noted in another context in \cite{Weyl grav as GR}.
The remaining $y$-dependence is the only obstruction to the Triviality
Theorem: if a conformal transformation could make $S_{abc}$ vanish
for \emph{all} $y_{\alpha}$, then the biconformal space would necessarily
be trivial.

Note that the field equations Eqs.(\ref{Field equations for co-torsion})
and the Bianchi identities Eqs.(\ref{Bianchi identities for co-torsion})
for the co-torsion are now all satisfied for any allowed $\mathbf{S}_{a}$.

\subsection{Collecting the results $(1+\chi)\protect\neq0$}

We have now solved for the full connection and satisfied all of the
field equations.
\begin{eqnarray*}
\boldsymbol{\omega}_{\;\;\;b}^{a} & = & \boldsymbol{\alpha}_{\;\;\;b}^{a}\left(x\right)-2\Delta_{db}^{ac}W_{c}\mathbf{e}^{d}\\
\mathbf{e}^{a} & = & e_{\alpha}^{\;\;a}\left(x\right)\mathbf{d}x^{\alpha}\\
\mathbf{f}_{a} & = & e_{a}^{\;\;\alpha}\mathbf{d}y_{\alpha}-\frac{1}{1+\chi}\left(\boldsymbol{\mathcal{R}}_{a}+\mathbf{D}_{\left(\alpha,x\right)}W_{a}+W_{a}\boldsymbol{\omega}-\frac{1}{2}\eta_{ab}W^{2}\mathbf{e}^{b}\right)\\
 & = & \mathbf{h}_{a}-\frac{1}{1+\chi}\boldsymbol{\mathscr{R}}_{a}\\
\boldsymbol{\omega} & = & -\left(1+\chi\right)y_{\alpha}\mathbf{d}x^{\alpha}
\end{eqnarray*}
where $\chi\equiv\frac{1}{n-1}\frac{1}{\left(n-1\right)\alpha-\beta}\Lambda$.
Notice that $\mathbf{d}\boldsymbol{\omega}\,=\left(1+\chi\right)\mathbf{e}^{c}\mathbf{f}_{c}$
defines a symplectic form.

The curvatures follow from the structure equations as

\begin{eqnarray}
\boldsymbol{\Omega}_{\;\;b}^{a} & = & \mathbf{C}_{\;\;b}^{a}+2\left(1+\chi\right)\Delta_{db}^{ac}\mathbf{h}_{c}\wedge\mathbf{e}^{d}\label{Generic curvature}\\
\mathbf{T}^{a} & = & 0\label{Generic torsion}\\
\mathbf{S}_{a} & = & -\frac{1}{1+\chi}\mathbf{D}_{\left(\alpha,x\right)}\boldsymbol{\mathcal{R}}_{a}+y_{b}\mathbf{C}_{\;\;a}^{b}\label{Generic co-torsion}\\
\boldsymbol{\Omega} & = & \chi\mathbf{e}^{a}\wedge\mathbf{h}_{a}\label{Generic dilatation}
\end{eqnarray}

The combination $\chi\mathbf{e}^{a}\wedge\mathbf{h}_{a}=\chi\mathbf{d}x^{\alpha}\wedge\mathbf{d}y_{\alpha}$
is also non-degenerate and closed, and therefore symplectic.

\subsection{The Lagrangian submanifold of spacetime}

The basis forms $\mathbf{h}_{a}=e_{a}^{\;\;\mu}\mathbf{d}y_{\mu}$
are manifestly involute. Holding $y_{\mu}$ constant so that $\mathbf{h}_{a}=0$,
the resulting vanishing of the symplectic form shows that the $\mathbf{h}_{a}=0$
submanifold is Lagrangian. The structure equations for the resulting
Lagrangian submanifold are
\begin{eqnarray}
\mathbf{d}\boldsymbol{\omega}_{\;b}^{a} & = & \boldsymbol{\omega}_{\;b}^{c}\wedge\boldsymbol{\omega}_{\;c}^{a}+2\left(1+\chi\right)\Delta_{db}^{ac}\mathbf{c}_{c}\wedge\mathbf{e}^{d}+\mathbf{C}_{\;\;b}^{a}\left(\alpha\right)\label{Spacetime curvature structure equation}\\
\mathbf{d}\mathbf{e}^{a} & = & \mathbf{e}^{b}\wedge\boldsymbol{\omega}_{\;\;b}^{a}+\boldsymbol{\omega}\wedge\mathbf{e}^{a}\label{Spacetime solder form structure equation}\\
\mathbf{d}\boldsymbol{\omega}\, & = & 0\nonumber 
\end{eqnarray}
and the remaining part of the co-solder equation is,
\begin{eqnarray}
\mathbf{S}_{a} & = & \mathbf{d}\mathbf{c}_{a}-\boldsymbol{\omega}_{\;a}^{b}\wedge\mathbf{c}_{b}-\mathbf{c}_{a}\wedge\boldsymbol{\omega}\;\;=\;\;\mathbf{D}_{\left(\omega,x\right)}\mathbf{c}_{a}\label{Spacetime co torsion}
\end{eqnarray}
With $y_{\mu}=y_{\mu}^{0}$ constant, the form of the connection is
\begin{eqnarray*}
\boldsymbol{\omega}_{\;b}^{a} & = & \boldsymbol{\alpha}_{\;b}^{a}\left(x\right)+2\left(1+\chi\right)\Delta_{db}^{ac}e_{c}^{\;\;\alpha}y_{\alpha}^{0}\mathbf{e}^{d}\\
\mathbf{e}^{a} & = & e_{\alpha}^{\;\;a}\left(x\right)\mathbf{d}x^{\alpha}\\
W_{\alpha} & = & -\left(1+\chi\right)y_{\alpha}^{0}
\end{eqnarray*}
Notice that the Weyl vector is now the gradient of $-\left(1+\chi\right)y_{\alpha}^{0}x^{\alpha}$
with respect to $x^{\alpha}$. There is one further consequence of
the curvature field equation,
\[
\mathbf{c}_{a}=-\frac{1}{1+\chi}\left(\boldsymbol{\mathcal{R}}_{a}+\mathbf{D}_{\left(\alpha,x\right)}W_{a}+W_{a}\boldsymbol{\omega}-\frac{1}{2}\eta_{ab}W^{2}\mathbf{e}^{b}\right)=-\frac{1}{1+\chi}\boldsymbol{\mathscr{R}}_{a}
\]
and the curvatures are as given in Eqs.(\ref{Generic curvature})-(\ref{Generic co-torsion})
with $y_{\mu}=y_{\mu}^{0}$, together with $\boldsymbol{\Omega}=0$.

\subsubsection{Interpreting $\mathbf{c}_{a}$}

Combining Eq.(\ref{Generic co-torsion}) at constant $y_{\mu}$ with
Eq.(\ref{Spacetime co torsion}), we expand the derivatives and replace
the Weyl curvature using the partition of the Riemann tensor, Eq.(\ref{Partition of Riemann}),
\begin{eqnarray*}
\mathbf{S}_{a} & = & -\frac{1}{1+\chi}\left(\mathbf{D}_{\left(\alpha,x\right)}\boldsymbol{\mathcal{R}}_{a}+\left(1+\chi\right)y_{b}^{0}\mathbf{C}_{\;\;a}^{b}\right)\\
\mathbf{D}_{\left(\alpha,x\right)}\mathbf{c}_{a}-W_{b}2\Delta_{da}^{bc}\mathbf{c}_{c}\wedge\mathbf{e}^{d} & = & -\frac{1}{1+\chi}\left(\mathbf{D}_{\left(\alpha,x\right)}\boldsymbol{\mathcal{R}}_{a}-W_{b}\mathbf{R}_{\;\;a}^{b}-W_{b}2\Delta_{da}^{bc}\boldsymbol{\mathcal{R}}_{c}\wedge\mathbf{e}^{d}\right)\\
-\left(1+\chi\right)\mathbf{D}_{\left(\alpha,x\right)}\mathbf{c}_{a}+2\left(1+\chi\right)W_{b}\Delta_{da}^{bc}\mathbf{c}_{c}\wedge\mathbf{e}^{d} & = & \mathbf{D}_{\left(\alpha,x\right)}\boldsymbol{\mathcal{R}}_{a}-W_{b}\mathbf{R}_{\;\;a}^{b}-W_{b}2\Delta_{da}^{bc}\boldsymbol{\mathcal{R}}_{c}\wedge\mathbf{e}^{d}
\end{eqnarray*}
and therefore
\begin{equation}
0=\mathbf{D}_{\left(\alpha,x\right)}\left(\boldsymbol{\mathcal{R}}_{a}+\left(1+\chi\right)\mathbf{c}_{a}\right)-W_{b}\mathbf{R}_{\;\;a}^{b}-W_{b}2\Delta_{da}^{bc}\left(\boldsymbol{\mathcal{R}}_{c}+\left(1+\chi\right)\mathbf{c}_{c}\right)\wedge\mathbf{e}^{d}\label{Scale covariant condition for Einstein equation}
\end{equation}
This is exactly the condition for the existence of a conformal transformation
to a spacetime satisfying the Einstein equation with matter sources,
found in \cite{WeylGeom}, where the matter source $\left(1+\chi\right)c_{ab}$
is given in terms of the energy-momentum tensor $T_{ab}$ by
\begin{equation}
\left(1+\chi\right)c_{ab}=-\frac{1}{n-2}\left(T_{ab}-\frac{1}{n-1}\eta_{ab}T\right)\label{Energy momentum tensor}
\end{equation}
Therefore, there exists an $x^{\alpha}$-dependent rescaling of the
solder form \textendash{} the Riemannian gauge \textendash{} such
that the co-torsion equation becomes
\begin{equation}
\boldsymbol{\mathcal{R}}_{a}=-\left(1+\chi\right)\mathbf{c}_{a}\label{Einstein equation in Riemannian gauge}
\end{equation}
which, in turn, is the Einstein equation with source given by $T_{ab}$.
Explicitly, substituting the definition of the Schouten tensor from
Eq.(\ref{Schouten tensor}) and the form of $c_{ab}$ from Eq.(\ref{Energy momentum tensor}),
\begin{eqnarray*}
\frac{1}{\left(n-2\right)}\left(R_{ab}-\frac{1}{2\left(n-1\right)}\eta_{ab}R\right) & = & \frac{1}{n-2}\left(T_{ab}-\frac{1}{n-1}\eta_{ab}T\right)
\end{eqnarray*}
we substitute for the trace of the energy momentum, $T=-\frac{1}{2}\left(n-2\right)R$.
Solving for the energy-momentum tensor, we find the usual form of
the Einstein equation,
\begin{eqnarray*}
R_{ab}-\frac{1}{2}\eta_{ab}R & = & T_{ab}
\end{eqnarray*}

The $\mathbf{h}_{a}=0$ submanifold is therefore a spacetime satisfying
the locally scale-covariant Einstein equation, including phenomenological
matter sources. Study is underway to determine whether the energy-momentum
of fundamental source fields automatically enter in this way in place
of $c_{ab}$, or if special couplings to matter are required in the
Lagrangian. Notice that in the Riemannian gauge Eq.(\ref{Scale covariant condition for Einstein equation})
reduces to
\begin{eqnarray*}
W_{b}\mathbf{R}_{\;\;a}^{b} & = & 0
\end{eqnarray*}
For a single vector to annihilate the curvature tensor can happen
only in the simplest Petrov type spacetimes ($O$ and $N$, and these
are already conformally Ricci flat; see \cite{Weyl grav as GR}) and
we conclude that, generically, the gauge transformation that makes
the spacetime Riemannian is simultaneously the one which makes the
Weyl vector vanish.

\subsubsection{Contractions of the Bianchi identity for the curvature on the Lagrangian
submanifold}

In components, the Bianchi identity for the Riemann-Weyl curvature
on the spacetime submanifold, Eq.(\ref{Reduced spacetime curvature Bianchi})
becomes
\begin{eqnarray*}
D_{a}^{\left(\alpha,x\right)}\mathcal{R}_{bc}-D_{c}^{\left(\alpha,x\right)}\mathcal{R}_{ba}-W_{e}C_{\;\;bac}^{e}+\left(1+\chi\right)S_{bac} & = & 0
\end{eqnarray*}
Taking the $bc$ contraction,
\begin{eqnarray*}
D_{a}^{\left(\alpha,x\right)}\mathcal{R}-D_{\left(\alpha,x\right)}^{c}\mathcal{R}_{ca}+\left(1+\chi\right)\eta^{bc}S_{bac} & = & 0
\end{eqnarray*}
In the Riemannian gauge, and therefore any gauge, the first two terms
cancel since
\[
D_{a}^{\left(\alpha,x\right)}\mathcal{R}=D_{\left(\alpha,x\right)}^{c}\mathcal{R}_{ca}
\]
follows from the Bianchi identity for the Riemannian curvature. Therefore,
\begin{eqnarray*}
\eta^{bc}S_{bac} & = & 0
\end{eqnarray*}

Now, expanding the co-torsion using Eq.(\ref{Spacetime co torsion})
\begin{eqnarray*}
\mathbf{S}_{a} & = & \mathbf{D}_{\left(\omega,x\right)}\mathbf{c}_{a}\\
 & = & \mathbf{D}_{\left(\alpha,x\right)}\mathbf{c}_{a}-2\Delta_{da}^{bc}W_{b}\mathbf{c}_{c}\wedge\mathbf{e}^{d}
\end{eqnarray*}
In components, this is $S_{abc}=D_{b}^{\left(\alpha,x\right)}c_{ac}-D_{c}^{\left(\alpha,x\right)}c_{ab}$,
so that
\begin{eqnarray*}
0 & = & \eta^{bc}S_{bac}\;\;=\;\;c_{;b}-c_{\;\;b;a}^{a}
\end{eqnarray*}
and substituting for $c_{ab}$ from Eq.(\ref{Energy momentum tensor})
in
\begin{eqnarray*}
0 & = & \frac{1}{\left(n-1\right)\left(n-2\right)}T_{;b}+\frac{1}{n-2}\left(T_{\;\;\;b;a}^{a}-\frac{1}{n-1}T_{;a}\right)\\
 & = & \frac{1}{n-2}T_{\;\;\;b;a}^{a}
\end{eqnarray*}
and we have established $c_{ab}$ to be both symmetric and conserved
on spacetime, in agreement with the requirments for the energy momentum
tensor. Naturally, this last condition also follows directly from
the vanishing divergence of the Einstein tensor.

\subsubsection{Metric on the Lagrangian submanifold}

While we have used the Killing form to motivate the choice of $h_{a}^{\;\;\;\mu}\left(x\right)$
in the basis form on the cotangent spaces, Eq.(\ref{Momentum submanifold basis form}),
the use of the Killing form is not necessary. Indeed, when general
relativity is developed as a gauge theory from the Poincaré group,
the Killing form vanishes when restricted to the base manifold. Instead,
the spacetime metric may be motivated by the spin connection, which
is compatible with Lorentzian signature. Ultimately, there is no inherent
group structure that requires the choice except this compatibility.
Similarly, in the biconformal gauging, we may introduce an $SO\left(p,q\right)$
compatible metric by hand on the Lagrangian submanifolds where the
restriction of the Killing form vanishes. For Lorentzian cases, with
an original $SO\left(n-1,1\right)$ spin connection, it is natural
to introduce the corresponding Minkowski metric on each Lagrangian
submanifold.

This choice is sufficient for the generic case, but since the restriction
of the conformal killing form to biconformal space is non-degenerate,
there are alternatives that trace their origin to the conformal group.
These have been explored in a variety of ways (\cite{WheelerQMComplex,Spencer Wheeler,Hazboun Wheeler,Hazboun Wheeler-1,LoveladyWheeler})
but these considerations take us too far beyond the scope of this
class of solutions.

\section{Non-abelian case: $1+\chi=0$ \label{sec:Non-abelian-case}}

We note that the condition $1+\chi=0$ becomes, in terms of the parameters
of the original action,
\begin{eqnarray*}
0 & = & n\gamma-\left(\left(n-1\right)\alpha-\beta\right)
\end{eqnarray*}
and this does not coincide with any other special conditions.

We now return to the form of the connection, structure equations,
and curvatures established at the end of Sec.(\ref{sec:The full space}),
and set $1+\chi=0$. The connection forms still take the form,
\begin{eqnarray*}
\boldsymbol{\omega}_{\;\;\;b}^{a} & = & \boldsymbol{\alpha}_{\;\;\;b}^{a}-2\Delta_{db}^{ac}W_{c}\mathbf{e}^{d}\\
\mathbf{e}^{a} & = & e_{\mu}^{\;\;\;a}\left(x\right)\mathbf{d}x^{\mu}\\
\mathbf{f}_{a} & = & h_{a}^{\;\;\mu}\left(x,y\right)\mathbf{d}y_{\mu}+c_{ab}\left(x,y\right)\mathbf{e}^{b}\\
 & \equiv & \mathbf{h}_{a}+\mathbf{c}_{a}\\
\boldsymbol{\omega} & = & W_{a}\left(x,y\right)\mathbf{e}^{a}
\end{eqnarray*}
The form of the spin connection immediately gives the solution for
$\Omega_{\;\;\;bcd}^{a}$ as the Riemann-Weyl curvature tensor of
an integrable Weyl geometry, given by Eq.(\ref{Riemann tensor of Weyl geometry}).
The field equation is the vanishing of the Weyl-Schouten tensor,
\begin{eqnarray*}
0 & = & \Omega_{\;\;bac}^{a}\\
 & = & \mathscr{R}_{bc}\\
 & = & \mathcal{R}_{bc}+\phi_{\left(b;c\right)}+\phi_{b}\phi_{c}-\frac{1}{2}\eta_{bc}\eta^{ad}\phi_{a}\phi_{d}
\end{eqnarray*}
and therefore vanishing Weyl-Ricci tensor. The field equation reduces
the full spacetime curvature to the Weyl curvature
\[
\Omega_{\;\;bcd}^{a}=C_{\;\;bcd}^{a}\left(\alpha\right)
\]
with $\boldsymbol{\alpha}_{\;\;\;b}^{a}$ the metric compatible connection.

The dilatation structure equation is now simply
\[
\mathbf{d}\boldsymbol{\omega}\,=0
\]
so to simplify the form of the field equations, we may gauge to $\boldsymbol{\omega}=0$.
In the $W_{a}=0$ gauge, the Weyl connection becomes Riemannian, $\boldsymbol{\omega}_{\;\;\;b}^{a}=\boldsymbol{\alpha}_{\;\;\;b}^{a}$
and the curvature is
\begin{eqnarray}
\frac{1}{2}\Omega_{\;\;bcd}^{a}\mathbf{e}^{c}\wedge\mathbf{e}^{d} & = & \mathbf{d}\boldsymbol{\alpha}_{\;b}^{a}-\boldsymbol{\alpha}_{\;b}^{c}\wedge\boldsymbol{\alpha}_{\;c}^{a}=\mathbf{R}_{\;\;b}^{a}\label{Special case curvature}
\end{eqnarray}
where $\boldsymbol{\alpha}_{\;\;b}^{a}$ is the metric compatible
connection, $\mathbf{d}\mathbf{e}^{a}=\mathbf{e}^{b}\wedge\boldsymbol{\alpha}_{\;\;b}^{a}$.
The curvature field equation is simply the vacuum Einstein equation.
The dilatation and curvature cross terms are now of unit magnitude,
\begin{eqnarray*}
\Omega_{\;\;b\quad d}^{a\quad c} & = & 2\Delta_{db}^{ac}\\
\Omega_{\;\;b}^{a} & = & \delta_{b}^{a}
\end{eqnarray*}

The only remaining field equations are those describing the co-torsion,
and the only remaining structure equation is the co-solder equation,
\[
\mathbf{d}\mathbf{f}_{a}=\boldsymbol{\alpha}_{\;a}^{b}\wedge\mathbf{f}_{b}+\frac{1}{2}S_{acd}\mathbf{e}^{c}\wedge\mathbf{e}^{d}+S_{a\quad d}^{\;\;c}\mathbf{f}_{c}\wedge\mathbf{e}^{d}-\frac{1}{2}c_{a}^{\;\;cd}\mathbf{f}_{c}\wedge\mathbf{f}_{d}
\]

When $1+\chi=0$, the remaining structure equation on the $x^{\mu}=constant,\:\mathbf{e}^{a}=0$
submanifold is given by Eq.(\ref{Maurer-Cartan for h}),
\begin{equation}
\mathbf{d}\mathbf{h}_{a}=-\frac{1}{2}c_{a}^{\;\;cd}\mathbf{h}_{c}\land\mathbf{h}_{d}\label{Maurer-Cartan equation for G-1}
\end{equation}
which is precisely the Maurer-Cartan equation for a Lie group $\mathcal{G}$
with structure constants $c_{a}^{\;\;cd}$.

Here we see the realization of one of the motivations for the use
of the conformal group as the starting point for Poincaré gravity,
and the subsequent motivation for the biconformal gauging. One anticipates
that by starting with the larger conformal group and taking the quotient
by the inhomogeneous Weyl group $\mathcal{C}/\mathcal{I}\mathcal{W}$
that the resulting additional symmetry might account for some known
or new fundamental interaction beyond gravity. This hope is frustrated
by the finding that the additional special conformal gauge fields
$\mathbf{f}_{a}$ are always auxiliary, and determined by the Ricci
tensor \cite{Auxiliary}. When these auxiliary gauge fields are substituted
back into the rest of the model, they serve to turn the Riemann curvature
tensor into the Weyl curvature tensor. As a result, though they enforce
conformal symmetry, they never provide an additional interaction.
As a way to avoid the elimination of $\mathbf{f}_{a}$, we are led
to the biconformal gauging, $\mathcal{C}/\mathcal{W}$, the idea being
that if both $\mathbf{e}^{a}$ and $\mathbf{f}_{b}$ together are
required to span the base manifold, then $\mathbf{f}_{b}$ cannot
possibly be removed as auxiliary \cite{Ivanov,NCG}. Although considerable
subsequent work continues to find $\mathbf{f}_{a}$ serving to turn
$\mathbf{R}_{\;\;b}^{a}$ to $\mathbf{C}_{\;\;b}^{a}$ as in subsection
(\ref{The Weyl curvature emerges}), the emergence of an additional
symmetry group is now realized in the $1+\chi$ subclass of cases.

The biconformal space comes equipped with the $SO\left(p,q\right)$
pseudo-rotation group of an $n$-dimensional space, but these rotations
and boosts acts on a $2n$-dimensional manifold. This is much less
than the $SO\left(n,n\right)$ one might expect. Indeed, as seen above,
the generic torsion-free solution dictates that half the space is
flat, so there is no curvature corresponding to the $\mathbf{d}y_{\mu}$
part of the spin connection. The spin connection reduces, essentially,
to the metric compatible connection of $\mathbf{e}^{a}$,
\[
\boldsymbol{\omega}_{\;\;b}^{a}=\omega_{\;\;b\mu}^{a}\left(x,y\right)\mathbf{d}x^{\mu}+\omega_{\;\;b}^{a\quad\mu}\left(x,y\right)\mathbf{d}y_{\mu}\Rightarrow\left(\alpha_{\;\;b\mu}^{a}\left(x\right)+2\Delta_{\mu b}^{ac}y_{c}\right)\mathbf{d}x^{\mu}
\]
which is fully expressed on the $n$-dimensional, constant $y_{\mu}$
Lagrangian submanifolds. This reduction of the spin connection reduces
the number of physical fields, but when $1+\chi=0$ the extra translational
gauge fields \textendash{} the co-solder form \textendash{} make up
for it by providing a new connection and field strength: there is
necessarily an $n$-dimensional Lie group $\mathcal{G}$ acting at
each $x^{\mu}$. Since the $n$-dimensions of this group are labeled
by an $SO\left(p,q\right)$ index, $SO\left(p,q\right)$ \emph{must
act on }$\mathcal{G}$\emph{.} We show in this section that this internal
symmetry group is gauged.

For a particularly pertinent example, suppose we have started with
Euclidean $4$-space. This does \emph{not} preclude a spacetime Lagrangian
submanifold, for it has been shown in \cite{Spencer Wheeler} that
time emerges uniquely from a Euclidean starting point, while \cite{Hazboun Wheeler}
shows that this emergence arises purely from properties of the conformal
group. With the $4$-dim Euclidean starting point, the spin connection
has symmetry $SO\left(4\right)=SU\left(2\right)\times SU\left(2\right)$.
The obvious $4$-dimensional subgroup is the electroweak symmetry,
$SU\left(2\right)\times U\left(1\right)$. In this case, the symmetry
breaking from a left-right symmetric electroweak theory to left-handed
representations of $SU\left(2\right)$ is forced by the requirement
of an $n$-dimensional subgroup. In a spinor representation of the
conformal group, the $P_{a}$ and $K^{a}$ ($x^{\mu}$ and $y_{\mu}$
submanifolds) are left- and right-handed, respectively. Details of
this case are currently under investigation.

It is important to note that although this symmetry $\mathcal{G}$
is restricted to be acted on by $SO\left(p,q\right)$ the connection
gauge field, structure constants and field strength arise completely
independently. The biconformal gauging has the additional fields required
for this further symmetry.

While a fiber bundle gives us a foliation, the converse is not always
the case. The central requirment to have a principal fiber bundle
is the existence of a projection from the bundle to the base manifold.
We establish this by showing a second involution. Separating the co-solder
equation into independent parts,
\begin{eqnarray*}
\mathbf{d}_{\left(x\right)}\mathbf{c}_{a} & = & \boldsymbol{\alpha}_{\;a}^{b}\wedge\mathbf{c}_{b}+\frac{1}{2}S_{acd}\mathbf{e}^{c}\wedge\mathbf{e}^{d}+S_{a\quad d}^{\;\;c}\mathbf{c}_{c}\wedge\mathbf{e}^{d}-\frac{1}{2}c_{a}^{\;\;cd}\mathbf{c}_{c}\wedge\mathbf{c}_{d}\\
\mathbf{d}_{\left(y\right)}\mathbf{c}_{a}+\mathbf{d}_{\left(x\right)}\mathbf{h}_{a} & = & \boldsymbol{\alpha}_{\;a}^{b}\wedge\mathbf{h}_{b}+S_{a\quad d}^{\;\;c}\mathbf{h}_{c}\wedge\mathbf{e}^{d}-c_{a}^{\;\;cd}\mathbf{h}_{c}\wedge\mathbf{c}_{d}\\
\mathbf{d}_{\left(y\right)}\mathbf{h}_{a} & = & -\frac{1}{2}c_{a}^{\;\;cd}\mathbf{h}_{c}\wedge\mathbf{h}_{d}
\end{eqnarray*}
we observe that the exterior $y_{\alpha}$-derivative of $\mathbf{c}_{a}$
must be linear in $\mathbf{d}y_{\alpha}$, and so linear in $\mathbf{h}_{a}$.
We may therefore write $\mathbf{d}_{\left(y\right)}\mathbf{c}_{a}=C_{a}^{\;\;\;b}\mathbf{h}_{b}\wedge\mathbf{e}^{a}$
and solve for $\mathbf{d}\mathbf{h}_{a}$ on the full biconformal
space,
\begin{eqnarray*}
\mathbf{d}\mathbf{h}_{a}=\mathbf{d}_{\left(x\right)}\mathbf{h}_{a}+\mathbf{d}_{\left(y\right)}\mathbf{h}_{a} & = & \boldsymbol{\alpha}_{\;a}^{b}\wedge\mathbf{h}_{b}+S_{a\quad d}^{\;\;c}\mathbf{h}_{c}\wedge\mathbf{e}^{d}-c_{a}^{\;\;cd}\mathbf{h}_{c}\wedge\mathbf{c}_{d}-C_{a}^{\;\;\;b}\mathbf{h}_{b}\wedge\mathbf{e}^{a}-\frac{1}{2}c_{a}^{\;\;cd}\mathbf{h}_{c}\wedge\mathbf{h}_{d}
\end{eqnarray*}
This shows that $\mathbf{h}_{a}$ is in involution. Setting $\mathbf{h}_{a}=0$
constitutes a projection to an $n$-dimensional submanifold spanned
by $\mathbf{e}^{a}$.

\subsection{Gauging $\mathcal{G}$}

We compare our usual gauging of $\mathcal{G}$ with the structures
already present in the biconformal geometry.

Our usual gauging of a symmetry is to take the Cartan generalization
of the Maurer-Cartan equation for $\mathcal{G}$, eq.(\ref{Maurer-Cartan equation for G-1}).
For this we replace the Maurer-Cartan connection $\mathbf{h}_{a}$
with a general connection, leading to the introduction of a field
strength. Taking $\mathbf{A}_{a}$ to be the generalization of $\mathbf{h}_{a}$,
the Maurer-Cartan equation becomes the Cartan equation,
\begin{equation}
\mathbf{d}\mathbf{A}_{a}=-\frac{1}{2}c_{a}^{\;\;cd}\mathbf{A}_{c}\wedge\mathbf{A}_{d}+\mathbf{F}_{a}\label{Cartan equation for internal group}
\end{equation}
where the field strength $\mathbf{F}_{a}$ is required to be horizontal
and the equation integrable. Horizontality demands
\begin{eqnarray}
\mathbf{F}_{a} & = & \frac{1}{2}F_{acd}\mathbf{e}^{c}\wedge\mathbf{e}^{d}\label{Horizontal field strength}
\end{eqnarray}
while integrability requires
\begin{eqnarray*}
0 & \equiv & \mathbf{d}^{2}\mathbf{A}_{a}\\
 & = & c_{a}^{\;\;cd}\mathbf{A}_{c}\wedge\mathbf{d}\mathbf{A}_{d}+\mathbf{d}\mathbf{F}_{a}\\
 & = & -\frac{1}{2}c_{d}^{\;\;[ef}c_{a}^{\;\;c]d}\mathbf{A}_{c}\wedge\mathbf{A}_{e}\wedge\mathbf{A}_{f}+c_{a}^{\;\;cd}\mathbf{A}_{c}\wedge\mathbf{F}_{d}+\mathbf{d}\mathbf{F}_{a}\\
 & = & \mathbf{D}_{\left(\mathcal{G}\right)}\mathbf{F}_{a}
\end{eqnarray*}
since $c_{d}^{\;\;[ef}c_{a}^{\;\;c]d}=0$ by the Jacobi identity for
$\mathcal{G}$.

Within the biconformal solution, we interpret the co-solder forms
$\mathbf{f}_{a}$ as these generalized potentials $\mathbf{A}_{a}$
for the $\mathcal{G}$-connection. The full structure equation for
$\mathbf{f}_{a}$, however, is not exactly what we expect for a typical
gauging. Instead we have
\[
\mathbf{d}\mathbf{f}_{a}=\boldsymbol{\omega}_{\;a}^{b}\wedge\mathbf{f}_{b}+\mathbf{f}_{a}\wedge\boldsymbol{\omega}+\frac{1}{2}S_{acd}\mathbf{e}^{c}\wedge\mathbf{e}^{d}+S_{a\quad d}^{\;\;c}\mathbf{f}_{c}\wedge\mathbf{e}^{d}-\frac{1}{2}c_{a}^{\;\;cd}\mathbf{f}_{c}\wedge\mathbf{f}_{d}
\]
The situation appears to be similar to what Kibble encountered in
writing general relativity as a gauge theory of the Poincaré group
\cite{Kibble1}. Kibble introduced Poincaré fibers over spacetime,
then ``soldered'' the translational gauge fields of the fiber symmetry
to the cotangent basis of the bundle. This identification avoided
double counting the translations. With the quotient method, such identification
is no longer needed since the quotient of the Poincaré group by the
Lorentz group automatically changes the translational symmetry into
the base manifold.

Here, we already have an $SO\left(p,q\right)$ symmetry on the fibers
and find that the base manifold has a similar, but restricted symmetry.
It is this latter, emergent symmetry we would like to use. The present
situation differs from the Poincaré case since it is the connection
and not the frame field that is doubled, with $\boldsymbol{\omega}_{\;a}^{d}\wedge\mathbf{f}_{d}$
and $\left(-\frac{1}{2}c_{a}^{\;\;cd}\mathbf{f}_{c}\right)\wedge\mathbf{f}_{d}$
both acting on the same index of $\mathbf{f}_{d}$. We cannot simply
solder the two together because they produce covariance with respect
to different symmetries. Moreover, we still need the original symmetry
to act on the remaining gauge fields in the usual way. There are a
number of possible resolutions to the difficulty. First, it might
be possible to build the gauging of the co-solder form into the original
quotient. However, though closer examination may reveal a natural
way to do this, it would seem to lose the appealing feature of an
emergent internal symmetry. A second approach is suggested by \cite{Hazboun Wheeler,Hazboun dissertation,Hazboun Wheeler-1},
in which an initially $SO\left(n\right)$ connection is written as
a Lorentz connection plus additional terms, which then introduce physical
fields. If a similar technique is applied here, perhaps along with
the transformation of \cite{Hazboun Wheeler}, it is possible that
the restriction of the connection can occur directly.

A third approach is to keep both connections but keep careful track
of which fields transform under which symmetry. This actually causes
no problem for additional fields we might wish to introduce, since
these will enter as representations of either the $SO\left(p,q\right)$
transformations, the $\mathcal{G}$ transformations, or both, leaving
no ambiguity about their transformations. The only potential conflict
arises with $\mathbf{f}_{a}\Leftrightarrow\mathbf{f}_{A}$ itself.
Thinking of $\mathcal{G}$ as a subgroup of $SO\left(p,q\right)$,
we must wonder whether a full $SO\left(p,q\right)$ transformation
would introduce sensible but distinct copies of the gauge potential.
In the case of electroweak symmetry, we could construct a theory with
$SO\left(4\right)$ breaking to $SU\left(2\right)\times U\left(1\right)$
on the fibers while the full $SO\left(4\right)$ is written following
\cite{Hazboun Wheeler}, as a Lorentz connection plus additional scalar
field and cosmological constant. It is not clear whether the resulting
weak fields would violate the causal structure of the Lorentz sector.

However, these conjectures will require\textendash and are the subject
of\textendash further study.

For the present we content ourselves with the following. First, we
satisfy the final field equation for the cross-term of the co-torsion,
\[
\Delta_{sb}^{ar}S_{c\quad a}^{\;\;\;b}=\Delta_{sc}^{ar}S_{a\quad b}^{\;\;b}
\]
by the sufficient condition $S_{c\quad a}^{\;\;\;b}=0$. The only
remaining field equation is $S_{a}^{\;\;\;ba}=-c_{a}^{\;\;\;ba}=0$,
the tracelessness of the structure constants. This leads to unit determinant
for elements of $\mathcal{G}$.

Having solved the field equations, we identify $\mathbf{f}_{a}$ as
the gauge field $\mathbf{A}_{a}$, we modify the indices to make it
clear which gauge group applies. For fields that transform under $SO\left(p,q\right)$,
we retain the lower case Latin indices, while for fields that transform
under $\mathcal{G}$ we replace the relevant indices by upper case
Latin. Both sets range from $1$ to $n$. The connection becomes a
pair,
\begin{eqnarray*}
\boldsymbol{\alpha}_{\;\;\;b}^{a} & \Rightarrow & \left(\begin{array}{cc}
\boldsymbol{\alpha}_{\;\;\;b}^{a} & 0\\
0 & 0
\end{array}\right)\\
\frac{1}{2}c_{a}^{\;\;\;cd}\mathbf{f}_{c} & \Rightarrow & \left(\begin{array}{cc}
0 & 0\\
0 & -\frac{1}{2}c_{B}^{\;\;\;CA}\mathbf{f}_{C}
\end{array}\right)
\end{eqnarray*}
so in the new notation, $\boldsymbol{\alpha}_{\;\;\;B}^{A}=0$ and
$\frac{1}{2}c_{a}^{\;\;cd}\mathbf{f}_{c}=0$.

The curvature and solder form equations are unchanged, but the co-torsion
equation becomes

\begin{eqnarray*}
\mathbf{d}\mathbf{f}_{A} & = & \boldsymbol{\alpha}_{\;A}^{B}\wedge\mathbf{f}_{B}+\frac{1}{2}S_{Acd}\mathbf{e}^{c}\wedge\mathbf{e}^{d}-\frac{1}{2}c_{B}^{\;\;\;CA}\mathbf{f}_{C}\wedge\mathbf{f}_{D}
\end{eqnarray*}
or, since $\boldsymbol{\alpha}_{\;A}^{B}=0$,
\begin{equation}
\frac{1}{2}S_{Acd}\mathbf{e}^{c}\wedge\mathbf{e}^{d}=\mathbf{d}\mathbf{f}_{A}+\frac{1}{2}c_{B}^{\;\;\;CA}\mathbf{f}_{C}\wedge\mathbf{f}_{D}\label{Co-torsion equation as Yang-Mills}
\end{equation}
This exactly reproduces the form of a Yang-Mills field, Eqs.(\ref{Cartan equation for internal group})
and (\ref{Horizontal field strength}).

Consistency of this restriction is now immediate because in the full
set of structure equations,
\begin{eqnarray*}
\mathbf{d}\boldsymbol{\alpha}_{\;b}^{a} & = & \boldsymbol{\alpha}_{\;b}^{c}\wedge\boldsymbol{\alpha}_{\;c}^{a}+\frac{1}{2}R_{\;\;bcd}^{a}\mathbf{e}^{c}\wedge\mathbf{e}^{d}\\
\mathbf{d}\mathbf{e}^{a} & = & \mathbf{e}^{b}\wedge\boldsymbol{\alpha}_{\;\;b}^{a}\\
\mathbf{d}\mathbf{f}_{A}+\frac{1}{2}c_{A}^{\;\;CD}\mathbf{f}_{C}\wedge\mathbf{f}_{D} & = & \frac{1}{2}S_{Acd}\mathbf{e}^{c}\wedge\mathbf{e}^{d}
\end{eqnarray*}
the internal symmetry $\mathcal{G}$ has fully decoupled from the
spactime geometry.

\subsubsection{Metric on the submanifolds in the nonabelian case}

The nonabelian case still allows us to simply choose the spacetime
metric as is done in general relativity and was used in the generic
case above. For the group submanifold, it is natural to choose the
Killing form of $\mathcal{G}$ if it is nondegenerate. While this
assignment is certainly possible for semisimple $\mathcal{G}$, the
details depend on the particular group and will be discussed elsewhere.
As with the generic case, the considerations of \cite{WheelerQMComplex,Spencer Wheeler,Hazboun Wheeler,Hazboun Wheeler-1,LoveladyWheeler}
may be relevant.

\subsection{Remaining issues}

While we have arrived at a satisfactory separation of a new internal
symmetry, we still lack both the field equation for $\mathbf{S}_{A}$
and the contribution of this additional field to the Einstein equation.
We consider three possible resolutions:
\begin{enumerate}
\item The absence of a source for gravity arising from $\mathcal{G}$ is
not surprising given the restriction of the action to linear curvature
terms. Including up to quadratic curvature terms in the original action
could provide both field equation and gravitational coupling.
\item Allow nonvanishing cross-term to the torsion. This preserves the involution
of the solder form and avoids spacetime torsion. The cross-term gives
extrinsic curvature to the embedding of the submanifolds into the
full biconformal space, and allows curvature of the group manifold
as in \cite{Milnor}. The cross-term of the torsion, driven by the
internal symmetry, then enters the spacetime curvature quadratically
and might supply the required gravitational source.
\item The gravitational instanton has been shown to introduce both field
equations and gravitational source in the usual form \cite{EguchiHanson,OhParkYang}.
\end{enumerate}
Leaving these considerations to further work, we make only the following
observation. It would be natural to introduce a quadratic term such
as $\int\mathbf{T}^{a}\wedge\,^{*}\mathbf{S}_{a}$ into the action.
With vanishing torsion, only the variation $\delta\mathbf{T}^{a}=\mathbf{D}\delta\mathbf{e}^{a}$
yields nonvanishing contributions
\begin{eqnarray*}
\delta_{e}\int\mathbf{T}^{a}\wedge\,^{*}\mathbf{S}_{a} & = & \int\delta\mathbf{e}^{a}\wedge\mathbf{D}{}^{*}\mathbf{S}_{a}+surface\:term+terms\:proportional\:to\:torsion
\end{eqnarray*}
This term leads to a divergence of $\mathbf{S}_{a}$ in the curvature
equation, but it because of the presence of $\mathbf{f}_{a}$ in the
volume form also gives terms quadratic in the components of $\mathbf{S}_{a}$.
However, with vanishing co-torsion cross-term the quadratic terms
added to the spacetime curvature field equations involve only products
of the spacetime and the momentum components,
\begin{eqnarray*}
\delta_{e2}S & \sim & \frac{n\left(n-2\right)}{12}\left(S_{acd}S_{b}^{\;\;\;cd}+S_{bcd}S_{a}^{\;\;\;cd}+S_{abc}S_{d}^{\;\;\;cd}+S_{abc}S_{d}^{\;\;\;cd}\right)\\
 & = & -\frac{n\left(n-2\right)}{12}\left(S_{acd}c_{b}^{\;\;\;cd}+S_{bcd}c_{a}^{\;\;\;cd}\right)
\end{eqnarray*}
and these are not in the form of energy momentum tensor. This suggests
that a combination of this quadratic term and nonvanishing cross-term
for torsion may provide a solution.

\section{Conclusions \label{sec:Conclusions}}

We have shown how general relativity emerges from the torsion-free
solutions to biconformal gravity. The derivations involve field equation
driven dimensional reduction and may therefore have relevance to dimensional
reduction of twistor string, or the reduction of Drinfeld doubles. 

\subsection{Results in biconformal gravity}

We began with the conformal group $\mathcal{C}_{p,q}$ of a compactified
space with $\left(p,q\right)$-signature metric in $n=p+q$ dimensions.
The quotient of $\mathcal{C}_{p,q}$ by its homogeneous Weyl subgroup
$\mathcal{W}$ gives a $2n$-dimensional Kahler manifold with local
$SO\left(p,q\right)$ and scale invariance. Generalizing this local
structure leads to a curved geometry characterized by $SO\left(p,q\right)$
curvature, torsion, co-torsion and dilatational curvature corresponding
to the generators of the conformal group. Throughout, $SO\left(p,q\right)$
may be replaced by $Spin\left(p,q\right)$ when a spinor representation
is desired. This biconformal space admits a scale invariant action
functional linear in the Cartan curvatures, Eq.(\ref{Action}). Varying
the action yields a gravity theory in $2n$ dimensions. All models
with $\left(n-2\right)\alpha-2\beta$ nonzero are considered. The
special case when $\left(\left(n-2\right)\alpha-2\beta\right)=0$
may include non-integrable Weyl geometries and, likely being unphysical,
these are not considered in detail.

We established the following distinct results for this model.
\begin{enumerate}
\item \textbf{Triviality with vanishing torsion and co-torsion.} If, in
addition to vanishing torsion, the co-torsion (the field strength
of special conformal transformations) is taken to vanish, the biconformal
space takes a trivial form, with the only nonvanishing components
of the $SO\left(p,q\right)$ and dilatational curvatures being constant
cross-terms.
\item \textbf{Foliation by a Lie subgroup}. We prove that half of the biconformal
space is foliated by an $n$-dimensional Lie group $\mathcal{G}$
with structure constants lying in a representation of $SO\left(p,q\right)$.
This result follows from the involution guaranteed by vanishing torsion,
together with the field equations. When $\alpha,\beta$ and $\gamma$
are chosen such that $\chi=\frac{1}{n-1}\left(1+\frac{n^{2}\gamma}{\left(n-1\right)\alpha-\beta}\right)=-1$,
$\mathcal{G}$ may be non-abelian, otherwise it is abelian. 
\item \textbf{Generic solution.} The generic solution assumes only that
$1+\chi$ is nonzero, together with vanishing torsion, $\mathbf{T^{a}}=0$.
The field equations reduce the only nontrivial dependence of all the
remaining curvatures \textendash{} the SO$\left(p,q\right)$ curvature,
co-torsion, and dilatation \textendash{} from $2n$ independent variables
$\left(x^{\alpha},y_{\beta}\right)$ down to $n$ variables $\left(x^{\alpha}\right)$
and reduce the number and form of the independent components. Explicitly,
the form and dependency of each curvature $\boldsymbol{\Omega}^{A}\in\left\{ \boldsymbol{\Omega}_{\quad b}^{a},\mathbf{S}_{a},\boldsymbol{\Omega}\right\} $
begins as three distinct tensors dependent on $2n$ coordinates,
\[
\boldsymbol{\Omega}^{A}=\frac{1}{2}\Omega_{\;\;\;cd}^{A}\left(x^{\alpha},y_{\beta}\right)\,\mathbf{e}^{c}\wedge\mathbf{e}^{d}+\Omega_{\quad d}^{Ac}\left(x^{\alpha},y_{\beta}\right)\,\mathbf{f}_{c}\wedge\mathbf{e}^{d}+\frac{1}{2}\Omega^{Acd}\left(x^{\alpha},y_{\beta}\right)\,\mathbf{f}_{c}\wedge\mathbf{f}_{d}
\]
and after application of the field equations, each is reduced as follows:
\end{enumerate}
\begin{quotation}
\begin{eqnarray*}
\boldsymbol{\Omega}_{\quad b}^{a} & = & \frac{1}{2}C_{\;\;bcd}^{a}\left(x\right)\mathbf{e}^{c}\wedge\mathbf{e}^{d}+2\left(1+\chi\right)\Delta_{cb}^{ac}\,\mathbf{f}_{c}\wedge\mathbf{e}^{d}\\
\mathbf{S}_{a} & = & -\frac{1}{1+\chi}\left(D_{c}^{\left(x\right)}\mathcal{R}_{ad}\left(x\right)-\frac{1}{2}\left(1+\chi\right)y_{b}C_{\;\;acd}^{b}\left(x\right)\right)\mathbf{e}^{c}\wedge\mathbf{e}^{d}\\
\boldsymbol{\Omega} & = & \chi\,\mathbf{e}^{c}\wedge\mathbf{f}_{c}
\end{eqnarray*}
where $\chi=\frac{1}{n-1}\left(1+\frac{n^{2}\gamma}{\left(n-1\right)\alpha-\beta}\right)$.
The resultant curvatures, $\frac{1}{2}C_{\;\;acd}^{b}\left(x\right)\mathbf{e}^{c}\wedge\mathbf{e}^{d}$
and $\mathcal{R}_{ab}\left(x\right)$ are the Weyl and Schouten parts
of the Riemann curvature tensor computed from the connection compatible
with the $n$-dimensional solder form, $\mathbf{e}^{a}\left(x\right)$.
Each of the $2$-forms 
\begin{eqnarray*}
\mathbf{d}\boldsymbol{\omega}\, & = & \left(1+\chi\right)\mathbf{e}^{c}\wedge\mathbf{f}_{c}\\
\boldsymbol{\Omega} & = & \chi\mathbf{e}^{a}\wedge\mathbf{h}_{a}\;\;=\;\;\chi\mathbf{d}x^{\alpha}\wedge\mathbf{d}y_{\alpha}
\end{eqnarray*}
 is closed and non-degenerate, hence symplectic on the full biconformal
space.

The basis forms $\mathbf{e}^{a},\mathbf{f}_{b}$ are given by
\begin{eqnarray*}
\mathbf{e}^{a} & = & e_{\alpha}^{\;\;a}\left(x\right)\mathbf{d}x^{\alpha}\\
\mathbf{f}_{a} & = & e_{a}^{\;\;\alpha}\mathbf{d}y_{\alpha}-\left(\frac{1}{1+\chi}\mathcal{R}_{ab}-D_{b}^{\left(\alpha,x\right)}y_{a}+\left(1+\chi\right)\left(y_{a}y_{b}-\frac{1}{2}\eta_{ab}\eta^{cd}y_{c}y_{d}\right)\right)\mathbf{e}^{b}\\
 & = & \mathbf{h}_{a}-\frac{1}{1+\chi}\boldsymbol{\mathscr{R}}_{a}
\end{eqnarray*}
and the manifest involution of $\mathbf{h}_{a}=e_{a}^{\;\;\alpha}\mathbf{d}y_{\alpha}$
shows that setting $y_{\alpha}=constant$ gives a Lagrangian submanifold
for spacetime. Conversely, setting $x^{\alpha}=constant$ gives conjugate
Lagrangian submanifolds, each the leaf of a foliation by flat manifolds.
The entire $2n$-dimensional biconformal spaces is therefore interpreted
as the cotangent bundle of spacetime. The spin connection and Weyl
vector are given by
\begin{eqnarray*}
\boldsymbol{\omega}_{\;b}^{a} & = & \boldsymbol{\alpha}_{\;b}^{a}\left(x\right)+2\left(1+\chi\right)\Delta_{db}^{ac}y_{c}\mathbf{e}^{d}\\
W_{\alpha} & = & -\left(1+\chi\right)y_{\alpha}
\end{eqnarray*}
where $\boldsymbol{\alpha}_{\;b}^{a}\left(x\right)$ is the metric
compatible connection, $\mathbf{d}\mathbf{e}^{a}=\mathbf{e}^{b}\wedge\boldsymbol{\alpha}_{\;b}^{a}\left(x\right)$.
Here the orthonormal frame field and spin connection pair $\left(\mathbf{e}^{a},\boldsymbol{\alpha}_{\;b}^{a}\right)$
is equivalent to the metric and Christoffel connection pair, $\left(g_{\mu\nu},\Gamma_{\;\;\mu\nu}^{\alpha}\right)$.

The sole remaining constraint on the system is the locally scale covariant
Einstein equation with source $c_{ab}$, Eq.(\ref{Scale covariant condition for Einstein equation}):
\[
0=\mathbf{D}_{\left(\alpha,x\right)}\left(\boldsymbol{\mathcal{R}}_{a}+\left(1+\chi\right)\mathbf{c}_{a}\right)-W_{b}\mathbf{R}_{\;\;a}^{b}-W_{b}2\Delta_{da}^{bc}\left(\boldsymbol{\mathcal{R}}_{c}+\left(1+\chi\right)\mathbf{c}_{c}\right)\wedge\mathbf{e}^{d}
\]
which has been shown in \cite{WeylGeom} to be the condition for the
existence of a conformal transformation to the sourced Einstein equation
when the source $\mathbf{c}_{a}=c_{ab}\mathbf{e}^{b}$ is written
as
\[
c_{ab}=-\frac{1}{n-2}\frac{1}{1+\chi}\left(T_{ab}-\frac{1}{n-1}\eta_{ab}T\right)
\]
We showed from the properties of $c_{ab}$ that $T_{ab}$ is symmetric
and divergence free, and in the Riemannian gauge (in which the Weyl
vector vanishes),
\[
R_{ab}-\frac{1}{2}\eta_{ab}R=T_{ab}
\]

We conclude that the generic case describes the locally scale covariant
$n$-dimensional Einstein equation sourced by a symmetric, divergence
free tensor and formulated on the co-tangent bundle of spacetime.
The reduction to $n$-dimensions is accomplished using only the field
equations with vanishing torsion.\smallskip{}

\hspace{-1cm}4. \textbf{The non-abelian cases.} When $\alpha,\beta$
and $\gamma$ are chosen such that $\chi=\frac{1}{n-1}\left(1+\frac{n^{2}\gamma}{\left(n-1\right)\alpha-\beta}\right)=-1$,
$\mathcal{G}$ may be non-abelian, and there are substantial differences
in the Cartan structure equations. For these cases, we again showed
the reduction to dependence on the spacetime solder form, $\mathbf{e}^{a}\left(x\right)$
, but now the final forms of the curvature and dilatation are
\begin{eqnarray*}
\boldsymbol{\Omega}_{\;\;b}^{a} & = & \frac{1}{2}C_{\;\;bcd}^{a}\left(\boldsymbol{\alpha}\right)\mathbf{e}^{c}\land\mathbf{e}^{d}+2\Delta_{db}^{ac}\mathbf{f}_{c}\land\mathbf{e}^{d}\\
\boldsymbol{\Omega} & = & \mathbf{e}^{a}\land\mathbf{f}_{a}
\end{eqnarray*}
The curvature is subject to the scale-covariant vacuum Einstein equation,
\[
\mathcal{R}_{ec}+D_{c}^{\left(x\right)}\phi_{e}+\phi_{e}\phi_{c}-\frac{1}{2}\eta_{ec}\phi^{d}\phi_{d}=0
\]
For the co-torsion, Lorentz transformations are suppressed while the
appearance of the $\mathcal{G}$-connection is automatic. This leads
to the co-solder form becoming the $\mathcal{G}$ gauge field and
the spacetime co-torsion becoming the usual Yang-Mills field strength,
\begin{eqnarray*}
\frac{1}{2}S_{Acd}\mathbf{e}^{c}\land\mathbf{e}^{d} & = & \mathbf{D}_{\left(W\right)}\mathbf{f}_{A}+\frac{1}{2}c_{A}^{\;\;CD}\mathbf{f}_{C}\land\mathbf{f}_{D}
\end{eqnarray*}
where $\mathbf{D}_{\left(W\right)}\mathbf{f}_{A}=\mathbf{d}\mathbf{f}_{A}-\mathbf{f}_{A}\wedge\boldsymbol{\omega}$
is the dilatation-covariant derivative. The biconformal space becomes
a principal $\mathcal{G}$-bundle over spacetime.

Effectively, the total principal bundle has homogeneous Weyl and $\mathcal{G}$
symmetry, $\mathcal{W}\times\mathcal{G}$. This is \emph{not} what
occurs if the original quotient is of the conformal group by inhomogeneous
Weyl, $C_{\left(p,q\right)}/\mathcal{IW}$, which essentially gives
Poincaré fibers over spacetime \cite{Auxiliary,Weyl grav as GR}.
The emergence of non-abelian symmetry from the degrees of freedom
of the special conformal gauge fields of the conformal group is a
new result. In $4$-dimensions, with $SO\left(p,q\right)=SO\left(4\right)$,
the maximal $\mathcal{G}$ is the electroweak symmetry with necessary
parity violation.
\end{quotation}

\subsection{A note on the metric and signature change}

To formulate general relativity as a gauge theory using the Cartan
techniques, we take the quotient of the Poincaré group by the Lorentz
group. The only guidance as to the metric is the presence of the Lorentzian
connection, which will leave the usual orthonormal metric, $\eta_{ab}=diag\left(-1,1,1,1\right)$
invariant. We then introduce the metric by hand using the orthonormal
frame field, $g_{\alpha\beta}=e_{\alpha}^{\;\;\;a}e_{\beta}^{\;\;\;b}\eta_{ab}$.
The situation is different in biconformal space, where there are natural
metric structures present. Of course, there is the $SO\left(p,q\right)$
connection, making it possible to introduce a $\left(p,q\right)$
signature metric by hand, just as we do in general relativity. However,
the Killing form of the conformal group has a non-degenerate restriction
to biconformal space, and we may use this instead. The resulting metric
on spacetime depends not only on the original signature, but also
on what submanifold is taken as spacetime.

Using the restricted Killing form, the metric is
\[
K_{AB}=\left(\begin{array}{cc}
0 & \delta_{b}^{a}\\
\delta_{a}^{b} & 0
\end{array}\right)
\]
and the restriction to the $\mathbf{e}^{a}$ or $\mathbf{f}_{b}$
subspaces has vanishing restriction. It is shown in \cite{Spencer Wheeler},
however, that if we seek orthogonal Lagrangian submanifolds on which
the Killing metric is non-degenerate, there are limited possibilities,
with initial spaces of signature $\left(n,0\right),\left(\frac{n}{2},\frac{n}{2}\right)$
or $\left(0,n\right)$ being the only consistent starting points,
and the two Euclidean cases leading uniquely to Lorentzian signature,
$\left(n-1,1\right)$ or $\left(1,n-1\right)$. This development of
a time direction from an initially Euclidean space is appealing.

There are other possibilities. If we drop the orthogonality requirement
from the theorem of \cite{Spencer Wheeler}, it becomes possible to
have different signature on the two Lagrangian submanifolds. This
too has its advantages, as we might arrange for Lorentzian signature
on one submanifold and Euclidean on the other, enabling an additional
compact internal symmetry. 

Some of these avenues have been explored. The Euclidean starting point
leading to Lorentzian signature on both Lagrangian submanifolds is
studied in \cite{Hazboun Wheeler}, in which all the results are seen
to depend only on structures inherent in the conformal group. In \cite{LoveladyWheeler}
connections of both types are introduced, and some possibilities are
explored in \cite{Hazboun}. Work is currently underway to examine
a $4+4$ dimensional model with mixed signatures, to take advantage
of the potential graviweak theory.

There is still another metric possibility, because the metric compatible
with the Kähler structure is different, having signature $\left(2p,2q\right)$
while the the Killing metric has signature $\left(n,n\right)$.

For the present results, it seems best to simply impose the metric
we choose. If we let the original $SO\left(p,q\right)$ be Lorentzian,
$SO\left(n-1,1\right)$ then the natural choice for spacetime is Lorentzian
(but see \cite{Hazboun Wheeler}).

\subsection{Discussion}

Biconformal spaces with appropriate signature give rise to general
relativity, generically formulated on the cotangent bundle of spacetime.
In a subclass of cases there may be an additional non-abelian internal
symmetry. While this internal symmetry ultimately arises from the
special conformal transformations, no previous gauging of the conformal
group has shown the direct possibility of a non-abelian symmetry.
This opens the possibility of a graviweak unification, which, while
still requiring additional structure for the strong force, holds out
the hope of a deeper understanding of parity violation and the breaking
of a left-right symmetric $SU\left(2\right)\times SU\left(2\right)$
model. This possibility is under current investigation.

As described in the introduction, these gravity models may provide
new insights into string theory. The existence of a conformal route
to general relativity, as opposed to fourth-order Weyl gravity, allows
for the consistent use of twistor string models. In addition, the
doubled dimension makes possible a compactification from $10$-dimensionsal
superstring theory to an $8$-dimensional biconformal space with an
immediate interpretation as $4$-dimensional general relativity. In
this way, the myriad $6$-dimensional compact spaces are avoided,
to be replaced by compactification of only $2$-dimensions and possibly
uniquely to a torus if other structures are to be maintained.

Finally, this reduction of the biconformal gauging shares many features
with Drinfeld doubles. The match between the Killing form and the
symmetric form of the Drinfeld product may suggest systematic ways
of reducing the doubles to their half-dimension. 

\section*{Acknowledgement}

The author thanks Jeffrey Shafiq Hazboun for his careful reading and
discussion of these results.

\section*{Appendix A: Compactification of spacetimes}

Consider a flat space $\mathcal{S}$ of signature $\left(p,q\right)$,
i.e., with indefinite metric
\[
\eta_{ab}=diag\left(\underbrace{1,\ldots,1}_{p},\underbrace{-1,\ldots,-1}_{q}\right)
\]
This space is noncompact due to curves heading off to infinite distance.
We complete the space by appending an inverse to every vector from
the origin.

Let $N$ be the set of null vectors from the origin, $N=\left\{ x^{\alpha}|x^{\beta}x_{\beta}=0,x^{\alpha}\in\mathcal{V}\right\} $,
and $\mathcal{C}N$ its complement, $\mathcal{C}N=\left\{ x^{\alpha}|x^{\beta}x_{\beta}\neq0,x^{\alpha}\in\mathcal{V}\right\} $.

For all points in $\mathcal{C}N$, we consider the set,
\[
\mathcal{W}_{0}=\left\{ \left.w^{\alpha}\right|w^{\alpha}=-\frac{x^{\alpha}}{\left|x\right|^{2}}\right\} 
\]
where $\left|x\right|^{2}=\eta_{\alpha\beta}x^{\alpha}x^{\beta}$.
Notice that $\mathcal{W}_{0}$ contains no null vectors since this
would require $\left|x\right|^{2}=0$. Inverting the transformation
we have
\begin{eqnarray*}
x^{\alpha} & = & -\frac{w^{\alpha}}{\left|w\right|^{2}}
\end{eqnarray*}
so we see the mapping between $x^{\alpha}$ and $w^{\alpha}$ is a
bijection.

We extend $\mathcal{W}_{0}$ by taking the union with a new set $\hat{\mathcal{W}}$
of points $\hat{w}^{\alpha}$ satisfying
\[
\hat{\mathcal{W}}=\left\{ \left.w^{\alpha}\right|\eta_{\alpha\beta}\hat{w}^{\alpha}\hat{w}^{\beta}=0\right\} 
\]
Clearly $\hat{\mathcal{W}}\cap\mathcal{W}_{0}=\phi$. We suggest that
\[
\mathcal{W}\equiv\hat{\mathcal{W}}\cup\mathcal{W}_{0}
\]
provides a compactification of the space.

If the signature is Euclidean, $\left(p,q\right)=\left(n,0\right)$,
then $N$ consists of the origin alone, and $\hat{\mathcal{W}}$ is
the 1-point compactification of $R^{n}$.

More generally, a detailed proof of compactness must rely on the specification
of a topology on indefinite spaces. For general spacetimes this relies
on introducing ideal points \cite{GerochKronheimerPenrose,HawkingEllis,Wald}.
While such methods should meet no obstruction in the flat, nonsingular
spaces considered here, the definition of the conformal group requires
only the existence of inverses. This much has already been accomplished
with the definition of $\mathcal{W}$. We therefore content ourselves
by defining a suitable extension and indicating compactness by studying
the resulting extensions of spacetime curves.

First, consider a straight line that runs off to infinity in spacetime
at half the speed of light,
\[
\mathcal{C}\left(\lambda\right)=\left\{ x^{\alpha}=\left(\lambda,\frac{\lambda}{2},0,0\right)\right\} 
\]
This curve starts at the origin and reaches no endpoint in the original
space, but in the compactification we may continue the curve to some
finite value of $\lambda=\lambda_{0}$ then change to $w^{\alpha}$
coordinates,
\begin{eqnarray*}
\mathcal{C}\left(\lambda_{0}\right) & = & \left\{ x^{\alpha}=\left(\lambda_{0},\frac{\lambda_{0}}{2},0,0\right)\right\} =\left\{ w^{\alpha}=\frac{4}{3}\left(\frac{1}{\lambda_{0}},\frac{1}{2\lambda_{0}},0,0\right)\right\} 
\end{eqnarray*}
Define a new parameter, $\xi=\frac{1}{\lambda}$ and continue the
curve
\begin{eqnarray*}
\mathcal{C}\left(\xi\right) & = & \left\{ w^{\alpha}=\frac{4}{3}\left(\xi,\frac{1}{2}\xi,0,0\right)\right\} 
\end{eqnarray*}
We continue the curve through $w^{\alpha}\left(0\right)$ and continue
$\xi$ to negative values. At some finite value, $\xi=-\xi_{0}<0$,
we are at the point
\begin{eqnarray*}
w^{\alpha} & = & -\frac{4}{3}\left(\xi_{0},\frac{1}{2}\xi_{0},0,0\right)
\end{eqnarray*}
corresponding to
\begin{eqnarray*}
x^{\alpha} & = & -\left(\frac{1}{\left(\frac{16}{9}\right)\left(-\xi_{0}^{2}+\frac{1}{4}\xi_{0}^{2}\right)}\right)-\frac{4}{3}\left(\xi_{0},\frac{1}{2}\xi_{0},0,0\right)\\
 & = & -\frac{4}{3}\left(\frac{1}{\frac{4}{3}\xi_{0}^{2}}\right)\left(\xi_{0},\frac{1}{2}\xi_{0},0,0\right)\\
 & = & -\left(\frac{1}{\xi_{0}},\frac{1}{2}\frac{1}{\xi_{0}},0,0\right)
\end{eqnarray*}
This is a point in the past light cone. Letting $\lambda'=\frac{1}{\xi}$,
we continue the path back to the origin as
\begin{eqnarray*}
x^{\alpha} & = & -\left(\lambda',\frac{1}{2}\lambda',0,0\right)
\end{eqnarray*}
and the curve is closed.

As an example of a curve reaching null infinity (hence some nonzero
null vector $\left|w\right|^{2}=0$), consider the accelerating curve
\[
\mathcal{C}\left(\lambda\right)=\left\{ x^{\alpha}=\left(a\cosh\lambda,a\sinh\lambda,0,0\right)\right\} 
\]
with timelike norm
\[
\left|x\right|^{2}=-a^{2}
\]
For this we choose a new parameter,
\[
\sigma=\tan^{-1}\left(\sinh\lambda\right)
\]
which approaches $\frac{\pi}{2}$ as $\lambda\rightarrow\infty$.
The curve takes the form
\begin{eqnarray*}
x^{\alpha} & = & a\left(\sqrt{1+\tan^{2}\sigma},\tan\sigma,0,0\right)
\end{eqnarray*}
At a value $\sigma=\frac{\pi}{2}$ we now have a null vector. We need
to reparameterize the curve before this happens, so set
\begin{eqnarray*}
w^{\alpha} & = & \frac{a}{\tan\sigma}\left(\sqrt{1+\tan^{2}\sigma},\tan\sigma,0,0\right)
\end{eqnarray*}
Now as $\sigma$ approaches $\frac{\pi}{2}$, $w^{\alpha}$ is given
by
\begin{eqnarray*}
w^{\alpha}\left(\sigma\right) & = & a\left(\sqrt{\frac{1}{\tan^{2}\sigma}+1},1,0,0\right)\\
 & \rightarrow & a\left(1,1,0,0\right)
\end{eqnarray*}
Let the parameterization of the curve in $w^{\alpha}$ coordinates
be $\xi=\frac{1}{\tan\sigma}$ so that
\[
w^{\alpha}\left(\xi\right)=a\left(\sqrt{1+\xi^{2}},1,0,0\right)
\]
As $\xi\rightarrow0$ this approaches a null vector, $w^{\alpha}=\left(a,a,0,0\right)$,
so we may now identify the limit $\sigma\rightarrow\infty$ with the
null vector $w^{\alpha}=\left(a,a,0,0\right)$. This vector is in
the space, and we may again continue the curve to negative $\xi$.

\section*{Appendix B: Variation of the spin connection}

Here we give details of the variation of the spin connection, since
some of the steps are novel. Because many of the expressions are long,
we introduce some notational conventions to make expressions more
compact and transparent. Specifically, since all differential forms
are rendered in boldface, there is no loss of information if we assume
wedge products between all adjacent forms, dropping the explicit wedge.
We further define a multi-index form,
\[
\boldsymbol{\omega}_{c\cdots d}\equiv\boldsymbol{\omega}_{c}\boldsymbol{\omega}_{c_{1}}\ldots\boldsymbol{\omega}_{d}\equiv\boldsymbol{\omega}_{c}\wedge\boldsymbol{\omega}_{c_{1}}\wedge\ldots\wedge\boldsymbol{\omega}_{d}
\]
for any number of basis 1-forms $\boldsymbol{\omega}_{c}$. It is
always possible to deduce the correct number of indices from the Levi-Civita
tensor.

The spin connection occurs only in the $SO\left(p,q\right)$ curvature,
$\boldsymbol{\Omega}_{\;\;b}^{a}$, so the spin connection variation
affects only the $\alpha$ term of the action,
\begin{eqnarray*}
\delta_{\omega_{b}^{a}}S & = & \delta_{\omega_{b}^{a}}\intop\left(\alpha\boldsymbol{\Omega}_{\;\;b}^{a}+\beta\delta_{\;\;b}^{a}\boldsymbol{\Omega}+\gamma\mathbf{e}^{a}\mathbf{f}_{b}\right)\mathbf{e}^{e\cdots f}\mathbf{f}_{c\cdots d}e_{\qquad ae\cdots f}^{bc\cdots d}\\
 & = & \alpha\intop\delta_{\omega_{b}^{a}}\left(\mathbf{d}\boldsymbol{\omega}_{\;b}^{a}-\boldsymbol{\omega}_{\;b}^{c}\boldsymbol{\omega}_{\;c}^{a}-2\Delta_{cb}^{ad}\mathbf{f}_{d}\mathbf{e}^{c}\right)\mathbf{e}^{e\cdots f}\mathbf{f}_{c\cdots d}e_{\qquad ae\cdots f}^{bc\cdots d}\\
 & = & \alpha\intop\left(\mathbf{d}\delta\boldsymbol{\omega}_{\;b}^{a}-\delta\boldsymbol{\omega}_{\;b}^{c}\boldsymbol{\omega}_{\;c}^{a}-\boldsymbol{\omega}_{\;b}^{c}\delta\boldsymbol{\omega}_{\;c}^{a}\right)\mathbf{e}^{e\cdots f}\mathbf{f}_{c\cdots d}e_{\qquad ae\cdots f}^{bc\cdots d}\\
 & = & \alpha\intop\mathbf{D}\left(\delta\boldsymbol{\omega}_{\;b}^{a}\right)\mathbf{e}^{e\cdots f}\mathbf{f}_{c\cdots d}e_{\qquad ae\cdots f}^{bc\cdots d}
\end{eqnarray*}
Integrating the covariant derivative by parts and discarding the surface
term,
\begin{eqnarray*}
0 & = & \alpha\intop\delta\boldsymbol{\omega}_{\;b}^{a}\mathbf{D}\left(\mathbf{e}^{e\cdots f}\mathbf{f}_{c\cdots d}e_{\qquad ae\cdots f}^{bc\cdots d}\right)\\
 & = & \alpha\intop\delta\boldsymbol{\omega}_{\;b}^{a}\left(\mathbf{T}^{e}\mathbf{e}^{e_{1}\cdots f}\mathbf{f}_{cc_{1}\cdots d}e_{\qquad aee_{1}\cdots f}^{bcc_{1}\cdots d}+\left(-1\right)^{n-1}\mathbf{e}^{ee_{1}\cdots f}\mathbf{S}_{c}\mathbf{f}_{c_{1}\cdots d}e_{\qquad aee_{1}\cdots f}^{bcc_{1}\cdots d}\right)\\
 & = & \alpha\intop\left(\delta A_{\;\;bc}^{a}\mathbf{e}^{c}+\delta B_{\;\;\;b}^{a\quad c}\mathbf{f}_{c}\right)\left(\mathbf{T}^{e}\mathbf{e}^{e_{1}\cdots f}\mathbf{f}_{cc_{1}\cdots d}e_{\qquad aee_{1}\cdots f}^{bcc_{1}\cdots d}+\left(-1\right)^{n-1}\mathbf{S}_{c}\mathbf{e}^{ee_{1}\cdots f}\mathbf{f}_{c_{1}\cdots d}e_{\qquad aee_{1}\cdots f}^{bcc_{1}\cdots d}\right)
\end{eqnarray*}
The covariant derivatives of the basis forms divide naturally into
two tensors, the torsion, $\mathbf{T}^{a}=\mathbf{D}\mathbf{e}^{a}$
and the \emph{co-torsion}, $\mathbf{S}_{a}=\mathbf{D}\mathbf{f}_{a}$.
We show below that if both of these vanish, the solution must be trivial
(i.e., non-gravitating) so it is important to realize when considering
torsion-free solutions that the co-torsion $\mathbf{S}_{a}$ remains
non-zero.

The variation must preserve the antisymmetry, $\eta_{bc}\eta^{ad}\boldsymbol{\omega}_{\;\;d}^{c}=-\boldsymbol{\omega}_{\;\;b}^{a}$,
of the spin connection, so , $\delta\boldsymbol{\omega}_{\;b}^{a}\Delta_{sb}^{ar}=\delta\boldsymbol{\omega}_{\;s}^{r}$,.
Therefore, the coefficients of the variation, $\delta A_{\;\;bc}^{a}$
and $\delta B_{\;\;b}^{a\quad c}$, are antisymmetric on the first
pair of indices. As a result, only the antisymmetric part of the rest
of the integrand vanishes, so we require the projection operator,
$\Delta_{db}^{ac}\equiv\frac{1}{2}\left(\delta_{d}^{a}\delta_{b}^{c}-\eta^{ac}\eta_{bd}\right)=\frac{1}{2}\eta^{an}\eta_{md}\left(\delta_{n}^{m}\delta_{b}^{c}-\delta_{n}^{c}\delta_{b}^{m}\right)$
with $\Delta_{sb}^{ar}\Delta_{nr}^{sm}=\Delta_{nb}^{am}$. This acts
to antisymmetrize $\left(\begin{array}{c}
1\\
1
\end{array}\right)$ tensors, $\Delta_{db}^{ac}T_{\;\;c}^{d}=\frac{1}{2}\eta^{an}\left(T_{nb}-T_{bn}\right)$.

With $\delta A_{\;\;bc}^{a}$ and $\delta B_{\;\;\;b}^{a\quad c}$
independent we find two equations,
\begin{eqnarray*}
0 & = & \alpha\Delta_{sb}^{ar}\mathbf{e}^{m}\left(\mathbf{T}^{e}\mathbf{e}^{e_{1}\cdots f}\mathbf{f}_{cc_{1}\cdots d}e_{\qquad aee_{1}\cdots f}^{bcc_{1}\cdots d}+\left(-1\right)^{n-1}\mathbf{e}^{ee_{1}\cdots f}\mathbf{S}_{c}\mathbf{f}_{c_{1}\cdots d}e_{\qquad aee_{1}\cdots f}^{bcc_{1}\cdots d}\right)\\
0 & = & \alpha\Delta_{sb}^{ar}\mathbf{f}_{m}\left(\mathbf{T}^{e}\mathbf{e}^{e_{1}\cdots f}\mathbf{f}_{cc_{1}\cdots d}e_{\qquad aee_{1}\cdots f}^{bcc_{1}\cdots d}+\left(-1\right)^{n-1}\mathbf{e}^{ee_{1}\cdots f}\mathbf{S}_{c}\mathbf{f}_{c_{1}\cdots d}e_{\qquad aee_{1}\cdots f}^{bcc_{1}\cdots d}\right)
\end{eqnarray*}
Next, we substitute the expansion of the torsion Eq.(\ref{Expanded torsion})
and use Eq.(\ref{Expanded curvature}) to write the co-torsion, finding
the one term of each which completes the volume form. Rearranging
the basis forms in standard order, we use the volume form replacement
of eq.(\ref{Volume form replacement}):
\begin{eqnarray*}
0 & = & \alpha\Delta_{sb}^{ar}\left(\mathbf{T}^{e}\mathbf{e}^{me_{1}\cdots f}\mathbf{f}_{cc_{1}\cdots d}e_{\qquad aee_{1}\cdots f}^{bcc_{1}\cdots d}+\left(-1\right)^{n-1}\mathbf{e}^{mee_{1}\cdots f}\mathbf{S}_{c}\mathbf{f}_{c_{1}\cdots d}e_{\qquad aee_{1}\cdots f}^{bcc_{1}\cdots d}\right)\\
 & = & \alpha\Delta_{sb}^{ar}\left(T_{\quad v}^{eu}\mathbf{f}_{u}\mathbf{e}^{v}\mathbf{e}^{me_{1}\cdots f}\mathbf{f}_{cc_{1}\cdots d}e_{\qquad aee_{1}\cdots f}^{bcc_{1}\cdots d}+\left(-1\right)^{n-1}\frac{1}{2}S_{c}^{\;\;uv}\mathbf{f}_{u}\mathbf{f}_{v}\mathbf{e}^{mee_{1}\cdots f}\mathbf{f}_{c_{1}\cdots d}e_{\qquad aee_{1}\cdots f}^{bcc_{1}\cdots d}\right)\\
 & = & \alpha\Delta_{sb}^{ar}\left(\left(-1\right)^{n}T_{\quad v}^{eu}\mathbf{e}^{vme_{1}\cdots f}\mathbf{f}_{ucc_{1}\cdots d}e_{\qquad aee_{1}\cdots f}^{bcc_{1}\cdots d}+\left(-1\right)^{n-1}\frac{1}{2}S_{c}^{\;\;uv}\mathbf{e}^{mee_{1}\cdots f}\mathbf{f}_{uvc_{1}\cdots d}e_{\qquad aee_{1}\cdots f}^{bcc_{1}\cdots d}\right)\\
 & = & \alpha\Delta_{sb}^{ar}\left(\left(-1\right)^{n}T_{\quad v}^{eu}e_{\qquad ucc_{1}\cdots d}^{vme_{1}\cdots f}e_{\qquad aee_{1}\cdots f}^{bcc_{1}\cdots d}+\left(-1\right)^{n-1}\frac{1}{2}S_{c}^{\;\;uv}e_{\qquad uvc_{1}\cdots d}^{mee_{1}\cdots f}e_{\qquad aee_{1}\cdots f}^{bcc_{1}\cdots d}\right)\boldsymbol{\Phi}
\end{eqnarray*}
Then, taking the dual to eliminate the forms and replacing the resulting
double Levi-Civita tensor using eq.(\ref{Contracted Levi-Civita tensors})
we arrive at the final field equation.
\begin{eqnarray*}
0 & = & \alpha\Delta_{sb}^{ar}\left(T_{\quad v}^{eu}e_{\qquad ucc_{1}\cdots d}^{vme_{1}\cdots f}e_{\qquad aee_{1}\cdots f}^{bcc_{1}\cdots d}-\frac{1}{2}S_{c}^{\;\;uv}e_{\qquad uvc_{1}\cdots d}^{mee_{1}\cdots f}e_{\qquad aee_{1}\cdots f}^{bcc_{1}\cdots d}\right)\\
 & = & \left(n-2\right)!\left(n-1\right)!\alpha\Delta_{sb}^{ar}\left(T_{\quad v}^{eu}\left(\delta_{a}^{v}\delta_{e}^{m}-\delta_{a}^{m}\delta_{e}^{v}\right)\delta_{u}^{b}-\frac{1}{2}S_{c}^{\;\;uv}\delta_{a}^{m}\left(\delta_{u}^{b}\delta_{v}^{c}-\delta_{v}^{b}\delta_{u}^{c}\right)\right)\\
 & = & \left(n-2\right)!\left(n-1\right)!\alpha\Delta_{sb}^{ar}\left(T_{\quad a}^{mb}-\delta_{a}^{m}T_{\quad e}^{eb}-\delta_{a}^{m}S_{c}^{\;\;bc}\right)
\end{eqnarray*}
so that
\begin{equation}
0=\alpha\Delta_{sb}^{ar}\left(T_{\quad a}^{cb}-\delta_{a}^{c}T_{\quad e}^{eb}-\delta_{a}^{c}S_{e}^{\;\;be}\right)\label{Third torsion equation}
\end{equation}

For the second equation, the same steps yield,
\begin{eqnarray*}
0 & = & \alpha\Delta_{sb}^{ar}\left(\left(-1\right)^{n}\mathbf{T}^{e}\mathbf{e}^{e_{1}\cdots f}\mathbf{f}_{mcc_{1}\cdots d}e_{\qquad aee_{1}\cdots f}^{bcc_{1}\cdots d}+\mathbf{e}^{ee_{1}\cdots f}\mathbf{S}_{c}\mathbf{f}_{mc_{1}\cdots d}e_{\qquad aee_{1}\cdots f}^{bcc_{1}\cdots d}\right)\\
 & = & \alpha\Delta_{sb}^{ar}\left(\left(-1\right)^{n}\frac{1}{2}T_{\;\;uv}^{e}\mathbf{e}^{uve_{1}\cdots f}\mathbf{f}_{mcc_{1}\cdots d}e_{\qquad aee_{1}\cdots f}^{bcc_{1}\cdots d}+S_{c\quad v}^{\;\;\;u}\mathbf{f}_{u}\mathbf{e}^{v}\mathbf{e}^{ee_{1}\cdots f}\mathbf{f}_{mc_{1}\cdots d}e_{\qquad aee_{1}\cdots f}^{bcc_{1}\cdots d}\right)\\
 & = & \alpha\Delta_{sb}^{ar}\left(\left(-1\right)^{n}\frac{1}{2}T_{\;\;uv}^{e}\mathbf{e}^{uve_{1}\cdots f}\mathbf{f}_{mcc_{1}\cdots d}e_{\qquad aee_{1}\cdots f}^{bcc_{1}\cdots d}+\left(-1\right)^{n}S_{c\quad v}^{\;\;\;u}\mathbf{e}^{vee_{1}\cdots f}\mathbf{f}_{umc_{1}\cdots d}e_{\qquad aee_{1}\cdots f}^{bcc_{1}\cdots d}\right)\\
 & = & \left(-1\right)^{n}\alpha\Delta_{sb}^{ar}\left(\frac{1}{2}T_{\;\;uv}^{e}e_{\qquad mcc_{1}\cdots d}^{uve_{1}\cdots f}e_{\qquad aee_{1}\cdots f}^{bcc_{1}\cdots d}+S_{c\quad v}^{\;\;\;u}e_{\qquad umc_{1}\cdots d}^{vee_{1}\cdots f}e_{\qquad aee_{1}\cdots f}^{bcc_{1}\cdots d}\right)\boldsymbol{\Phi}
\end{eqnarray*}
Then
\begin{eqnarray*}
0 & = & \alpha\Delta_{sb}^{ar}\left(\frac{1}{2}T_{\;\;uv}^{e}e_{\qquad mcc_{1}\cdots d}^{uve_{1}\cdots f}e_{\qquad aee_{1}\cdots f}^{bcc_{1}\cdots d}+S_{c\quad v}^{\;\;\;u}e_{\qquad umc_{1}\cdots d}^{vee_{1}\cdots f}e_{\qquad aee_{1}\cdots f}^{bcc_{1}\cdots d}\right)\\
 & = & \left(n-2\right)!\left(n-1\right)!\alpha\Delta_{sb}^{ar}\left(\frac{1}{2}T_{\;\;uv}^{e}\delta_{m}^{b}\left(\delta_{a}^{u}\delta_{e}^{v}-\delta_{a}^{v}\delta_{e}^{u}\right)+S_{c\quad v}^{\;\;\;u}\delta_{a}^{v}\left(\delta_{u}^{b}\delta_{m}^{c}-\delta_{u}^{b}\delta_{m}^{c}\right)\right)\\
 & = & \left(n-2\right)!\left(n-1\right)!\alpha\Delta_{sb}^{ar}\left(T_{\;\;ae}^{e}\delta_{m}^{b}+S_{m\quad a}^{\;\;\;b}-\delta_{m}^{b}S_{u\quad a}^{\;\;\;u}\right)
\end{eqnarray*}
so finally,
\begin{equation}
\alpha\Delta_{sb}^{ar}\left(\delta_{c}^{b}T_{\;\;ae}^{e}+S_{c\quad a}^{\;\;\;b}-\delta_{c}^{b}S_{e\quad a}^{\;\;\;e}\right)=0\label{Fourth torsion equation}
\end{equation}

\section*{Appendix C: The curvature Bianchi identity \label{sec:Appendix-B-CurvatureBianchi}}

Here we give details of the development of the curvature Bianchi identity
leading from Eq.(\ref{Curvature Bianchi eee}) to Eq.(\ref{Spacetime curvature Bianchi}).

When we combine the solution for $\boldsymbol{\Omega}_{\;\;b}^{a}$
with the spacetime part of the curvature Bianchi identity, we have
Eq.(\ref{Curvature Bianchi eee}),
\begin{eqnarray*}
D_{[e}^{\left(\omega,x\right)}\Omega_{\;\;\left|b\right|cd]}^{a}+2\left(1+\chi\right)\Delta_{\left[e\right|b}^{af}S_{f\left|cd\right]} & = & 0\\
D_{e}^{\left(\omega,x\right)}C_{\;\;bcd}^{a}+D_{c}^{\left(\omega,x\right)}C_{\;\;bde}^{a}+D_{d}^{\left(\omega,x\right)}C_{\;\;bec}^{a} & = & -2\left(1+\chi\right)\left(\Delta_{eb}^{af}S_{fcd}+\Delta_{cb}^{af}S_{fde}+\Delta_{db}^{af}S_{fec}\right)
\end{eqnarray*}
With $\frac{1}{2}\Omega_{\;\;bcd}^{a}\mathbf{e}^{c}\land\mathbf{e}^{d}=\frac{1}{2}C_{\;\;bcd}^{a}\mathbf{e}^{c}\land\mathbf{e}^{d}=\mathbf{C}_{\;\;b}^{a}$,
the covariant exterior derivative of the Weyl curvature is
\begin{eqnarray*}
\mathbf{D}_{\left(\omega,x\right)}\left(\frac{1}{2}C_{\;\;bcd}^{a}\mathbf{e}^{c}\land\mathbf{e}^{d}\right) & \equiv & \mathbf{d}_{\left(x\right)}\left(\frac{1}{2}C_{\;\;bcd}^{a}\mathbf{e}^{c}\land\mathbf{e}^{d}\right)+\left(\frac{1}{2}C_{\;\;bef}^{c}\mathbf{e}^{e}\land\mathbf{e}^{f}\right)\land\boldsymbol{\omega}_{\;c}^{a}-\left(\frac{1}{2}C_{\;\;cef}^{a}\mathbf{e}^{e}\land\mathbf{e}^{f}\right)\land\boldsymbol{\omega}_{\;b}^{c}\\
 & = & \mathbf{D}_{\left(\alpha,x\right)}\left(\frac{1}{2}C_{\;\;bcd}^{a}\mathbf{e}^{c}\land\mathbf{e}^{d}\right)+\left(\frac{1}{2}C_{\;\;bef}^{c}\mathbf{e}^{e}\land\mathbf{e}^{f}\right)\land\left(\boldsymbol{\alpha}_{\;c}^{a}+\boldsymbol{\beta}_{\;c}^{a}\right)-\left(\frac{1}{2}C_{\;\;cef}^{a}\mathbf{e}^{e}\land\mathbf{e}^{f}\right)\land\left(\boldsymbol{\alpha}_{\;b}^{c}+\boldsymbol{\beta}_{\;b}^{c}\right)\\
 & = & \mathbf{D}_{\left(\alpha,x\right)}\mathbf{C}_{\;\;b}^{a}+\mathbf{C}_{\;\;b}^{a}\land\boldsymbol{\beta}_{\;c}^{a}-\mathbf{C}_{\;\;c}^{a}\land\boldsymbol{\beta}_{\;b}^{c}
\end{eqnarray*}
But $\mathbf{C}_{\;\;b}^{a}$ is the usual traceless part of the Riemann
curvature, which satisfies
\begin{eqnarray*}
0 & \equiv & \mathbf{D}_{\left(\alpha,x\right)}\mathbf{R}_{\;\;b}^{a}\\
 & = & \mathbf{D}_{\left(\alpha,x\right)}\mathbf{C}_{\;\;b}^{a}-2\Delta_{db}^{ae}\mathbf{D}_{\left(\alpha,x\right)}\boldsymbol{\mathcal{R}}_{e}\mathbf{e}^{d}
\end{eqnarray*}

Restoring the basis, the Bianchi identity is
\begin{eqnarray*}
0 & = & \mathbf{D}_{\left(\omega,x\right)}\mathbf{C}_{\;\;b}^{a}+2\left(1+\chi\right)\Delta_{cb}^{ad}\mathbf{S}_{d}^{\left(ee\right)}\land\mathbf{e}^{c}\\
 & = & \mathbf{D}_{\left(\alpha,x\right)}\mathbf{C}_{\;\;b}^{a}+\mathbf{C}_{\;\;b}^{c}\land\boldsymbol{\beta}_{\;c}^{a}-\mathbf{C}_{\;\;c}^{a}\land\boldsymbol{\beta}_{\;b}^{c}+2\left(1+\chi\right)\Delta_{cb}^{ad}\mathbf{S}_{d}^{\left(ee\right)}\land\mathbf{e}^{c}\\
 & = & 2\Delta_{db}^{ae}\mathbf{D}_{\left(\alpha,x\right)}\boldsymbol{\mathcal{R}}_{e}\mathbf{e}^{d}+\mathbf{C}_{\;\;b}^{c}\land\left(-2\Delta_{dc}^{ae}W_{e}\mathbf{e}^{d}\right)-\mathbf{C}_{\;\;c}^{a}\land\left(-2\Delta_{db}^{ce}W_{e}\mathbf{e}^{d}\right)+2\left(1+\chi\right)\Delta_{cb}^{ad}\mathbf{S}_{d}^{\left(ee\right)}\land\mathbf{e}^{c}\\
 & = & 2\Delta_{db}^{ae}\mathbf{D}_{\left(\alpha,x\right)}\boldsymbol{\mathcal{R}}_{e}\mathbf{e}^{d}-2\Delta_{dc}^{ae}\mathbf{C}_{\;\;b}^{c}\land W_{e}\mathbf{e}^{d}+2\Delta_{db}^{ce}\mathbf{C}_{\;\;c}^{a}\land W_{e}\mathbf{e}^{d}+2\left(1+\chi\right)\Delta_{db}^{ae}\mathbf{S}_{e}^{\left(ee\right)}\land\mathbf{e}^{d}\\
 & = & 2\Delta_{df}^{ge}\left(\delta_{b}^{f}\delta_{g}^{a}\mathbf{D}_{\left(\alpha,x\right)}\boldsymbol{\mathcal{R}}_{e}-\delta_{g}^{a}W_{e}\mathbf{C}_{\;\;b}^{f}+\delta_{g}^{c}\delta_{b}^{f}W_{e}\mathbf{C}_{\;\;c}^{a}+\left(1+\chi\right)\delta_{g}^{a}\delta_{b}^{f}\mathbf{S}_{e}^{\left(ee\right)}\right)\land\mathbf{e}^{d}
\end{eqnarray*}
Now expand $\Delta$ projections and carry out the simplifications
\begin{eqnarray*}
0 & = & 2\Delta_{db}^{ae}\mathbf{D}_{\left(\alpha,x\right)}\boldsymbol{\mathcal{R}}_{e}\mathbf{e}^{d}-2\Delta_{dc}^{ae}\mathbf{C}_{\;\;b}^{c}\land W_{e}\mathbf{e}^{d}+2\Delta_{db}^{ce}\mathbf{C}_{\;\;c}^{a}\land W_{e}\mathbf{e}^{d}+2\left(1+\chi\right)\Delta_{db}^{ae}\mathbf{S}_{e}^{\left(ee\right)}\land\mathbf{e}^{d}\\
 & = & \left(\delta_{d}^{a}\delta_{b}^{e}-\eta^{ae}\eta_{bd}\right)\mathbf{D}_{\left(\alpha,x\right)}\boldsymbol{\mathcal{R}}_{e}\mathbf{e}^{d}-\left(\delta_{d}^{a}\delta_{c}^{e}-\eta^{ae}\eta_{cd}\right)\mathbf{C}_{\;\;b}^{c}\land W_{e}\mathbf{e}^{d}+\left(\delta_{d}^{c}\delta_{b}^{e}-\eta^{ce}\eta_{bd}\right)\mathbf{C}_{\;\;c}^{a}\land W_{e}\mathbf{e}^{d}+\left(1+\chi\right)\left(\delta_{d}^{a}\delta_{b}^{e}-\eta^{ae}\eta_{bd}\right)\mathbf{S}_{e}^{\left(ee\right)}\land\mathbf{e}^{d}\\
 & = & \mathbf{D}_{\left(\alpha,x\right)}\boldsymbol{\mathcal{R}}_{b}\land\mathbf{e}^{a}-\eta_{bd}\mathbf{D}_{\left(\alpha,x\right)}\boldsymbol{\mathcal{R}}^{a}\land\mathbf{e}^{d}-W_{e}\mathbf{C}_{\;\;b}^{e}\land\mathbf{e}^{a}+\eta^{ae}W^{a}\mathbf{C}_{db}\land\mathbf{e}^{d}\\
 &  & +W_{b}\mathbf{C}_{\;\;c}^{a}\land\mathbf{e}^{c}+\eta^{ca}\eta_{bd}W_{e}\mathbf{C}_{\;\;c}^{e}\land\mathbf{e}^{d}+2\left(1+\chi\right)\mathbf{S}_{b}^{\left(ee\right)}\land\mathbf{e}^{a}-\left(1+\chi\right)\eta^{ae}\eta_{bd}\mathbf{S}_{e}^{\left(ee\right)}\land\mathbf{e}^{d}\\
 & = & \left(\mathbf{D}_{\left(\alpha,x\right)}\boldsymbol{\mathcal{R}}_{b}-W_{e}\mathbf{C}_{\;\;b}^{e}+2\left(1+\chi\right)\mathbf{S}_{b}^{\left(ee\right)}\right)\land\mathbf{e}^{a}-\eta_{bd}\eta^{ac}\left(\mathbf{D}_{\left(\alpha,x\right)}\boldsymbol{\mathcal{R}}_{c}-W_{e}\mathbf{C}_{\;\;c}^{e}+\left(1+\chi\right)\mathbf{S}_{c}^{\left(ee\right)}\right)\land\mathbf{e}^{d}\\
 & = & \delta_{d}^{a}\delta_{b}^{c}\left(\mathbf{D}_{\left(\alpha,x\right)}\boldsymbol{\mathcal{R}}_{c}-W_{e}\mathbf{C}_{\;\;c}^{e}+2\left(1+\chi\right)\mathbf{S}_{c}^{\left(ee\right)}\right)\land\mathbf{e}^{d}-\eta_{bd}\eta^{ac}\left(\mathbf{D}_{\left(\alpha,x\right)}\boldsymbol{\mathcal{R}}_{c}-W_{e}\mathbf{C}_{\;\;c}^{e}+\left(1+\chi\right)\mathbf{S}_{c}^{\left(ee\right)}\right)\land\mathbf{e}^{d}
\end{eqnarray*}
since the first Bianchi gives $\mathbf{C}_{db}\land\mathbf{e}^{d}=0$.
Then we have 
\begin{eqnarray}
0 & = & \Delta_{db}^{ac}\left(\mathbf{D}_{\left(\alpha,x\right)}\boldsymbol{\mathcal{R}}_{c}-W_{e}\mathbf{C}_{\;\;c}^{e}+2\left(1+\chi\right)\mathbf{S}_{c}^{\left(ee\right)}\right)\land\mathbf{e}^{d}\label{Spacetime curvature Bianchi-1}
\end{eqnarray}

\subsection*{Resolving the projection}

We now show that we can eliminate the $\Delta$-projection and the
wedge product with the solder form $\mathbf{e}^{d}$. Extracting the
basis forms, we antisymmetrize,

\begin{eqnarray*}
0 & = & \Delta_{db}^{ac}\left(D_{f}^{\left(\alpha,x\right)}\mathcal{R}_{cg}-\frac{1}{2}W_{e}C_{\;\;cfg}^{e}+\left(1+\chi\right)S_{cfg}^{\left(ee\right)}\right)\mathbf{e}^{f}\land\mathbf{e}^{g}\land\mathbf{e}^{d}\\
0 & = & \Delta_{db}^{ac}\left(D_{f}^{\left(\alpha,x\right)}\mathcal{R}_{cg}-W_{e}C_{\;\;cfg}^{e}+2\left(1+\chi\right)S_{cfg}^{\left(ee\right)}\right)+\Delta_{fb}^{ac}\left(D_{g}^{\left(\alpha,x\right)}\mathcal{R}_{cd}-W_{e}C_{\;\;cgd}^{e}+2\left(1+\chi\right)S_{cgd}^{\left(ee\right)}\right)\\
 &  & +\Delta_{gb}^{ac}\left(D_{d}^{\left(\alpha,x\right)}\mathcal{R}_{cf}-W_{e}C_{\;\;cdf}^{e}+2\left(1+\chi\right)S_{cdf}^{\left(ee\right)}\right)-\Delta_{db}^{ac}D_{g}^{\left(\alpha,x\right)}\mathcal{R}_{cf}-\Delta_{fb}^{ac}D_{d}^{\left(\alpha,x\right)}\mathcal{R}_{cg}-\Delta_{gb}^{ac}D_{f}^{\left(\alpha,x\right)}\mathcal{R}_{cd}
\end{eqnarray*}
Now, contract $ad$,
\begin{eqnarray*}
0 & = & \left(n-1\right)\left(D_{f}^{\left(\alpha,x\right)}\mathcal{R}_{bg}-W_{e}C_{\;\;bfg}^{e}+2\left(1+\chi\right)S_{bfg}^{\left(ee\right)}\right)+2\Delta_{fb}^{ac}\left(D_{g}^{\left(\alpha,x\right)}\mathcal{R}_{ca}-W_{e}C_{\;\;cga}^{e}+2\left(1+\chi\right)S_{cga}^{\left(ee\right)}\right)\\
 &  & +2\Delta_{gb}^{ac}\left(D_{a}^{\left(\alpha,x\right)}\mathcal{R}_{cf}-W_{e}C_{\;\;caf}^{e}+2\left(1+\chi\right)S_{caf}^{\left(ee\right)}\right)-\left(n-1\right)D_{g}^{\left(\alpha,x\right)}\mathcal{R}_{bf}-2\Delta_{fb}^{ac}D_{a}^{\left(\alpha,x\right)}\mathcal{R}_{cg}-2\Delta_{gb}^{ac}D_{f}^{\left(\alpha,x\right)}\mathcal{R}_{ca}\\
 & = & \left(n-1\right)D_{f}^{\left(\alpha,x\right)}\mathcal{R}_{bg}-\left(n-1\right)W_{e}C_{\;\;bfg}^{e}+2\left(n-1\right)\left(1+\chi\right)S_{bfg}^{\left(ee\right)}\\
 &  & +\left(D_{g}^{\left(\alpha,x\right)}\mathcal{R}_{bf}-W_{e}C_{\;\;bgf}^{e}+2\left(1+\chi\right)S_{bgf}^{\left(ee\right)}\right)-\eta^{ac}\eta_{bf}D_{g}^{\left(\alpha,x\right)}\mathcal{R}_{ca}-2\left(1+\chi\right)\eta^{ac}\eta_{bf}S_{cga}^{\left(ee\right)}\\
 &  & +D_{g}^{\left(\alpha,x\right)}\mathcal{R}_{bf}-W_{e}C_{\;\;bgf}^{e}+2\left(1+\chi\right)S_{bgf}^{\left(ee\right)}-\eta^{ac}\eta_{bg}D_{a}^{\left(\alpha,x\right)}\mathcal{R}_{cf}-2\left(1+\chi\right)\eta^{ac}\eta_{bg}S_{caf}^{\left(ee\right)}\\
 &  & -\left(n-1\right)D_{g}^{\left(\alpha,x\right)}\mathcal{R}_{bf}-D_{f}^{\left(\alpha,x\right)}\mathcal{R}_{bg}+\eta_{bf}\eta^{ac}D_{a}^{\left(\alpha,x\right)}\mathcal{R}_{cg}-D_{f}^{\left(\alpha,x\right)}\mathcal{R}_{bg}+\eta^{ac}\eta_{bg}D_{f}^{\left(\alpha,x\right)}\mathcal{R}_{ca}\\
 & = & -\left(n-3\right)W_{e}C_{\;\;bfg}^{e}+2\left(1+\chi\right)\left(\left(n-3\right)S_{bfg}^{\left(ee\right)}-\eta_{bf}\eta^{ac}S_{cga}^{\left(ee\right)}+\eta_{bg}\eta^{ac}S_{cfa}^{\left(ee\right)}\right)\\
 &  & +\left(n-3\right)D_{f}^{\left(\alpha,x\right)}\mathcal{R}_{bg}-\left(n-3\right)D_{g}^{\left(\alpha,x\right)}\mathcal{R}_{bf}+\eta_{bf}\eta^{ac}D_{a}^{\left(\alpha,x\right)}\mathcal{R}_{cg}-\eta_{bg}\eta^{ac}D_{a}^{\left(\alpha,x\right)}\mathcal{R}_{cf}+\eta_{bg}D_{f}^{\left(\alpha,x\right)}\mathcal{R}-\eta_{bf}D_{g}^{\left(\alpha,x\right)}\mathcal{R}
\end{eqnarray*}
Write this as
\begin{eqnarray*}
\left(n-3\right)\left(D_{g}^{\left(\alpha,x\right)}\mathcal{R}_{bf}-D_{f}^{\left(\alpha,x\right)}\mathcal{R}_{bg}+W_{e}C_{\;\;bfg}^{e}\right) & = & 2\left(1+\chi\right)\left(\left(n-3\right)S_{bfg}^{\left(ee\right)}-\eta_{bf}\eta^{ac}S_{cga}^{\left(ee\right)}+\eta_{bg}\eta^{ac}S_{cfa}^{\left(ee\right)}\right)\\
 &  & +\eta_{bf}\eta^{ac}D_{a}^{\left(\alpha,x\right)}\mathcal{R}_{cg}-\eta_{bg}\eta^{ac}D_{a}^{\left(\alpha,x\right)}\mathcal{R}_{cf}+\eta_{bg}D_{f}^{\left(\alpha,x\right)}\mathcal{R}-\eta_{bf}D_{g}^{\left(\alpha,x\right)}\mathcal{R}
\end{eqnarray*}
If we trace $bg$,
\begin{eqnarray*}
\left(n-3\right)\eta^{bg}\left(D_{g}^{\left(\alpha,x\right)}\mathcal{R}_{bf}-D_{f}^{\left(\alpha,x\right)}\mathcal{R}_{bg}\right) & = & 2\left(1+\chi\right)\left(\left(n-3\right)\eta^{bg}S_{bfg}^{\left(ee\right)}-\eta^{ac}S_{cfa}^{\left(ee\right)}+n\eta^{ac}S_{cfa}^{\left(ee\right)}\right)\\
 &  & +\eta^{ac}D_{a}^{\left(\alpha,x\right)}\mathcal{R}_{cf}-n\eta^{ac}D_{a}^{\left(\alpha,x\right)}\mathcal{R}_{cf}+nD_{f}^{\left(\alpha,x\right)}\mathcal{R}-D_{f}^{\left(\alpha,x\right)}\mathcal{R}\\
4\left(1+\chi\right)\left(n-2\right)\eta^{bg}S_{bfg}^{\left(ee\right)} & = & 2\left(n-2\right)\eta^{ac}D_{a}^{\left(\alpha,x\right)}\mathcal{R}_{cf}-2\left(n-2\right)D_{f}^{\left(\alpha,x\right)}\mathcal{R}\\
2\left(1+\chi\right)\eta^{bg}S_{bfg}^{\left(ee\right)} & = & \eta^{ac}D_{a}^{\left(\alpha,x\right)}\mathcal{R}_{cf}-D_{f}^{\left(\alpha,x\right)}\mathcal{R}
\end{eqnarray*}
Substituting back into the original,
\begin{eqnarray*}
\left(n-3\right)\left(D_{g}^{\left(\alpha,x\right)}\mathcal{R}_{bf}-D_{f}^{\left(\alpha,x\right)}\mathcal{R}_{bg}+W_{e}C_{\;\;bfg}^{e}\right) & = & 2\left(1+\chi\right)\left(\left(n-3\right)S_{bfg}^{\left(ee\right)}-\eta_{bf}\eta^{ac}S_{cga}^{\left(ee\right)}+\eta_{bg}\eta^{ac}S_{cfa}^{\left(ee\right)}\right)\\
 &  & -\eta_{bg}\left(\eta^{ac}D_{a}^{\left(\alpha,x\right)}\mathcal{R}_{cf}-D_{f}^{\left(\alpha,x\right)}\mathcal{R}\right)+\eta_{bf}\left(\eta^{ac}D_{a}^{\left(\alpha,x\right)}\mathcal{R}_{cg}-D_{g}^{\left(\alpha,x\right)}\mathcal{R}\right)\\
 & = & 2\left(1+\chi\right)\left(\left(n-3\right)S_{bfg}^{\left(ee\right)}-\eta_{bf}\eta^{ac}S_{cga}^{\left(ee\right)}+\eta_{bg}\eta^{ac}S_{cfa}^{\left(ee\right)}\right)\\
 &  & -\eta_{bg}2\left(1+\chi\right)\eta^{bg}S_{bfg}^{\left(ee\right)}+\eta_{bf}2\left(1+\chi\right)\eta^{bc}S_{bgc}^{\left(ee\right)}\\
 & = & 2\left(1+\chi\right)\left(n-3\right)S_{bfg}^{\left(ee\right)}
\end{eqnarray*}
Therefore, in dimensions greater than 3,
\begin{eqnarray*}
2\left(1+\chi\right)S_{bfg}^{\left(ee\right)} & = & D_{g}^{\left(\alpha,x\right)}\mathcal{R}_{bf}-D_{f}^{\left(\alpha,x\right)}\mathcal{R}_{bg}+W_{e}C_{\;\;bfg}^{e}
\end{eqnarray*}
Restoring two basis forms, we may write this as Eq.(\ref{Spacetime curvature Bianchi}),
\begin{eqnarray*}
\mathbf{D}^{\left(\alpha,x\right)}\boldsymbol{\mathcal{R}}_{b}-W_{e}\mathbf{C}_{\;\;b}^{e}+2\left(1+\chi\right)\mathbf{S}_{b}^{\left(ee\right)} & = & 0
\end{eqnarray*}
which solves the full Bianchi relation, Eq.(\ref{Spacetime curvature Bianchi-1})
as desired. We note that this proves:
\begin{description}
\item [{Theorem:}] In any torsion-free biconformal space with integrable
Weyl vector, $W_{\alpha}=\partial_{\alpha}\phi$, and $1+\chi\neq0$,
the spacetime co-torsion is the obstruction to conformal Ricci flatness.
\end{description}

\end{document}